\pdfoutput=1

\documentclass[annual]{acmsiggraph}

\usepackage{subfig}
\usepackage{algorithm}
\usepackage[noend]{algpseudocode}

\TOGvolume{0}
\TOGnumber{0}
\TOGarticleDOI{1111111.2222222}

\TOGprojectURL{}
\TOGvideoURL{}
\TOGdataURL{}
\TOGcodeURL{}

\title{Partial Procedural Geometric Model Fitting for Point Clouds}

\author{
Zongliang Zhang$^{1}$\thanks{e-mail: zhangzongliang@stu.xmu.edu.cn} \qquad Jonathan Li$^1$\thanks{e-mail: junli@xmu.edu.cn (Corresponding author)} $^{,2}$ \qquad Yulan Guo$^3$ \qquad Yangbin Lin$^1$\qquad Ming Cheng$^1$ \qquad Cheng Wang$^1$\\
$^1$Fujian Key Laboratory of Sensing and Computing for Smart Cities, Xiamen Key Laboratory of \\Geospatial Sensing and Computing, Innovation Center of Sensing and Computing for Smart Cities, \\School of Information Science and Engineering, Xiamen University, Xiamen, Fujian 361005, China\\$^2$Mobile Sensing and Geodata Science Lab, Department of Geography and Environmental Management, \\University of Waterloo, Waterloo, ON N2L 3G1, Canada\\$^3$College of Electronic Science and Engineering, \\National University of Defense Technology, Changsha, Hunan 410073, China
}
\pdfauthor{Zongliang Zhang}


\begin{document}


 \teaser{
\subfloat[]{\label{fig:xiangan-evolve1}\includegraphics[height=1.0in]{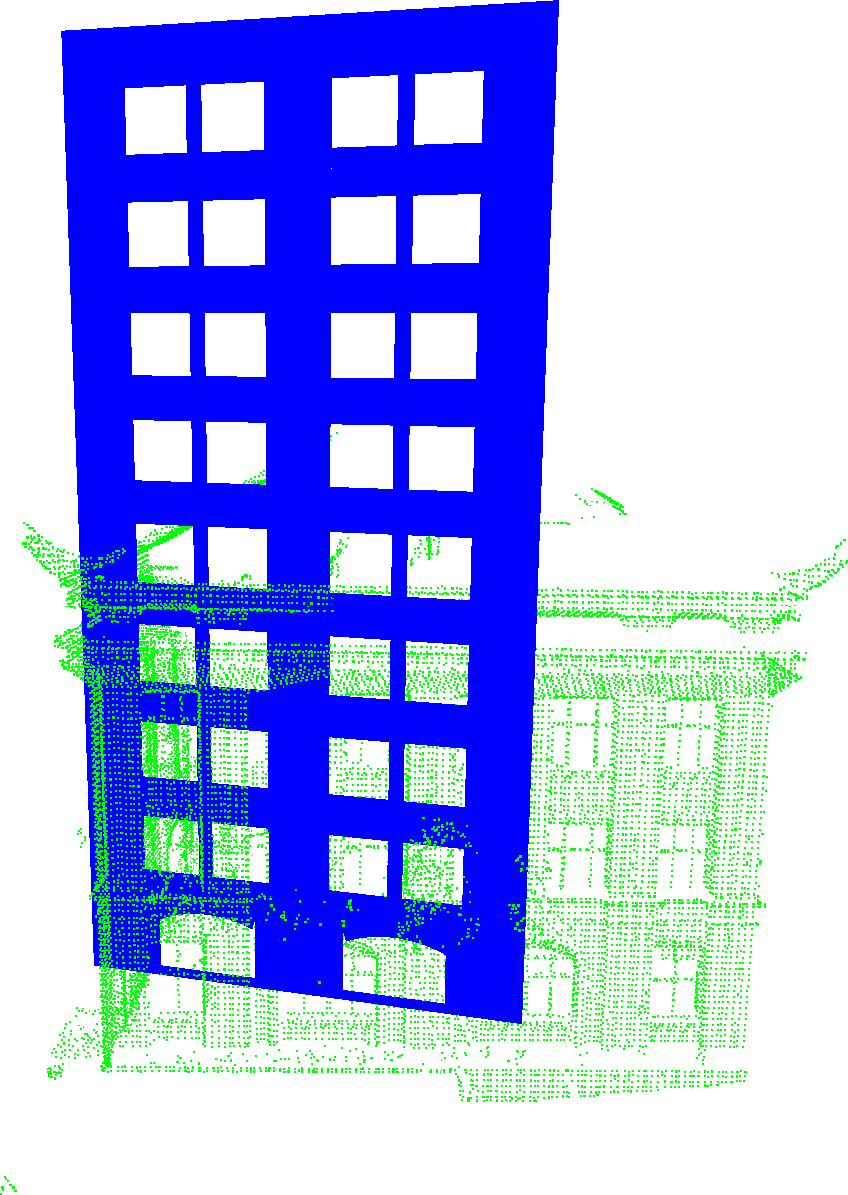}}
\hfill
\subfloat[]{\label{fig:xiangan-evolve2}\includegraphics[height=1.0in]{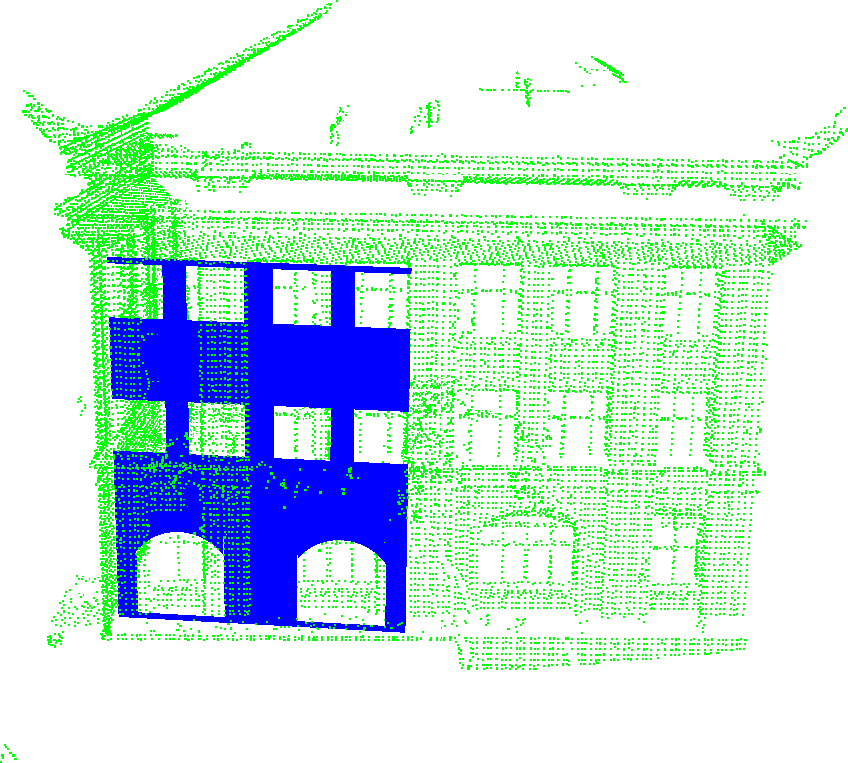}}
\hfill
\subfloat[]{\label{fig:xiangan-evolve3}\includegraphics[height=1.0in]{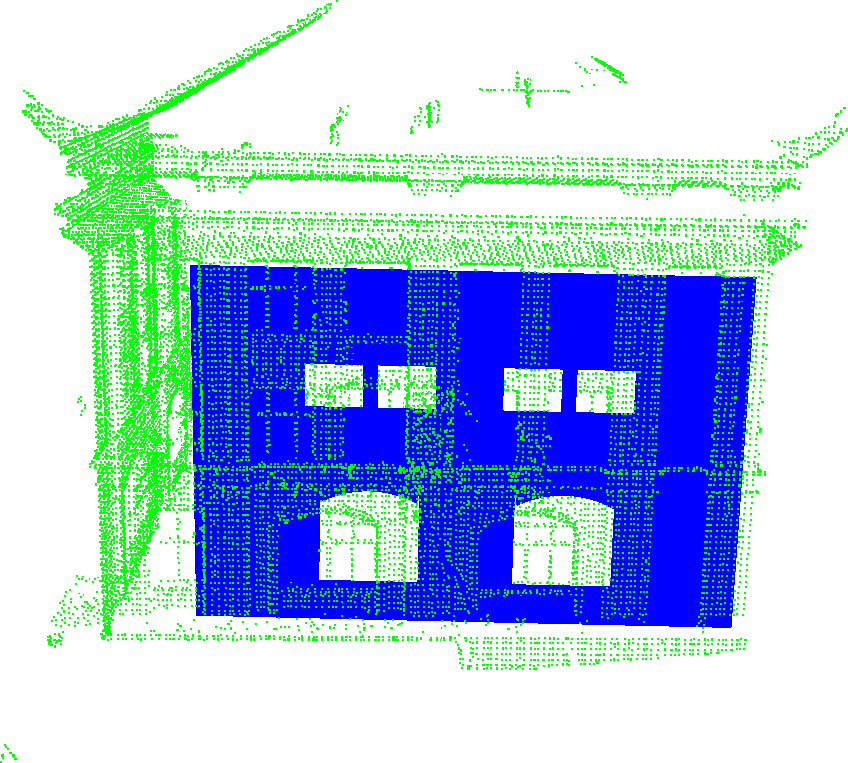}}
\hfill
\subfloat[]{\label{fig:xiangan-evolve4}\includegraphics[height=1.0in]{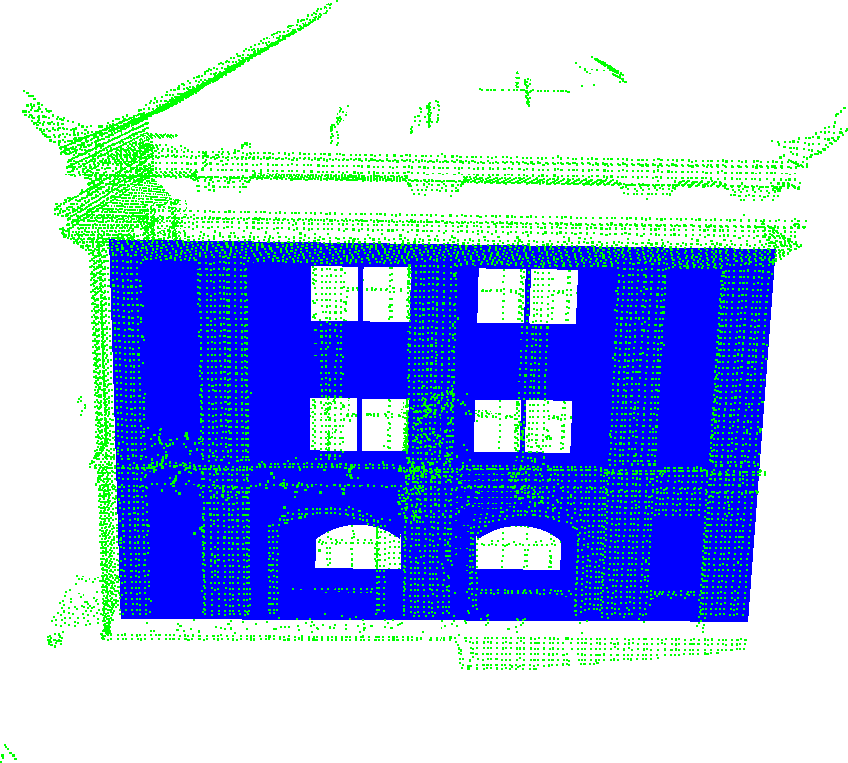}}
\hfill
\subfloat[]{\label{fig:xiangan-evolve5}\includegraphics[height=1.0in]{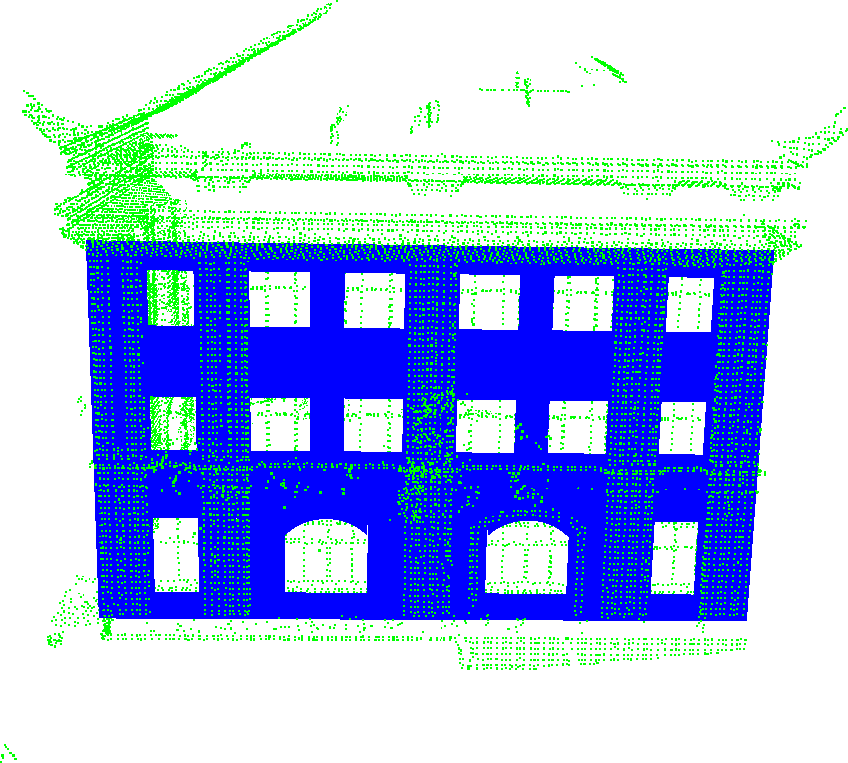}}
\hfill
\subfloat[]{\label{fig:xiangan-evolve6}\includegraphics[height=1.0in]{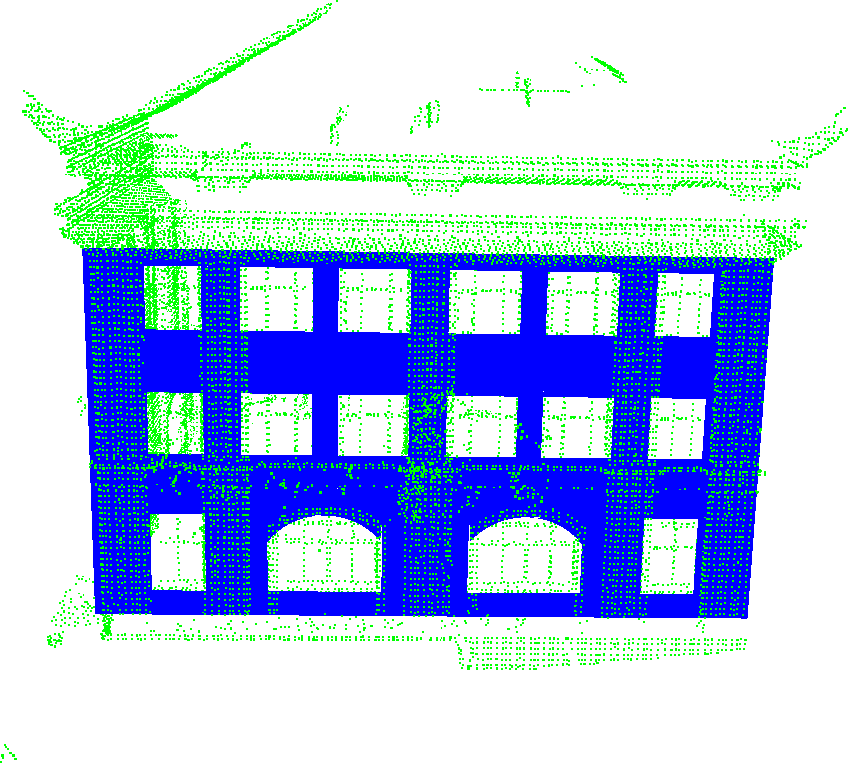}}
   \caption{Model evolution of fitting a procedural geometric model $\mathcal{M}_8$ to a point cloud $Q_{11}$ (green) (Section \ref{sec:laser}). From left to right: fitted models at different iterations (time in seconds): 0 (0), 683 (979.4786), 1782 (2458.3916), 5566 (5968.5444), 16666 (11487.1016) and 43299 (23074.8750), respectively.}
   \label{fig:xianganevolution}
 }

\maketitle


\begin{abstract}

Geometric model fitting is a fundamental task in computer graphics and computer vision. However, most geometric model fitting methods are unable to fit an arbitrary geometric model (e.g. a surface with holes) to incomplete data, due to that the similarity metrics used in these methods are unable to measure the rigid partial similarity between arbitrary models. This paper hence proposes a novel rigid geometric similarity metric, which is able to measure both the full similarity and the partial similarity between arbitrary geometric models. The proposed metric enables us to perform partial procedural geometric model fitting (PPGMF).

The task of PPGMF is to search a procedural geometric \textit{model space} for the model rigidly similar to a \textit{query} of non-complete point set. Models in the procedural \textit{model space} are generated according to a set of parametric modeling rules. A typical \textit{query} is a point cloud. PPGMF is very useful as it can be used to fit arbitrary geometric models to non-complete (incomplete, over-complete or hybrid-complete) point cloud data. For example, most laser scanning data is non-complete due to occlusion. Our PPGMF method uses Markov chain Monte Carlo technique to optimize the proposed similarity metric over the model space. To accelerate the optimization process, the method also employs a novel coarse-to-fine model dividing strategy to reject dissimilar models in advance. Our method has been demonstrated on a variety of geometric models and non-complete data. Experimental results show that the PPGMF method based on the proposed metric is able to fit non-complete data, while the method based on other metrics is unable. It is also shown that our method can be accelerated by several times via early rejection.

\end{abstract}


\begin{CRcatlist}
  \CRcat{I.3.5}{Computer Graphics}{Computational Geometry and Object Modeling}{Geometric algorithms, languages, and systems;}
\end{CRcatlist}


\keywordlist






\section{Introduction}

A geometric model is a continuous point set (e.g. a surface). A geometric model space is a set of geometric models. A procedural model space is defined by a set of parametric modeling rules \cite{dang2015interactive}. Retrieving desired model from procedural space is called inverse procedural modeling (IPM). As an important but challenging problem in computer graphics and computer vision, IPM has been actively studied in recent years \cite{musialski2013survey}. PPGMF is a special case of IPM, as it searches a procedural geometric model space for the model which is rigidly similar to a query of non-complete point set. PPGMF can be used in a number of applications including pattern recognition, shape matching, geometric modeling and point cloud reconstruction.

The task of basic geometric model fitting (BGMF) is to rigidly fit basic geometric models to the input geometric data (i.e. a point set).  The most famous BGMF method is RANSAC \cite{fischler1981random}, which can be used to fit several basic geometric models such as planes and spheres. Following RANSAC, a lot of BGMF methods have been proposed \cite{isack2011energy-based}. However, BGMF methods cannot be used to fit arbitrary models. To fit arbitrary models, the rigid geometric similarity between arbitrary models has to be calculated. A rigid geometric similarity metric should ensure that a model is most similar to itself than any other models. To the best of our knowledge, symmetric Hausdorff distance (SHD) is the only rigid geometric similarity metric. However, it is time consuming to calculate SHD, making it difficult to perform arbitrary model fitting. This paper hence proposes a novel efficient rigid geometric similarity metric to perform arbitrary model fitting.

Given the similarity metric, the remaining task of arbitrary model fitting is to optimize over a given arbitrary model space. An arbitrary model space is usually defined by a procedural modeling approach. That is, a set of procedural modeling rules are used to generate models \cite{smelik2014survey}. Retrieving desired models from procedural model space is called IPM. Many IPM methods have been proposed using different retrieval criteria such as indicator satisfying \cite{vanegas2012inverse} and image resembling \cite{teboul2013parsing} \cite{lake2015human}. In this paper, we focus on geometric criterion based IPM (GIPM) methods, which aim at fitting procedural geometric models to the input geometric data. As one geometric criterion, voxel difference (VD) has been investigated in GIPM methods \cite{talton_metropolis_2011} \cite{ritchie2015controlling}. However, VD is an approximate geometric similarity criterion. It is obvious that part information is lost by voxelization, as the voxelization resolution cannot be as small as $0$. Theoretically, our metric does not rely on a resolution. In other words, VD-based GIPM method cannot be used for rigid model fitting, which requires the calculation of rigid geometric similarity. To the best of our knowledge, our PPGMF method is the first GIPM method which can be used for arbitrary rigid model fitting. 

PPGMF aims at rigidly fitting procedural geometric models to a query of non-complete (incomplete, over-complete or hybrid-complete) geometric object. ``Hybrid-complete'' means hybrid incomplete and over-complete. Typical queries are point clouds. There are two key processes in a PPGMF method. The first key process is to calculate the rigid geometric similarity between a model and the query. We have found that a SHD or VD based method is unable to fit non-complete point clouds. We hence propose a novel partial rigid geometric similarity metric to fit non-complete data. The second key process is to optimize over the procedural model space. Optimizing over the procedural space is challenging due to the hierarchical and recursive nature of modeling rules. Markov chain Monte Carlo (MCMC) technique is used to perform optimization. Although the similarity calculation based on our metric is faster than SHD, it is still too slow for practical applications. We hence propose a novel coarse-to-fine model dividing strategy to reject dissimilar models in advance to accelerate this optimization.

PPGMF is of important significance, and is very useful as it can be used to fit arbitrary geometric models to non-complete data. For example, most laser scanning data is non-complete (cluttered) due to occlusion \cite{guo20143d}. We have tested our metric and PPGMF method on a variety of geometric models and non-complete data. Experimental results show that the PPGMF method based on the proposed metric is able to fit non-complete data, while the SHD or VD based method fails. Experimental results also show that our method can be accelerated by several times using early rejection. 

In summary, our contributions are: (1) A novel rigid geometric similarity metric is proposed to measure the similarity between two geometric models. (2) An effective method is proposed to rigidly fit arbitrary geometric models to non-complete data. (3) A coarse-to-fine geometric model dividing strategy is proposed to reject dissimilar models in advance for the acceleration of PPGMF.

The rest of this paper is organized as follows. Sections \ref{sec:related} and \ref{sec:overview} present related work and  the overview of our method, respectively. Section \ref{sec:similarity} introduces our rigid geometric similarity metric. Section \ref{sec:optimize} presents the MCMC-based optimization approach and our coarse-to-fine model dividing strategy. Sections \ref{sec:result} and \ref{sec:conclude} present experimental results and conclusion, respectively.

\section{Related Work}\label{sec:related}

Most GIPM methods take either a particular type of geometric model or geometric data as input. BGMF methods such as \cite{fischler1981random} \cite{isack2011energy-based} work on basic geometric models. \cite{debevec1996modeling} \cite{mathias2011procedural} rely on image information to achieve IPM while our work does not rely on images. \cite{ullrich2008semantic} assumes the number of model parameters is fixed. \cite{bokeloh2010a} takes symmetry as an assumption. \cite{wan2012grammar} is limited to facade point clouds and split grammar. \cite{vanegas2012automatic} takes Manhattan-World as an assumption. \cite{boulch2013semantizing} is limited to constrained attribute grammar. \cite{toshev2010detecting} \cite{lafarge2010structural} \cite{huang2013generative} work well on airborne laser scanning data, however, it is hard to extend them to other types of data. \cite{stava2014inverse} works on tree models. \cite{demir2015procedural} relys on semi-automatic segmentation operations. Our method is full-automatic and makes no assumption about the type of input geometric model and geometric data. Consequently, similar to \cite{talton_metropolis_2011} \cite{ritchie2015controlling}, our method can be used for general-purpose GIPM.

It is worth noting that there are a lot of other types of geometric data reconstruction methods but not GIPM methods such as \cite{pu2009knowledge} \cite{zheng2010non-local} \cite{nan2010smartboxes} \cite{li2011globfit} \cite{lafarge2012creating} \cite{poullis2013a} \cite{lin2013semantic} \cite{lin2015line} \cite{monszpart2015rapter} \cite{wang2016shape}. A GIPM method results in procedural models, while other types of methods usually result in mesh models. Procedural model is more powerful than mesh model \cite{weissenberg2013there}, as it perceives the abstract causal structure of input data \cite{lake2015human}.

\section{Method Overview}\label{sec:overview}

\begin{figure}[ht!]
\begin{center}
		\includegraphics[width=0.9\columnwidth]{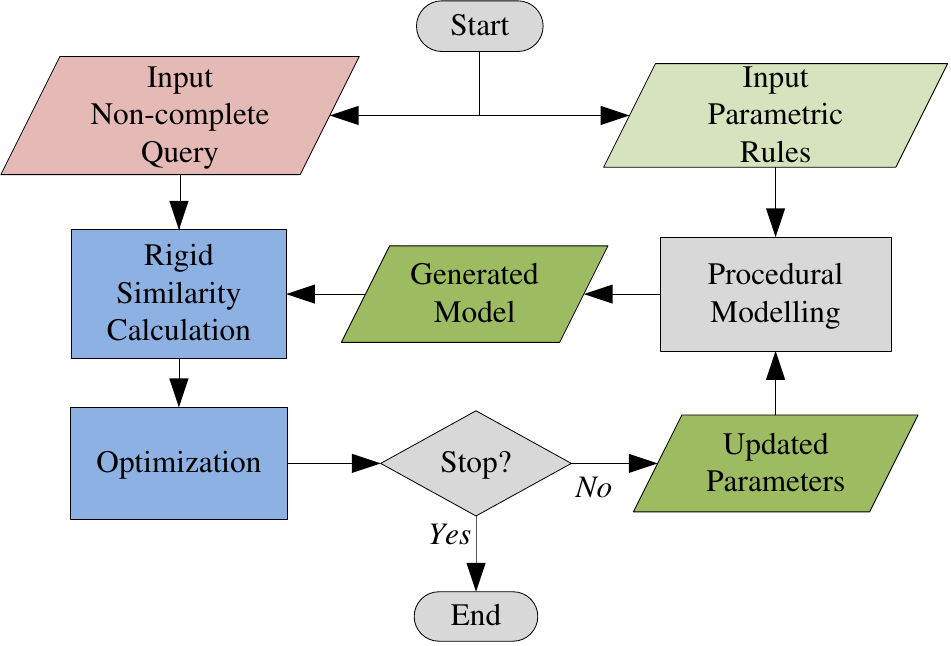}
	\caption{Our PPGMF pipeline.}
	\label{fig:flowchart}
\end{center}
\end{figure}

Figure \ref{fig:flowchart} shows our PPGMF pipeline. The task of PPGMF is to search a procedural geometric model space for the model which is rigidly geometrically similar to a query of non-complete point set. Hence the input of a PPGMF method consists of an non-complete query and a set of parametric geometric procedural modeling rules, which defines the target model space. A query can be a continuous or discrete point set (i.e. a point cloud). In this paper, we focus on point cloud queries.

An example set of procedural modeling rules is shown in Table \ref{tab:rules}, where $p(\cdot)$ denotes parameter prior. There are 3 rules $A$, $B$ and $D$ in this example. The axiom rule $A$ manages a non-recursive parameter $\mathbf{x}_D$ and a recursive parameter $\mathbf{x}_B$. It is straightforward to optimize for non-recursive parameters. However, it is challenging to deal with recursive parameters. When the rules are executed, a recursive parameter will spawn a family of non-recursive parameters under the same name of this recursive parameter. We have to individually identify every non-recursive parameter spawned from the same recursive parameter. To this end, a calling trace can be used. For example, we can identify a parameter $\mathbf{x}_B$ in a calling level 3 like this $\mathbf{x}_B$: $A\_B2\_B1\_B2$.

\begin{table}[ht!]
	\centering
		\begin{tabular}{|l|l|l|}\hline
			 \textbf{rule} $A()$&\textbf{rule} $B(\mathbf{x}_B)$&\textbf{rule} $D(\mathbf{x}_D)$\\
			 \,Sample ${\mathbf{x}_B} \sim p({\mathbf{x}_B})$&\,Sample ${\mathbf{x}_B} \sim p({\mathbf{x}_B})$&\,...\\
			 \,Call $B(\mathbf{x}_B)$ &\,Call $B(\mathbf{x}_B)$&\textbf{end rule}\\
			 \,Sample ${\mathbf{x}_D} \sim p({\mathbf{x}_D})$ &\,Sample ${\mathbf{x}_B} \sim p({\mathbf{x}_B})$&\\
			 \,Call $D(\mathbf{x}_D)$ &\,Call $B(\mathbf{x}_B)$&\\
			 \,... &\,...&\\
			 \textbf{end rule} &\textbf{end rule}&\\\hline
		\end{tabular}
	\caption{An example set of rules.}
	\label{tab:rules}
\end{table}

As shown in Fig. \ref{fig:flowchart}, given the query and the rules with parameter ${\mathbf{x}}$, a similarity calculation procedure is used to calculate the rigid geometric similarity between the query and the model generated by the procedural modeling procedure according to the rules. Based on the calculated similarity, ${\mathbf{x}}$ is iteratively updated by the optimization procedure. It is worth noting that the number of parameters may vary during the optimization. Based on Bayesian inference theory, the optimization problem can be formulated as follows:
\begin{equation}\label{eq:maximization}
\mathop {\max }\limits_{\mathbf{x}} p\left( {{\mathbf{x}}|Q} \right) \propto L(Q|{\mathbf{x}}){\mkern 1mu} p({\mathbf{x}})
\end{equation}
where $Q$ is the query, $p( \cdot | \cdot )$ is the posterior of the parameters given the query, $L( \cdot | \cdot )$ is the likelihood of the query given the parameters, and $p( \cdot )$ is the parameter prior.

The prior is directly drawn from the input modeling rules. Assume that $M^{\mathbf{x}}$ represents the model corresponding to $\mathbf{x}$, the likelihood depends on the rigid similarity between $M^{\mathbf{x}}$ and $Q$ (Section \ref{sec:similarity}), the optimization process is then performed (Section \ref{sec:optimize}).

\section{Rigid Geometric Similarity}\label{sec:similarity}
A rigid geometric similarity metric should ensure that a geometric model is most rigidly similar to itself than any other models. Let $\mathcal{M}$ be the universal geometric model space, the self rigidly similar property is formally stated as: 
\begin{equation}\label{eq:self-similar}
\forall M \in \mathcal{M},\forall M' \ne M,s(M',M) < s(M,M) 
\end{equation}
where $s( \cdot , \cdot )$ denotes the similarity metric.

\subsection{Full Similarity}

SHD is the only metric satisfying the self rigidly similar property. The SHD between two point sets $P$ and $P'$ is defined as:
\begin{equation}\label{eq:symmetrical}
{d_s}(P,P') = \max \left\{ {d(P,P'),d(P',P)} \right\}
\end{equation}
where $d( \cdot , \cdot )$ is the one-sided Hausdorff distance (OHD):
\begin{equation}\label{eq:one-sided}
d(P,P') = \mathop {\max }\limits_{a \in P} \mathop {\min }\limits_{b \in P'} \left\| {a - b} \right\|
\end{equation}
where $\left\|  \cdot  \right\|$ is Euclidean norm. Note that, in general, $d(P,P') \ne d(P',P)$.

Usually, we can use SHD to exactly measure the rigid similarity between two point sets $P$ and $P'$. If ${d_s}({P},{P'})$ is $0$, then $P$ and $P'$ are the same. However, to calculate SHD, we have to compute OHD two times, i.e., one time from $P$ to $P'$ and another time from $P'$ to $P$. In other words, only one OHD is insufficient for similarity assessment \cite{aspert2002mesh}. If only one OHD calculation is required, the computational cost for similarity calculation can be reduced. Fortunately, one of the point sets involved in our similarity calculation is a geometric model. The measure of the model allows us to compute OHD only once to assess similarity. It is worth noting that different types of models have different types of measures. For example, the measure of a curve is its length, the measure of a surface is its area.

\begin{figure}[ht!]
\centering
\subfloat[]{\label{fig:partial-similarity:a}\includegraphics[height=0.4in]{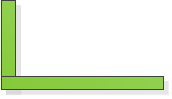}}
\hfill
\subfloat[]{\label{fig:partial-similarity:b}\includegraphics[height=0.4in]{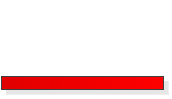}}
\hfill
\subfloat[]{\label{fig:partial-similarity:c}\includegraphics[height=0.4in]{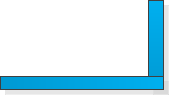}}
\hfill
\subfloat[]{\label{fig:partial-similarity:d}\includegraphics[height=0.4in]{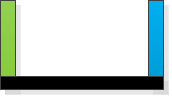}}
\caption{An illustration of rigid similarity. \protect\subref{fig:partial-similarity:a} Curve $C_1$, \protect\subref{fig:partial-similarity:b} Curve $C_2$, \protect\subref{fig:partial-similarity:c} Curve $C_3$, and \protect\subref{fig:partial-similarity:d} the overlap of $C_1$, $C_2$ and $C_3$. The overlapping part (black) shows that $C_2$ is a part of $C_1$ or $C_3$.}
	\label{fig:partial-similarity}
\end{figure}

Our insight is that, in real world, two models $M$ and $M'$ are identical if and only if every point of $M$ is in $M'$ (i.e. $d(M,M') = 0$) and the measure of $M$ is equal to the measure of $M'$. That means the measure can be used for similarity assessment. We hence propose a Mean Measure (MM) to represent the rigid similarity between a model $M$ and a query point set $Q$. Formally, MM is defined as the ratio of the measure of $M$ to the OHD from $M$ to $Q$:
\begin{equation}\label{eq:r}
r(M,Q) = \frac{{\left| M \right|}}{{\epsilon + d(M,Q)}}
\end{equation}
where $\left| \cdot \right|$ denotes the measure of $M$, $\epsilon$ is a small positive number used to derive different similarities for the models with different measures but the same OHD of $0$. For example, as shown in Fig. \ref{fig:partial-similarity}, both $d(C_1,C_1)$ and $d(C_2,C_1)$ are equal to $0$. If $\epsilon$ is $0$, then both $r(C_1,C_1)$ and $r(C_2,C_1)$ are infinite despite $C_1$ is more similar to $C_1$ than $C_2$. Theoretically, when $\epsilon$ is sufficiently small, the similarity metric defined by MM can ensure that a model is most similar to itself than any other models. 

\textbf{Theorem}. MM is a rigid geometric similarity metric for real-world geometric models. Note that, a real-world model $M$ has a positive finite measure, i.e. $0 < \left| M \right| < +\infty $. Let $\mathcal{M}_W$ be the real-world geometric model space, this theorem is proved as follows. 

\textit{Proof}. $\forall M \in \mathcal{M}_W, \forall M' \ne M$, (1) If $ M' \subset M $, then $\left| {M'} \right| < \left| M \right|$ and $d(M',M) = d(M,M)$, so $r(M',M) < r(M,M)$ because $\epsilon >  0 $; (2) If $M' \not\subset M$, then $r(M',M) <  + \infty $ because $\left| M' \right| < +\infty $ and $d(M',M) > 0$. Meanwhile, $r(M,M) \to  + \infty $ because, $d(M,M)=0$, $0 < \left| M \right|$ and $\epsilon$ is assumed to be sufficiently small, i.e. $\epsilon \to  0 $. So $r(M',M) < r(M,M)$. So MM is a rigid similarity metric for real-world models according to Eq. \eqref{eq:self-similar}.  

Note that, the values of MM are comparable for the same query, but are incomparable for different queries. That is, it makes no sense to compare the MM values across different queries. For example, as shown in Fig. \ref{fig:partial-similarity}, it makes no sense to compare $r(C_2,C_2)$ and $r(C_2,C_1)$, although $r(C_2,C_2)=r(C_2,C_1)$. It is worth noting that, the query is unnecessary to have a geometric measure. That is, the query can be a discrete point set, i.e., a point cloud. If the query $Q$ is discrete, then $\epsilon$ is trivial because $d(M,Q)$ is always larger than $0$. In practice, we use squared mean measure (SMM), which is a variant of MM and is defined as:
\begin{equation}\label{eq:rs}r_s(M,Q) = \frac{{\left| M \right|}}{{\epsilon + d^2(M,Q)}}
\end{equation}

\subsection{Partial Similarity}
SHD and MM are defined as full similarity metrics as they assume that the query is complete. However, if the query is non-complete, we have to calculate partial similarity, which is challenging. Partial similarity is not straightforward and is fundamentally different from full similarity. If two point sets have a common part, then these two point sets are partially similar. As shown in Fig. \ref{fig:partial-similarity}, each pair of $C_1$, $C_2$ and $C_3$ are partially similar, while they are not fully similar. We expect that the partial similarity between $C_1$ and $C_2$ is equal to the partial similarity between $C_2$ and $C_2$. Because the common part between $C_1$ and $C_2$ is the same as the common part between $C_2$ and $C_2$. 

Consequently, we propose a Weighted Mean Measure (WMM) to represent the partial similarity between a geometric model $M$ and a query point set $Q$. We divide $M$ into $N$ non-overlapping sub-models: $M = {\kern 1pt} \mathop  \cup \limits_{i = 1}^N {M_i}$, and define WMM as:
\begin{equation}\label{eq:wme}
{r_w}(M,Q) = \frac{{\sum\limits_{i = 1}^N {{w_i}\left| {{M_i}} \right|} }}{{\epsilon + {d_w}(M,Q)}}
\end{equation}
where $w_i$ is the weight: ${w_i} = \exp \left( { - d({M_i},Q){\mkern 1mu} \, h} \right)$, where $h$ is the weighting factor, which is a non-negative number. When $h$ is $0$, WMM becomes a full similarity metric. ${d_w}(\cdot,\cdot)$ is the weighted mean error \cite{aspert2002mesh}:
\begin{equation}\label{eq:mean-OHD}
{d_w}(M,Q) = \frac{{\sum\limits_{i = 1}^N {{w_i}d({M_i},Q)} }}{{\sum\limits_{i = 1}^N {{w_i}} }}
\end{equation}

By weighting, the sub-models of $M$ far away from $Q$ have less contribution to the computation of WMM than those close sub-models. In other words, the common part of $M$ and $Q$ makes major contribution to WMM, making WMM plausible to measure partial similarity. One merit of WMM is that it has only one argument $h$, as $\epsilon$ is trivial. Similar to MM, it can be easily proved that WMM is a rigid geometric similarity metric.

Let $\mathcal{C}_1$, $\mathcal{C}_2$ and $\mathcal{C}_3$ be model spaces containing models $C_1$, $C_2$ and $C_3$ (Fig. \ref{fig:partial-similarity}), respectively. There are 3 cases of partial similarity for PPGMF. (1) Target model is a part of query. For example, target model $C_2$ is a part of query $C_1$. This case corresponds to partially fitting $\mathcal{C}_2$ to over-complete data $C_1$. (2) Query is a part of target model. For example, query $C_2$ is a part of target model $C_1$. This case corresponds to partially fitting $\mathcal{C}_1$ to incomplete data $C_2$. (3) Target model and query have common part. For example, target model $C_1$ and query $C_3$ have common part. This case corresponds to partially fitting $\mathcal{C}_1$ to hybrid-complete data $C_3$.
\subsection{Similarity Calculation}
To compute MM (SMM or WMM) between a model and a point set, we have to compute OHD from the model to the point set, which consists of two steps. First, the model is uniformly divided into sub-models, and the center points of the sub-models are sampled (Section \ref{sec:multi-level}). Second, the nearest point is searched in the point set for a query point. This is time-consuming if the point set contains a large number of points (e.g. a laser scanning point cloud consisting of millions of points). We employ the FLANN \cite{muja2014scalable} algorithm to perform nearest neighbour searching. The computational complexity for computing MM depends on the number of points sampled from the model and the size of the point set.

\section{Optimization}\label{sec:optimize}
Given the rigid geometric similarity defined by MM, we empirically define the likelihood in the optimization problem (see Eq. \eqref{eq:maximization}) as: 
\begin{equation}\label{eq:likelihood}
L(Q|{\mathbf{x}}){\mkern 1mu}  = \exp \left( {\sqrt {{r}({M^{\mathbf{x}}},Q)} } \right)
\end{equation} 
Eq. \eqref{eq:maximization} defines a derivative-free optimization problem, for which traditional mathematical optimization methods are not applicable. We use the Metropolis-Hastings (MH) algorithm \cite{metropolis1953equation} \cite{hastings1970monte} to solve Eq. \eqref{eq:maximization}. MH algorithm is a general and popular MCMC optimization algorithm \cite{talton_metropolis_2011}.

\subsection{Metropolis-Hastings Algorithm}
Let $x^i$ be the value of variable $x$ in iteration $i$, the MH algorithm works as follows. First, $x$ is randomly initialized as $x^0$. To determine $x^{i+1}$ in each iteration, $\tilde x$ is sampled from a proposal density function $p(x|{x^i})$. The probability of accepting $\tilde{x}$ as $x^{i+1}$ is defined as:
\begin{equation}\label{eq:acceptance}
\alpha ({x^i} \to \tilde x) = \mathop {\min} \left\{ {1,\frac{{p(\tilde x|Q)}}{{p({x^i}|Q)}}\frac{{p({x^i}|\tilde x)}}{{p(\tilde x|{x^i})}}} \right\}
\end{equation}
That is, the probability for ${x^{i + 1}} = \tilde x$ is $\alpha$, and the probability for ${x^{i + 1}} = {x^{i}}$ is $(1-\alpha)$.

We now define the proposal function for the modeling parameter ${\mathbf{x}}$. Similar to \cite{vanegas2012inverse}, each parameter $x  \in {\mathbf{x}}$ is required to be within a range of $[x_{min},x_{max}]$. For a continuous parameter, we randomly select one of the following two proposal functions, i.e., local move function and global move function in each iteration. The local move function is a Gaussian function, that is, $\tilde x \sim \mathcal{N}({x^i},{\sigma _x^2})$, where ${\sigma _x} = \sigma ({x_{\max }}-{x_{\min }})$, and $\sigma$ is the standard deviation ratio.
The global move function is a uniform function, that is, $\tilde{x} \sim [x_{min},x_{max}]$. We use $\beta $ to denote the probability for selecting local move function, and $1-\beta$ to denote the probability for selecting global move function. For a discrete parameter, we always perform global move. Since both local move and global move functions are symmetric, the probability of accepting $\tilde{x}$ is simplified as:
\begin{equation}\label{eq:reduced-acceptance}
\alpha ({x^i} \to \tilde x) = \mathop {\min} \left\{ {1,\frac{{p(\tilde x|Q)}}{{p({x^i}|Q)}}} \right\}
\end{equation}

\subsection{Early Rejection}\label{sec:multi-level}
\begin{figure}[ht!]
\centering
\subfloat[]{\label{fig:accept-carefully:a}\includegraphics[height=0.6in]{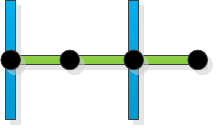}}
\hfill
\subfloat[]{\label{fig:accept-carefully:b}\includegraphics[height=0.6in]{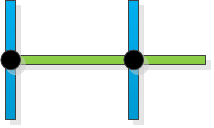}}
\hfill
\subfloat[]{\label{fig:accept-carefully:c}\includegraphics[height=0.6in]{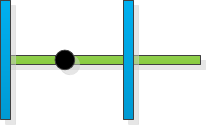}}

\caption{Overlap between Curves $C_4$ (green) and $C_5$ (blue). Black dots represent the points sampled from $C_4$. \protect\subref{fig:accept-carefully:a}, \protect\subref{fig:accept-carefully:b} and \protect\subref{fig:accept-carefully:c} show that 4, 2 and 1 point(s) are sampled, respectively.}
\label{fig:reject-or-accept}
\end{figure}

The acceptance probability indicates that the proposed model with a larger similarity is more likely to be accepted than the model with a smaller similarity. More time will be consumed to obtain more accurate similarity since more points have to be sampled from the model. However, we observe that, it is sufficient to determine the dissimilarity by sampling only one point from the model. As shown in Fig. \ref{fig:reject-or-accept}, Curve $C_4$ consists of one horizontal line segment, and Curve $C_5$ consists of two vertical line segments, these two curves are dissimilar. The similarities computed by sampling one point (Fig. \ref{fig:accept-carefully:c}) and four points (Fig. \ref{fig:accept-carefully:a}) are the same and equal to the true similarity. However, if two points are sampled (Fig. \ref{fig:accept-carefully:b}), the computed similarity will be incorrect as it shows that $C_4$ and $C_5$ are similar. It can be inferred that a small similarity between two objects means that these two objects are dissimilar. However, a large similarity between two objects does not mean that these two objects are really similar. In other words, a proposed model should be accepted carefully but rejected boldly.

\begin{figure}[ht!]
\centering
\subfloat[]{\label{fig:coarse-to-fine:a}\includegraphics[width=0.2\columnwidth]{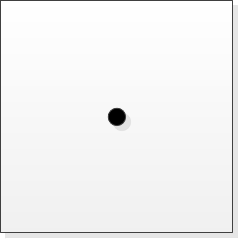}}
\hfill
\subfloat[]{\label{fig:coarse-to-fine:b}\includegraphics[width=0.2\columnwidth]{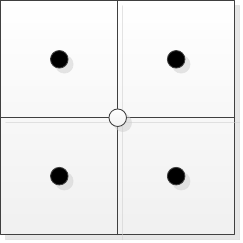}}
\hfill
\subfloat[]{\label{fig:coarse-to-fine:c}\includegraphics[width=0.2\columnwidth]{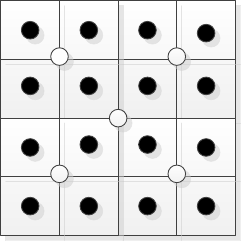}}
\caption{A square surface to illustrate the coarse-to-fine model dividing. From left to right, the dividing level is 0, 1 and 2, respectively. The black dots represents the points sampled in the current level, and the white dots represents the points sampled in previous levels.}
\label{fig:coarse-to-fine}
\end{figure}

Consequently, to reduce computational time, we propose a coarse-to-fine model dividing strategy for similarity calculation to reject dissimilar models in advance. We take a square surface for example (as shown in Fig. \ref{fig:coarse-to-fine}), and the conclusions can be easily adapted to other types of geometric models. Assuming that the length of the square surface is $\gamma $, given a predefined minimal dividing resolution $\delta $, the top dividing level is: 
\begin{equation}\label{eq:divdinglevel}
{\eta _{top}} = {\log _2}({\gamma }/{\delta } + 1)
\end{equation} 
At each level $\eta$, we uniformly divide the surface into ${2^{2\eta }}$ sub-surfaces, and sample only one point (center point) from each sub-surface to calculate OHD. The similarity is then calculated to decide whether to accept or reject the proposed surface. If it is accepted, then the surface is divided into more sub-surfaces and more points are sampled at a higher level to obtain more accurate similarity. Otherwise, a new surface is proposed.

\subsection{Pseudo Code}\label{sec:code}

The pseudo code of our MH-PPGMF method is presented in Algorithm \ref{alg:algorithm}, where ${{p_\eta}({\cdot}|\cdot)}$ denotes the posterior computed at dividing level $\eta$. The minimal model dividing resolution $\delta$ should be set as small as possible to obtain accurate similarity. To achieve better performance, parallel tempering with the same configuration as in \cite{talton_metropolis_2011} is also used. That is, $10$ Markov chains are run with different temperatures and the chains are randomly swapped.

\begin{algorithm}[ht!]
\caption{MH-PPGMF with early rejection}
\label{alg:algorithm}
\begin{algorithmic}
\State\textbf{input}: a set of modeling rules with parameter $\mathbf{x}$, query $Q$, posterior function $p({\mathbf{x}}|Q) $, computational budget $I$, standard deviation ratio $\sigma$, local move probability $\beta$, and minimal model dividing resolution $\delta$.
\State\textbf{output}: a maximum a posteriori estimate of $\mathbf{x}$: $\mathbf{x}^*$.
\State Randomly initialize ${{\mathbf{x}}^0}$, ${{\mathbf{x}}^*} \leftarrow {{\mathbf{x}}^0}$
\For {$i=0$ to $I$}
\State  Randomly select a parameter $x  \in {\mathbf{x}^i}$
\State Sample $t \sim [0,1]$
\If {$t < \beta $ \textbf{and} $x$ is continuous} \State Sample $\tilde x \sim \mathcal{N} ( {x^i},\sigma _x^2 )$
\Else \, Sample $\tilde{x} \sim [x_{min},x_{max}]$
\EndIf
\State Compute ${\eta _{top}}$ of $M^{{\mathbf{\tilde x}}}$ according to $\delta$
\For {$\eta = 0$ to $\eta _{top}$}
\State  $\alpha  \leftarrow \min \left\{ {1,\frac{{{p_\eta}({\mathbf{\tilde x}}|Q)}}{{p({{\mathbf{x}}^i}|Q)}}} \right\}$
\State  Sample $t \sim [0,1]$
\If {$t<\alpha$}  ${{{\mathbf{x}}^{i+1}}}\leftarrow \mathbf{\tilde x}$
\Else  \,\,${{{\mathbf{x}}^{i+1}}}\leftarrow {{{\mathbf{x}}^{i}}}$, \textbf{break}
\EndIf
\EndFor
\If {$p({{\mathbf{x}}^{i + 1}}|Q)>  p({\mathbf{x}}^{*}|Q) $} ${\mathbf{x}^{* }} \leftarrow {\mathbf{x}^{i+1}}$
\EndIf
\EndFor
\end{algorithmic}
\end{algorithm}

\section{Results}\label{sec:result}
We implemented our method in C++ and conducted our experiments on a machine running Ubuntu 14.04 with Intel Core i5-3470 3.20GHz CPU and 12GB RAM. In all experiments, we set $\epsilon  = {10^{ - 8}}$, $\beta=0.8$, $\sigma=0.05$. $\delta$ should be at least 2 times smaller than query resolution.
\subsection{Metric Comparison}
\begin{figure*}[ht!]
\centering
\subfloat[]{\label{fig:m1}\includegraphics[width=0.35\columnwidth]{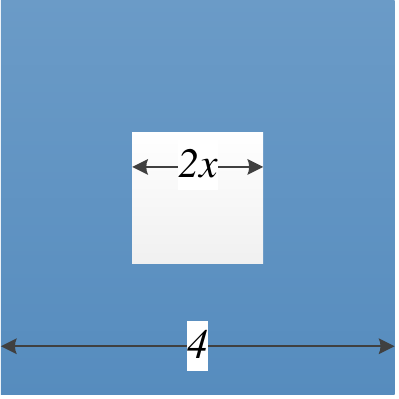}}
\hfill
\subfloat[]{\label{fig:m2}\includegraphics[width=0.35\columnwidth]{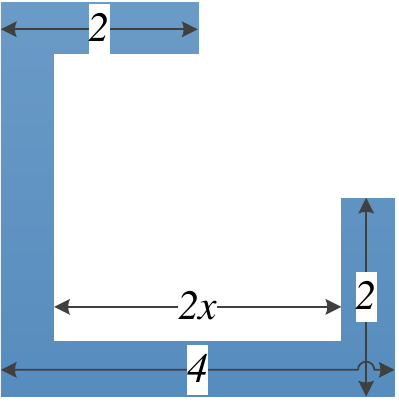}}
\hfill
\subfloat[]{\label{fig:m3}\includegraphics[width=0.35\columnwidth]{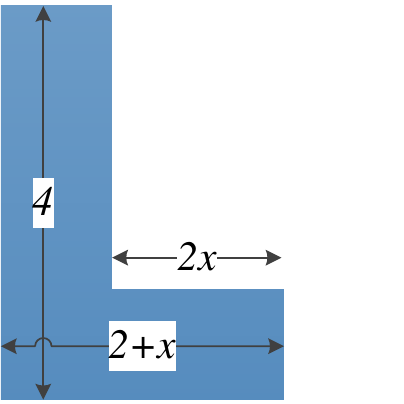}}
\hfill
\subfloat[]{\label{fig:m4}\includegraphics[width=0.35\columnwidth]{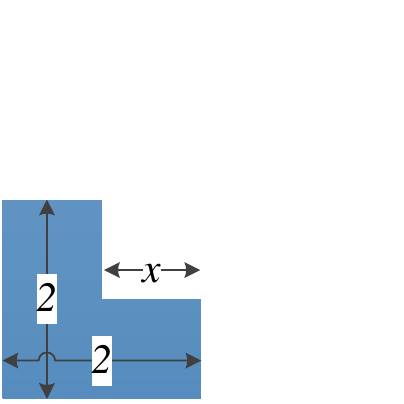}}
\caption{Model spaces. From left to right: Model spaces $\mathcal{M}_{1}$, $\mathcal{M}_{2}$, $\mathcal{M}_{3}$ and $\mathcal{M}_{4}$. Each of these 4 spaces has only one parameter $x \in \left[ {0,2} \right]$.}
\label{fig:m1-4}
\end{figure*}

\begin{figure}[ht!]
\centering
\subfloat[]{\label{fig:q1}\includegraphics[height=0.8in]{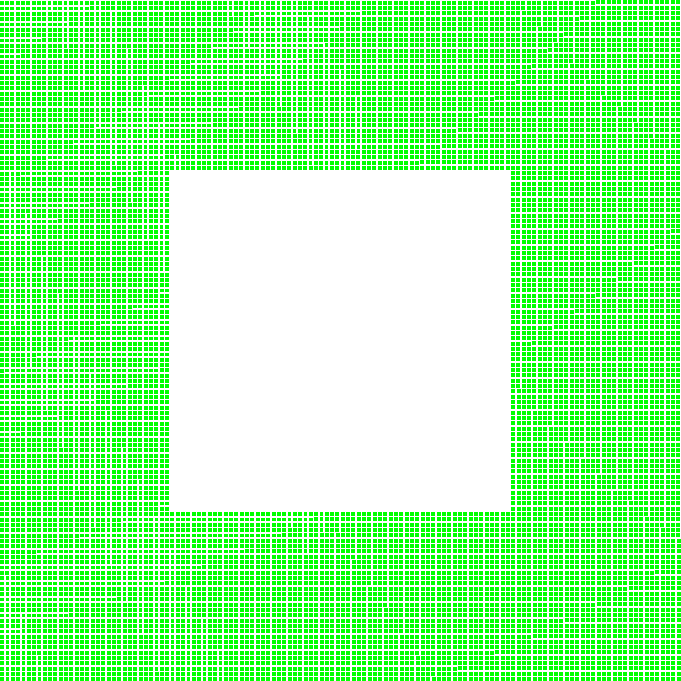}}
\hfill
\subfloat[]{\label{fig:q2}\includegraphics[height=0.8in]{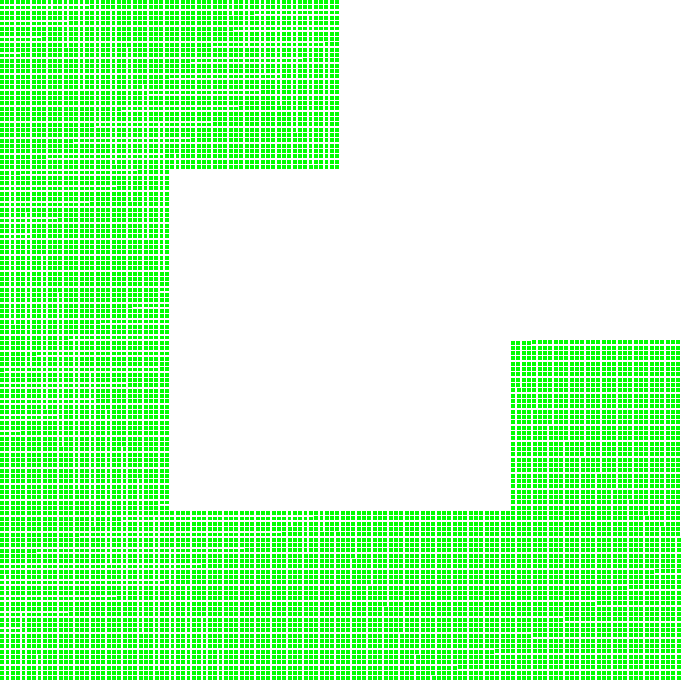}}
\hfill
\subfloat[]{\label{fig:q3}\includegraphics[height=0.8in]{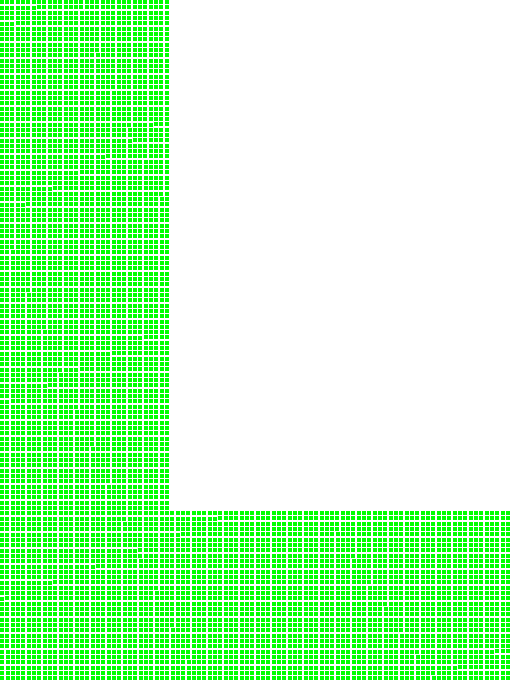}}
\hfill
\subfloat[]{\label{fig:q4}\includegraphics[height=0.8in]{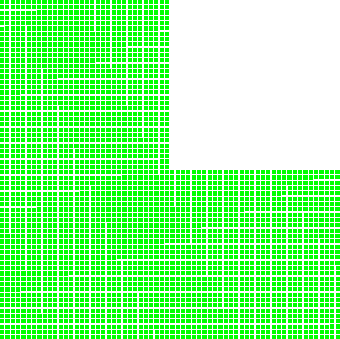}}
\caption{Queries. From left to right: Queries $Q_1$, $Q_2$, $Q_3$ and $Q_4$. For $i=1$ to $4$, $Q_i$ is a point cloud uniformly sampled from Model ${M}^{x=1}_{i}$ with $0.02$ resolution. $Q_1$, $Q_2$, $Q_3$ and $Q_4$ consist of 12288, 9216, 6144 and 3072 points, respectively.}
\label{fig:q1-4}
\end{figure}

We compare several metrics with our WMM metric by fitting 4 models (Fig. \ref{fig:m1-4}) to 4 queries (Fig. \ref{fig:q1-4}). Model ${M}^x_{1}\in \mathcal{M}_1$ is a ring-like surface between an outer square and an inner square. The outer and inner squares have the same center. The length of the outer and inner squares are 4 and $2x$, respectively. Models ${M}^x_{2}\in \mathcal{M}_2$, ${M}^x_{3}\in \mathcal{M}_3$ and ${M}^x_{4}\in \mathcal{M}_4$ are 0.75, 0.5 and 0.25 part of ${M}^x_{1}$, respectively. As shown in Fig. \ref{fig:q1-4}, for $i=1$ to $4$, the ground-truth model of $Q_i$ is $M^{x=1}_i$. In this paper, we refer to the target model of a query as the model which is partially similar to the ground-truth model. Therefore, for each query in Fig. \ref{fig:q1-4}, there is an target model existing in each model space (as shown in Fig. \ref{fig:m1-4}). That is, for $i=1$ to $4$ and $j=1$ to $4$, the target model of $Q_i$ in $\mathcal{M}_j$ is $M^{x=1}_j$.

\begin{figure*}[ht!]
\centering
\subfloat[]{\label{fig:q1m1}\includegraphics[height=1.3in]{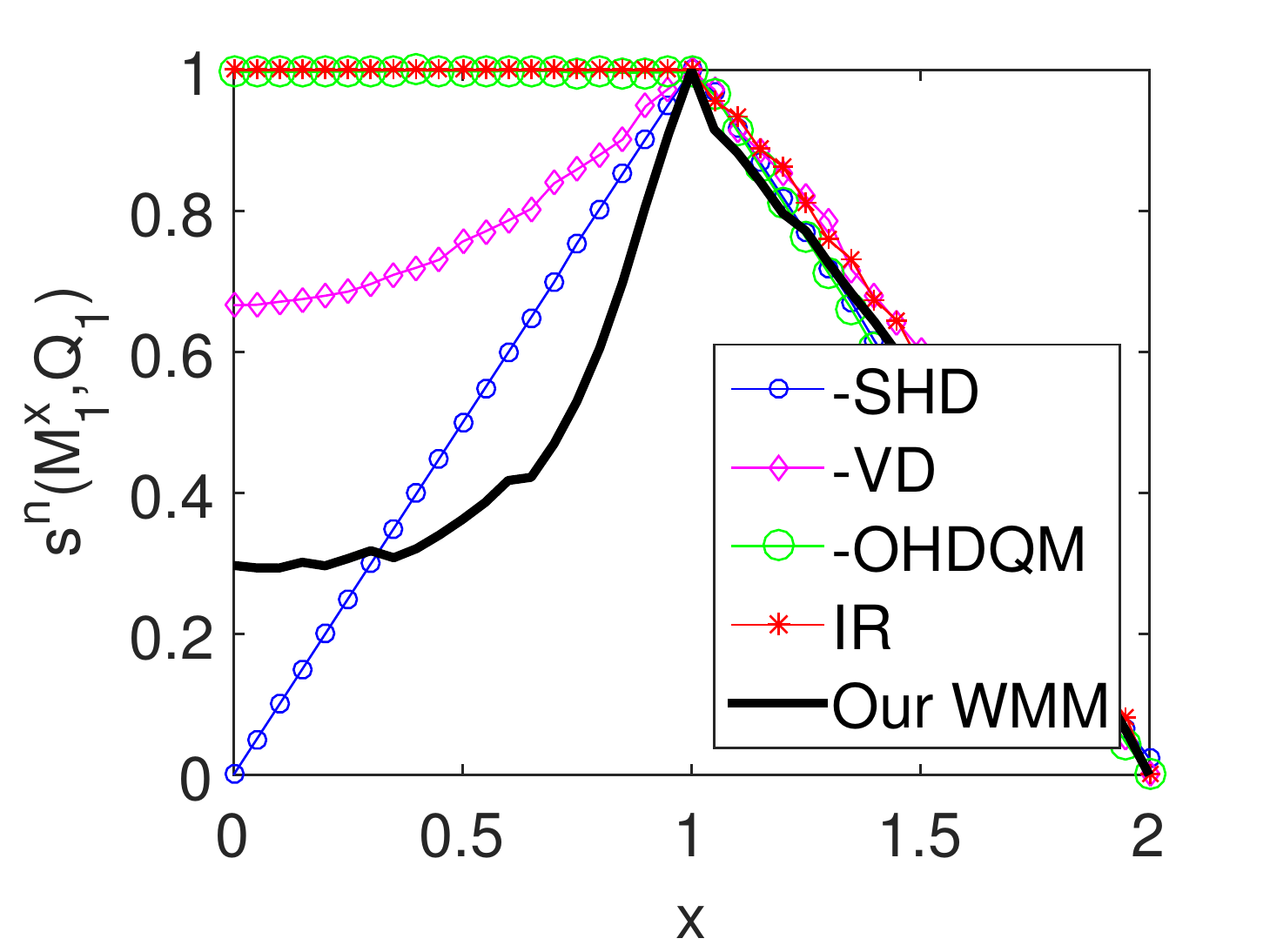}}
\hfill
\subfloat[]{\label{fig:q1m2}\includegraphics[height=1.3in]{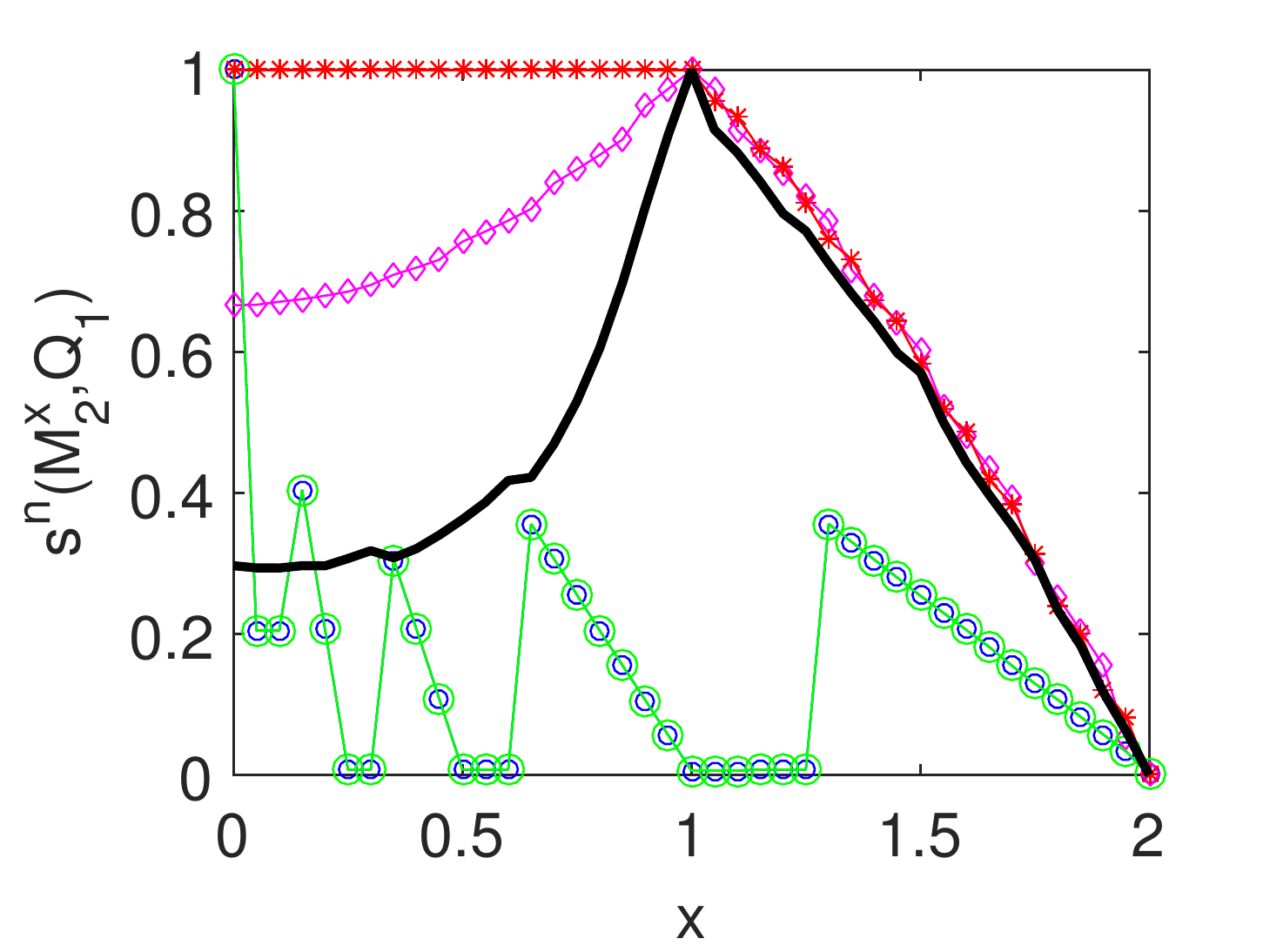}}
\hfill
\subfloat[]{\label{fig:q1m3}\includegraphics[height=1.3in]{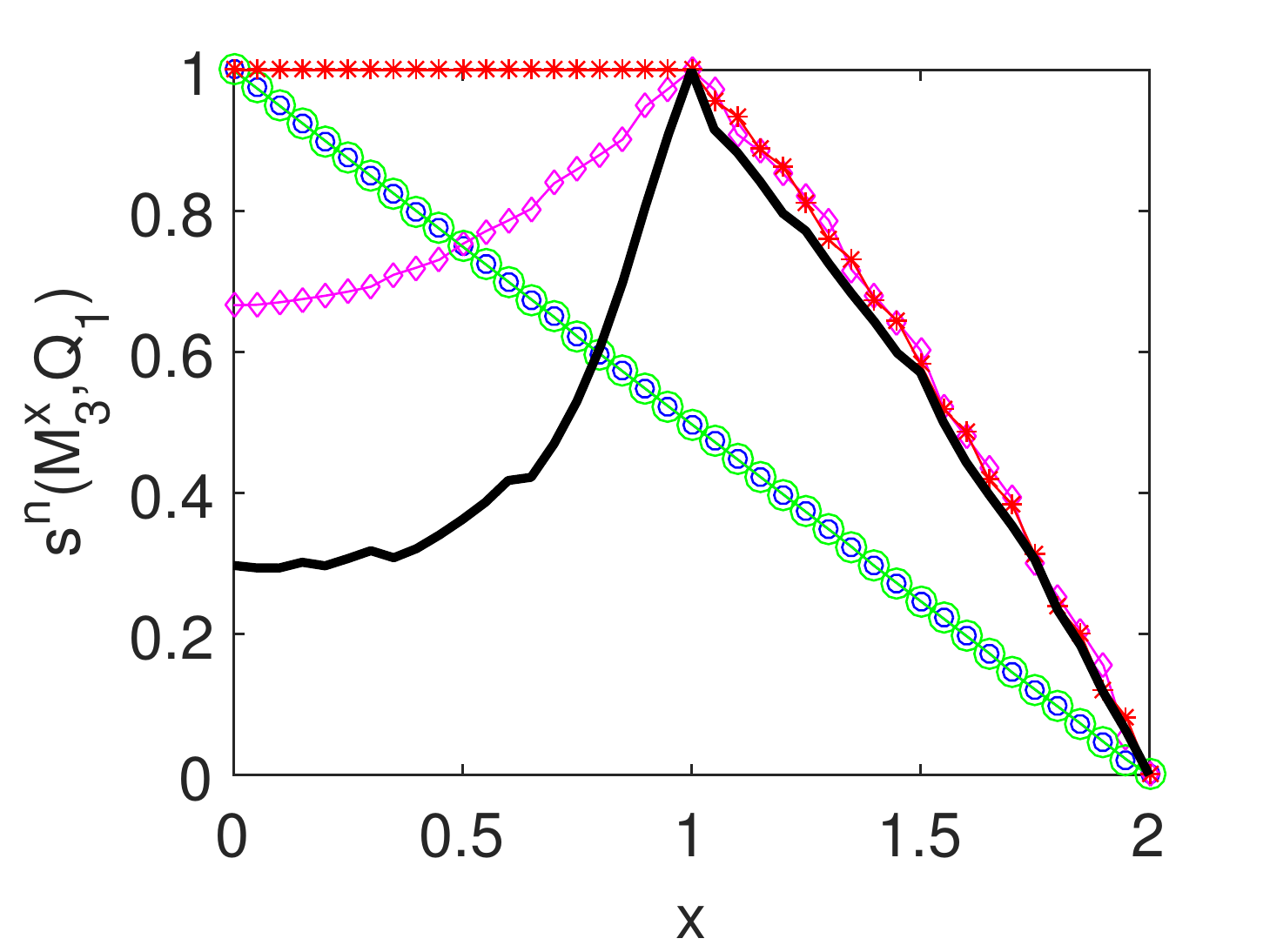}}
\hfill
\subfloat[]{\label{fig:q1m4}\includegraphics[height=1.3in]{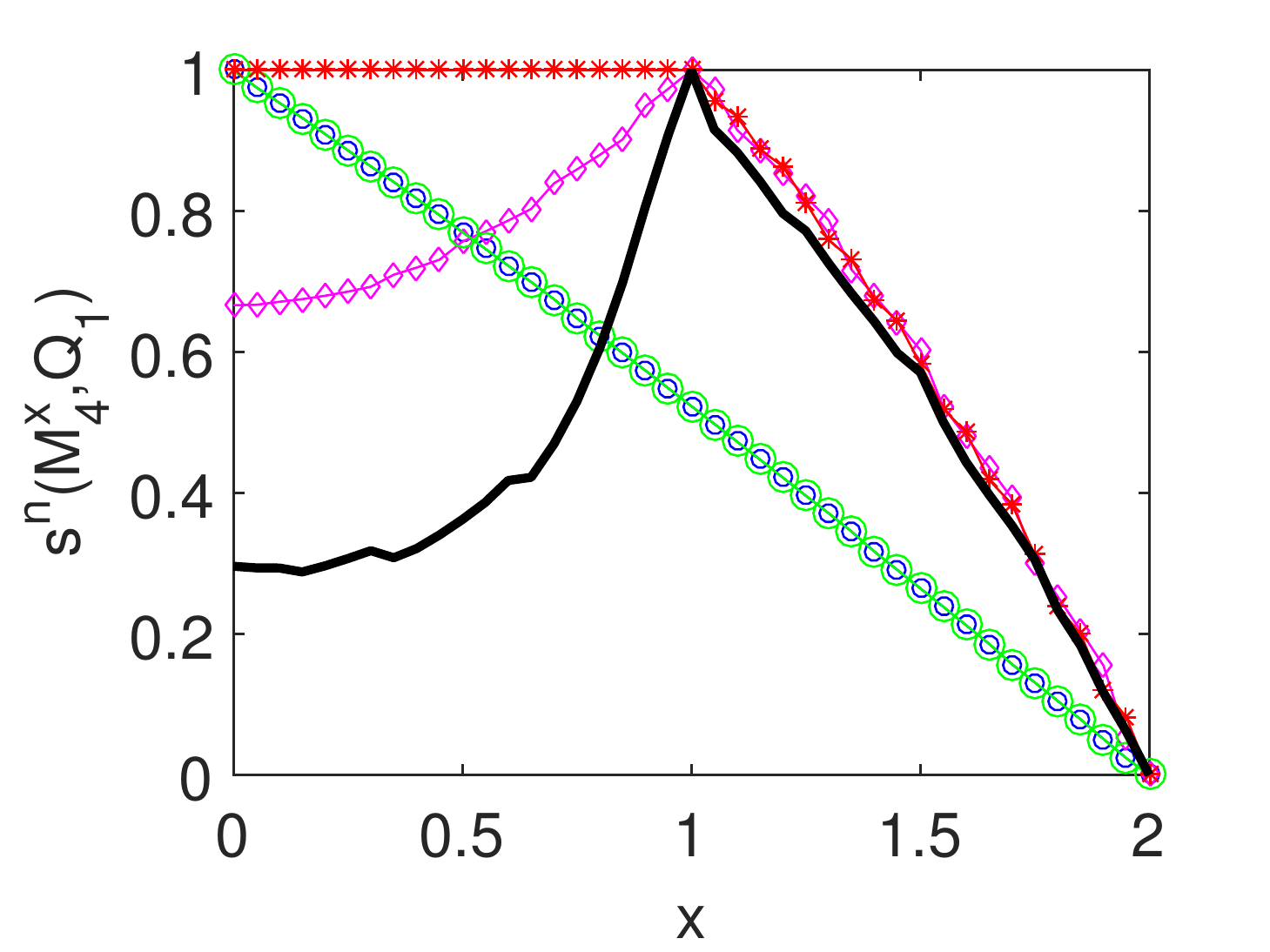}}\\
\subfloat[]{\label{fig:q2m1}\includegraphics[height=1.3in]{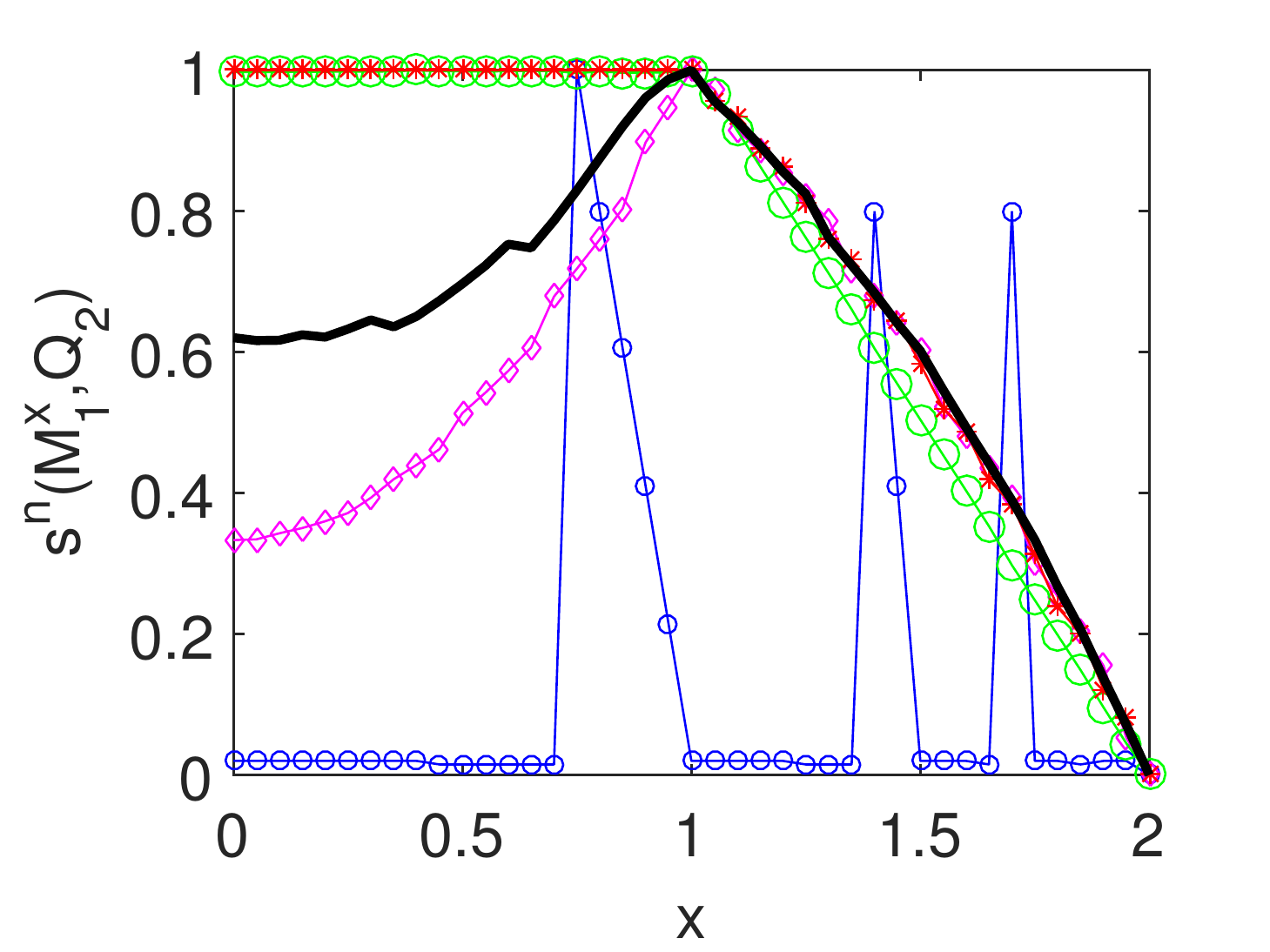}}
\hfill
\subfloat[]{\label{fig:q2m2}\includegraphics[height=1.3in]{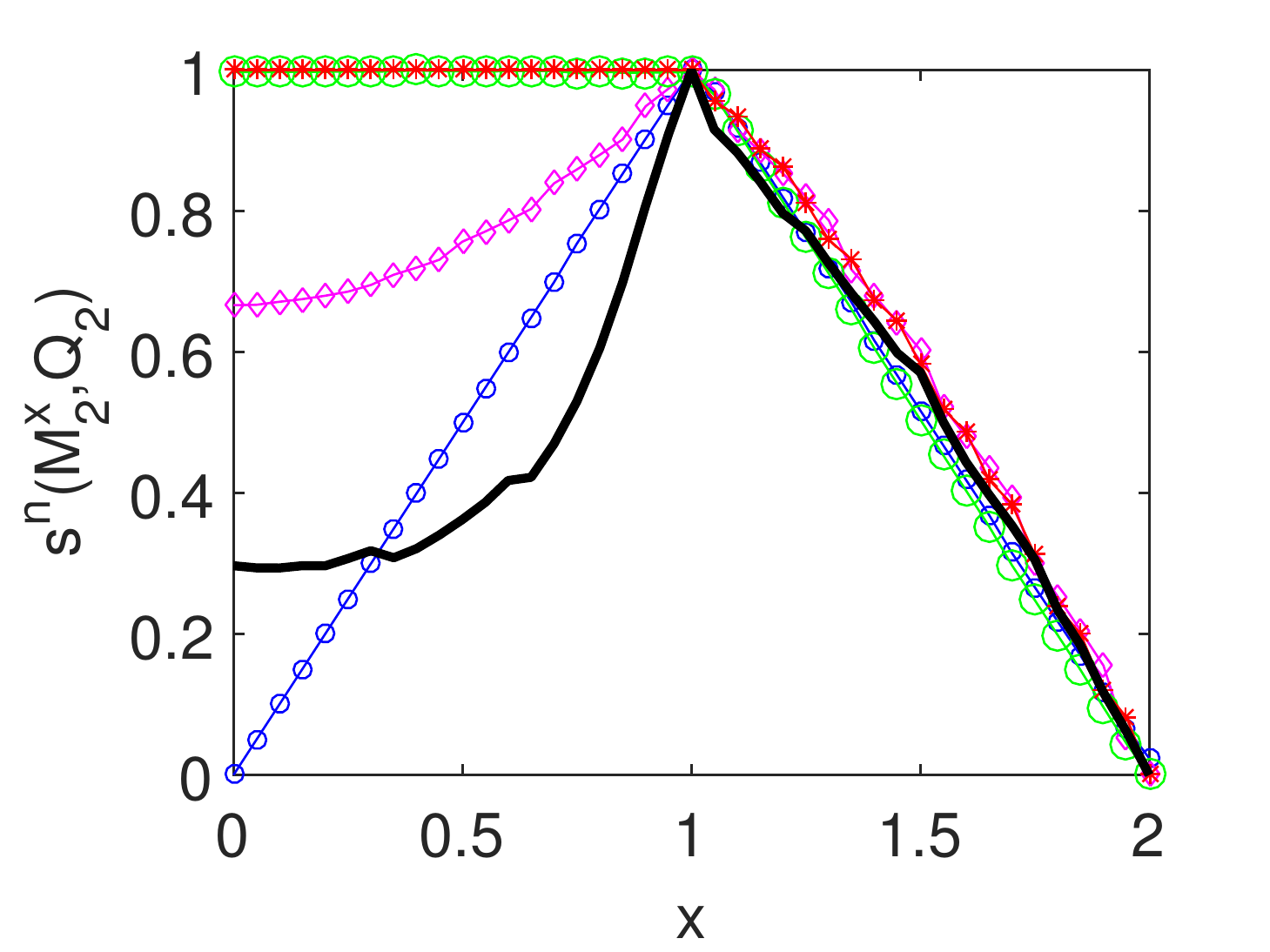}}
\hfill
\subfloat[]{\label{fig:q2m3}\includegraphics[height=1.3in]{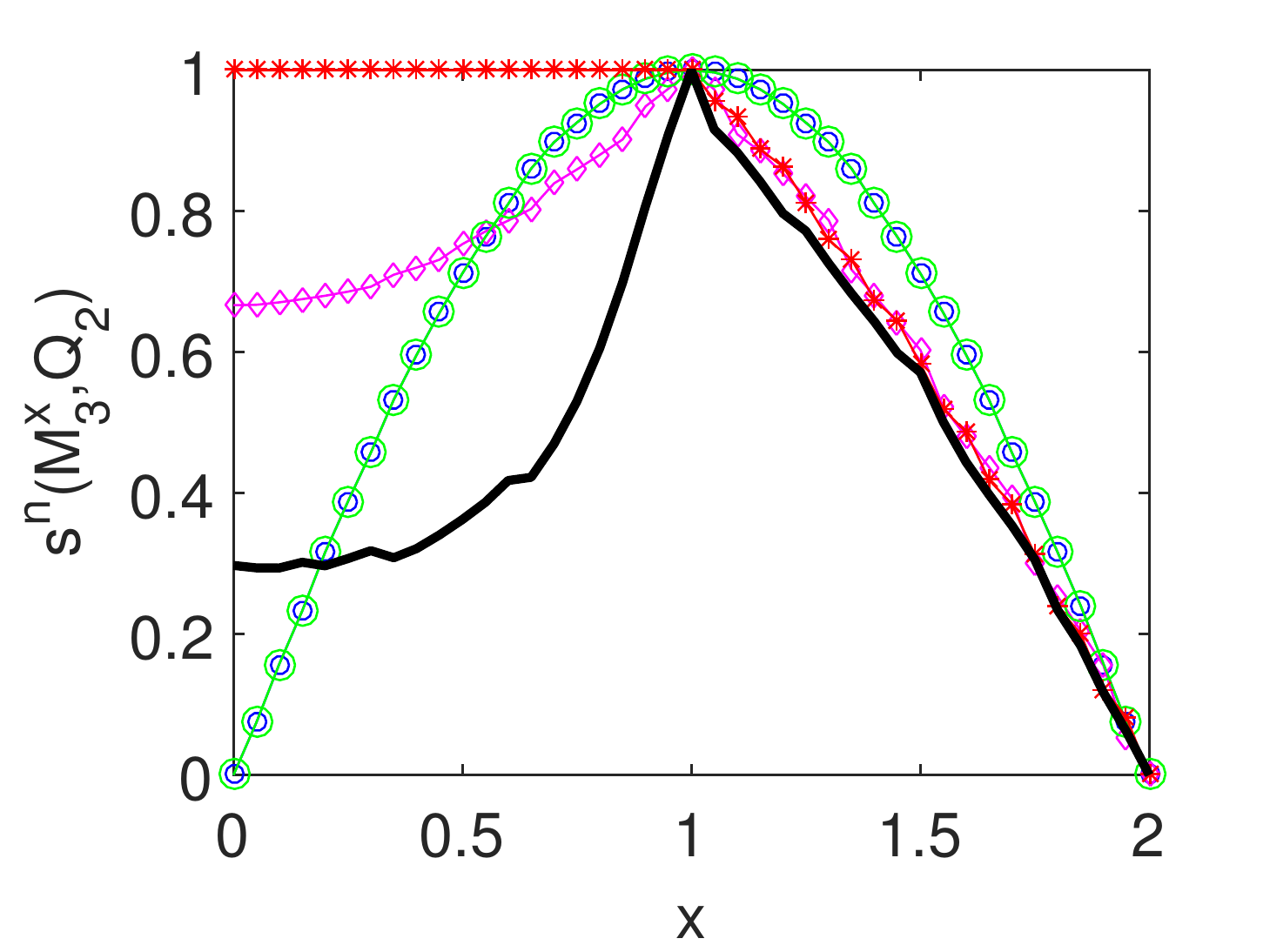}}
\hfill
\subfloat[]{\label{fig:q2m4}\includegraphics[height=1.3in]{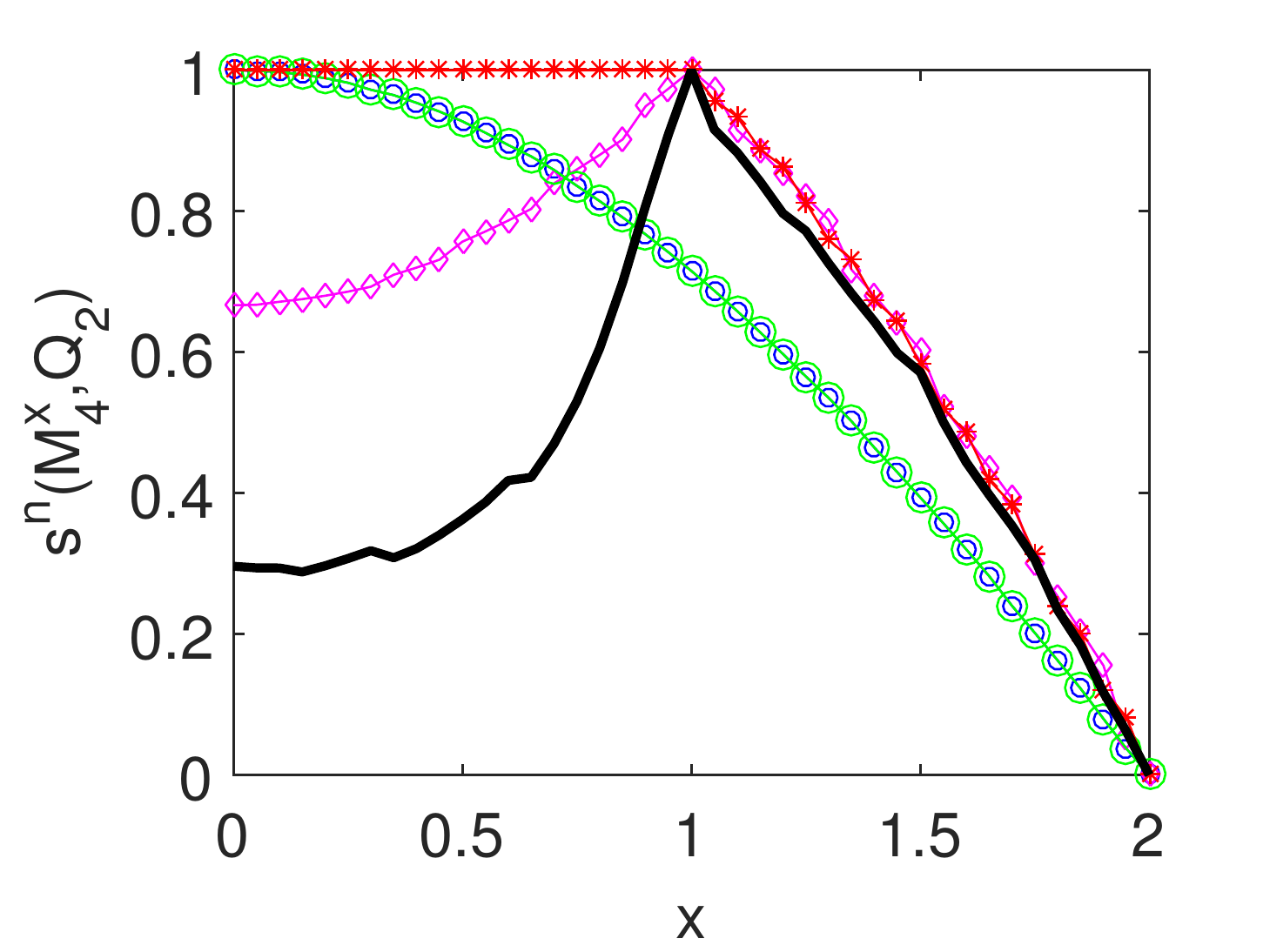}}\\
\subfloat[]{\label{fig:q3m1}\includegraphics[height=1.3in]{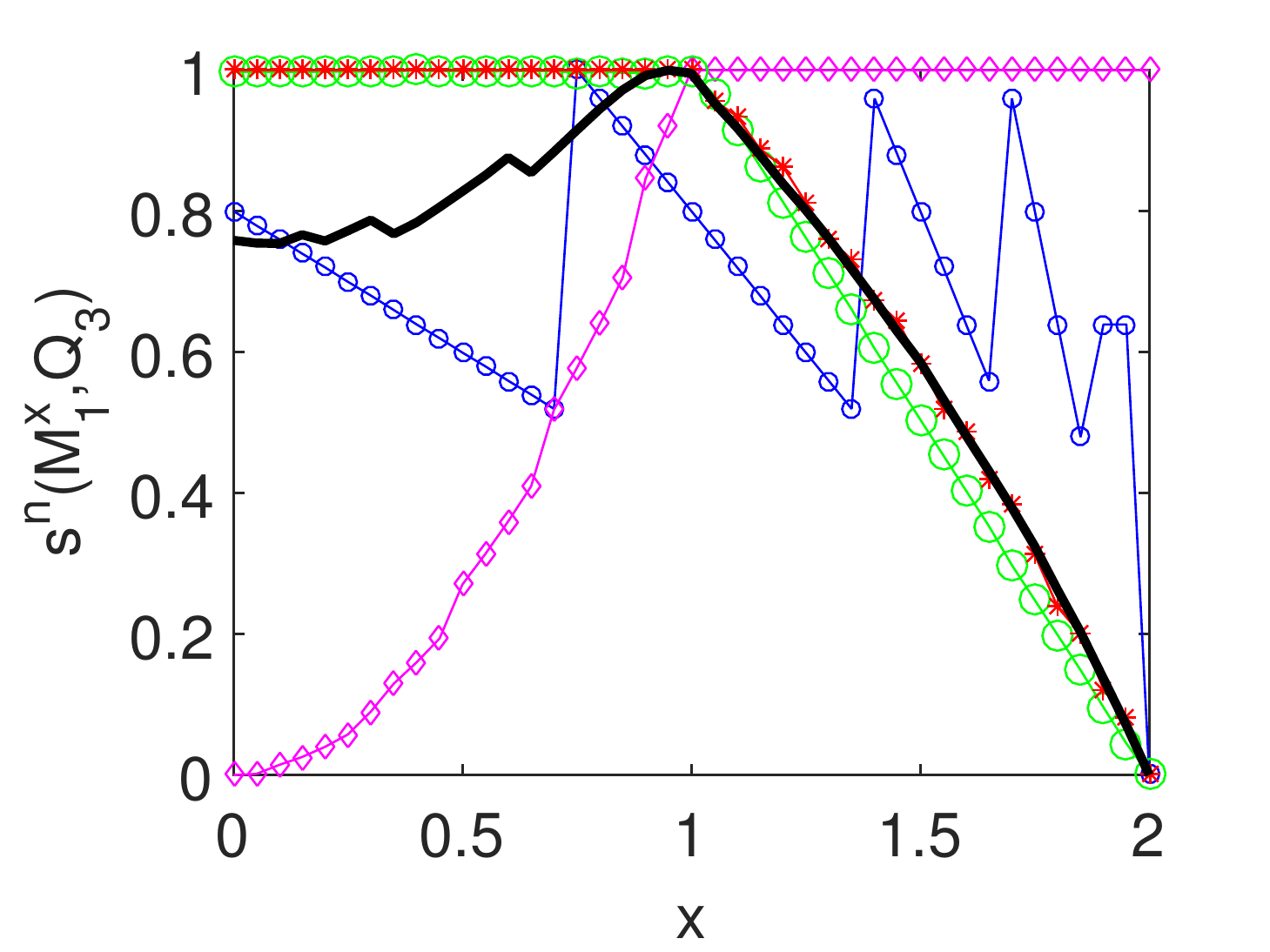}}
\hfill
\subfloat[]{\label{fig:q3m2}\includegraphics[height=1.3in]{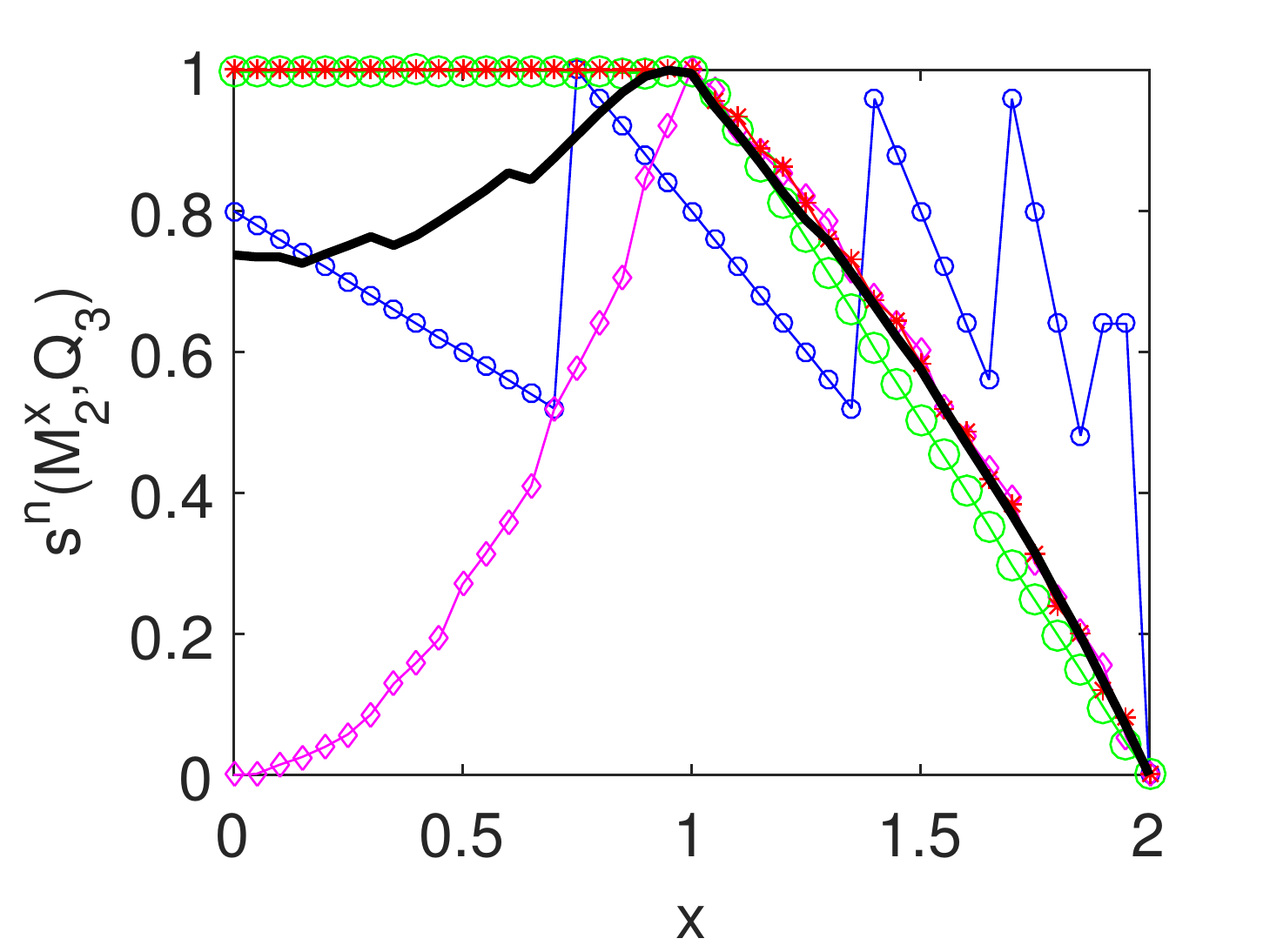}}
\hfill
\subfloat[]{\label{fig:q3m3}\includegraphics[height=1.3in]{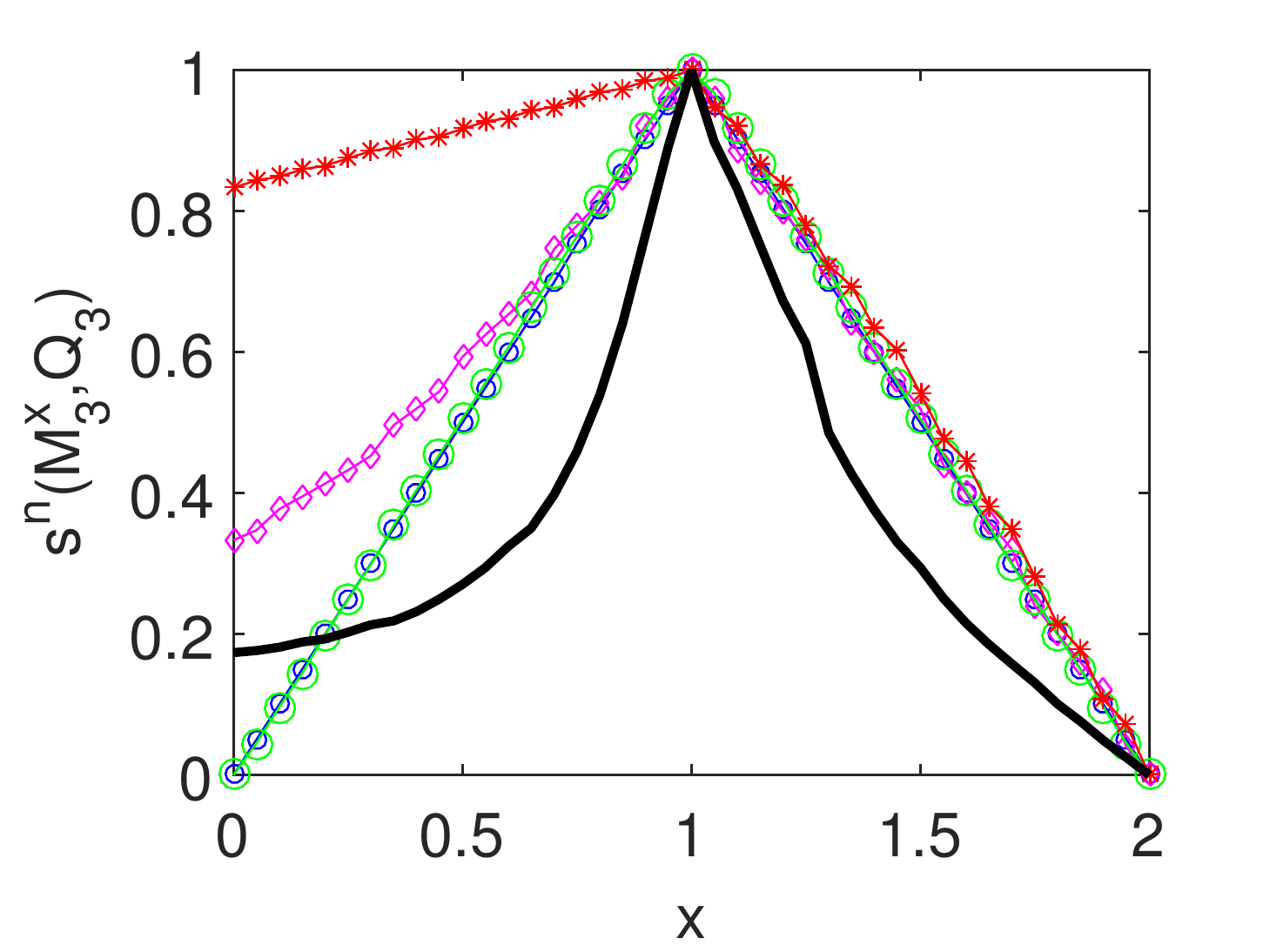}}
\hfill
\subfloat[]{\label{fig:q3m4}\includegraphics[height=1.3in]{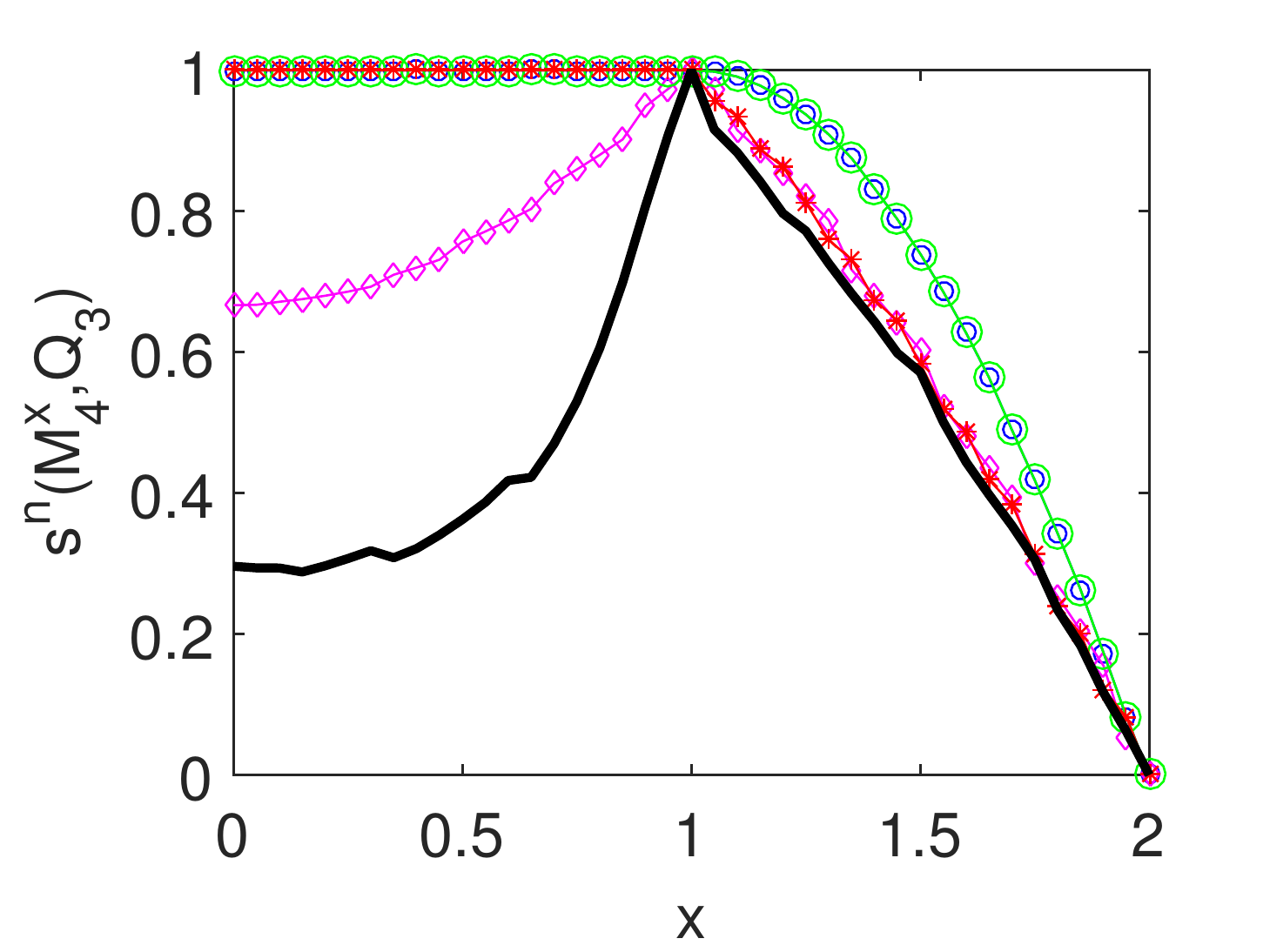}}\\
\subfloat[]{\label{fig:q4m1}\includegraphics[height=1.3in]{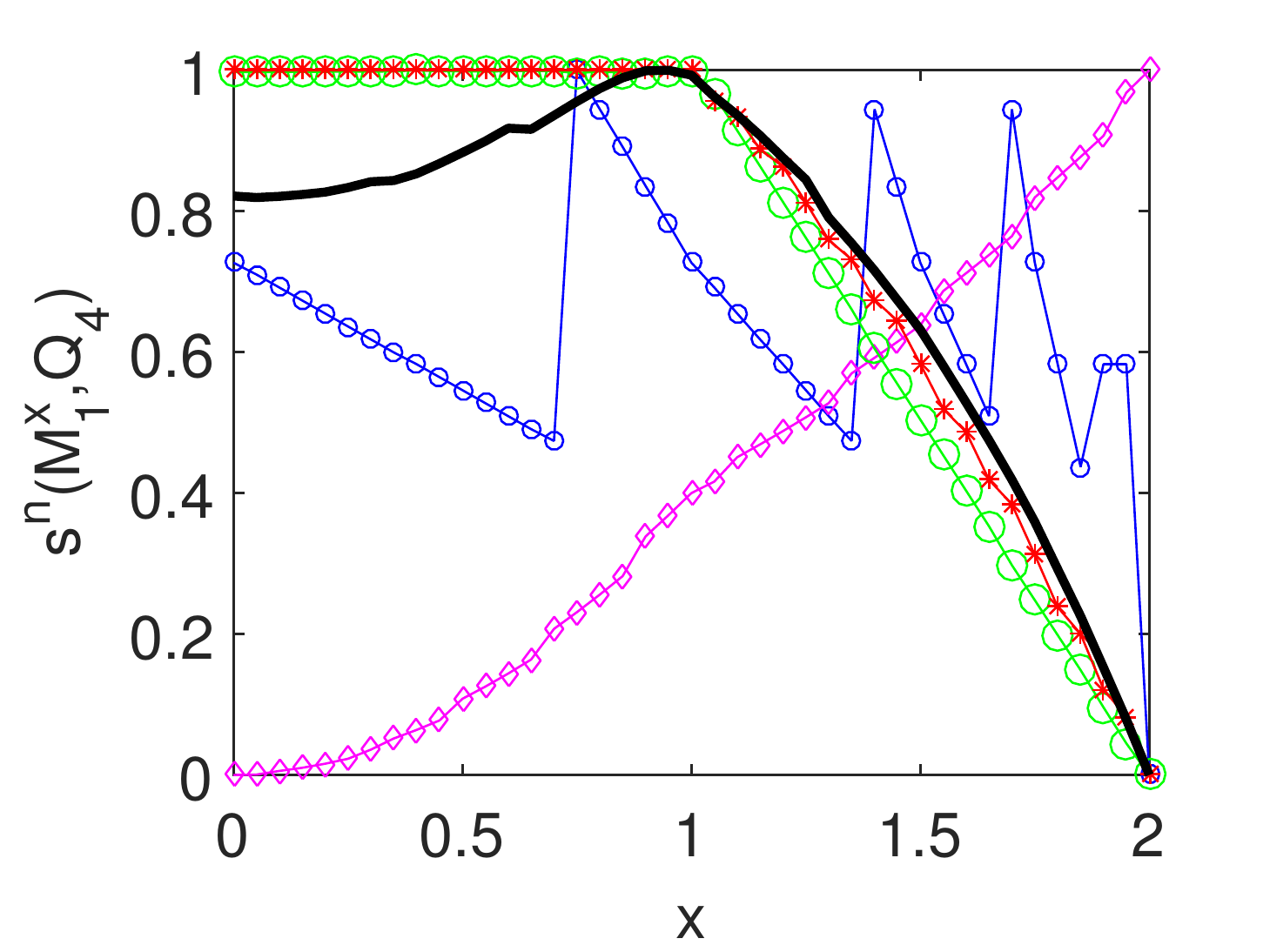}}
\hfill
\subfloat[]{\label{fig:q4m2}\includegraphics[height=1.3in]{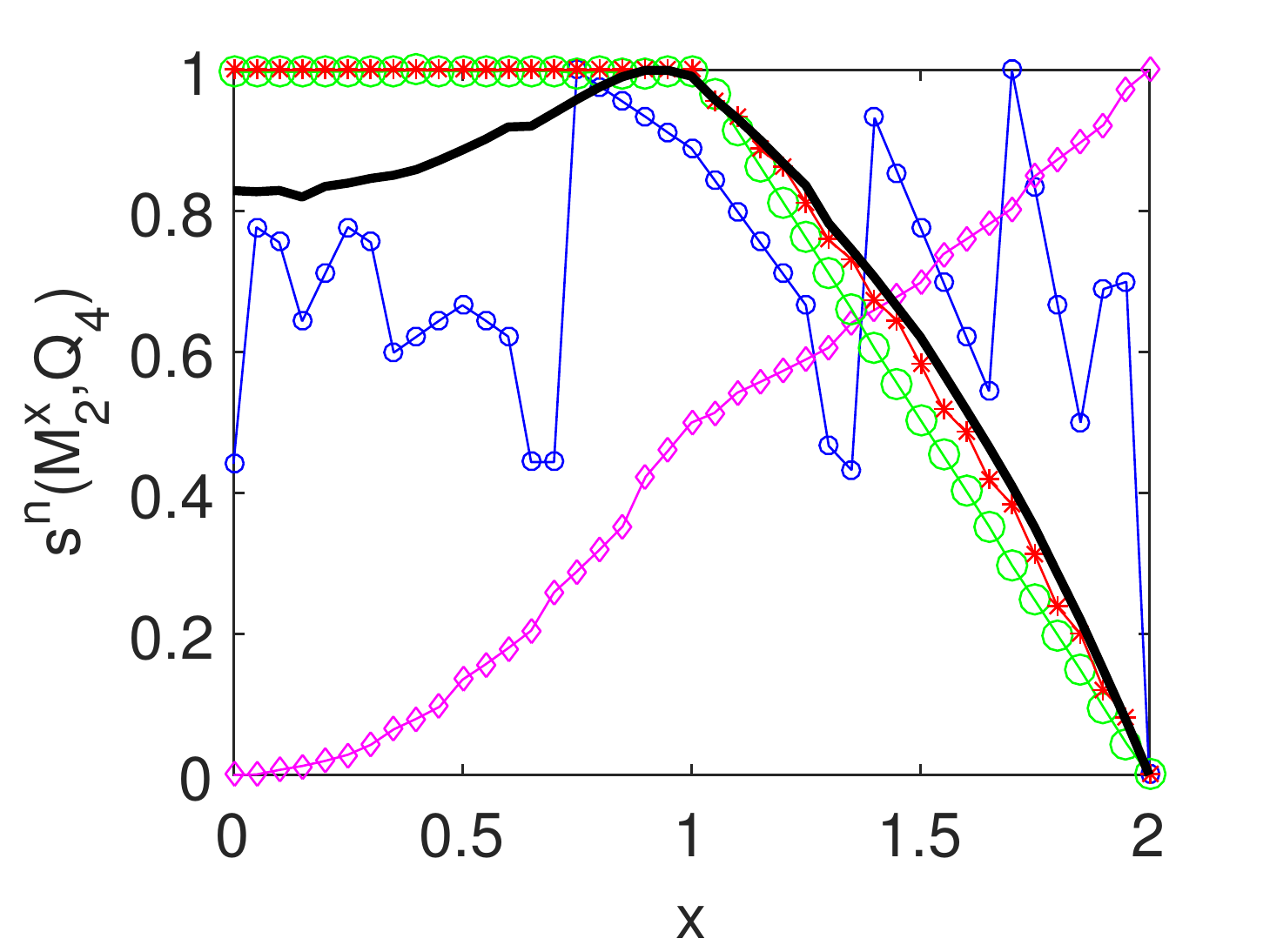}}
\hfill
\subfloat[]{\label{fig:q4m3}\includegraphics[height=1.3in]{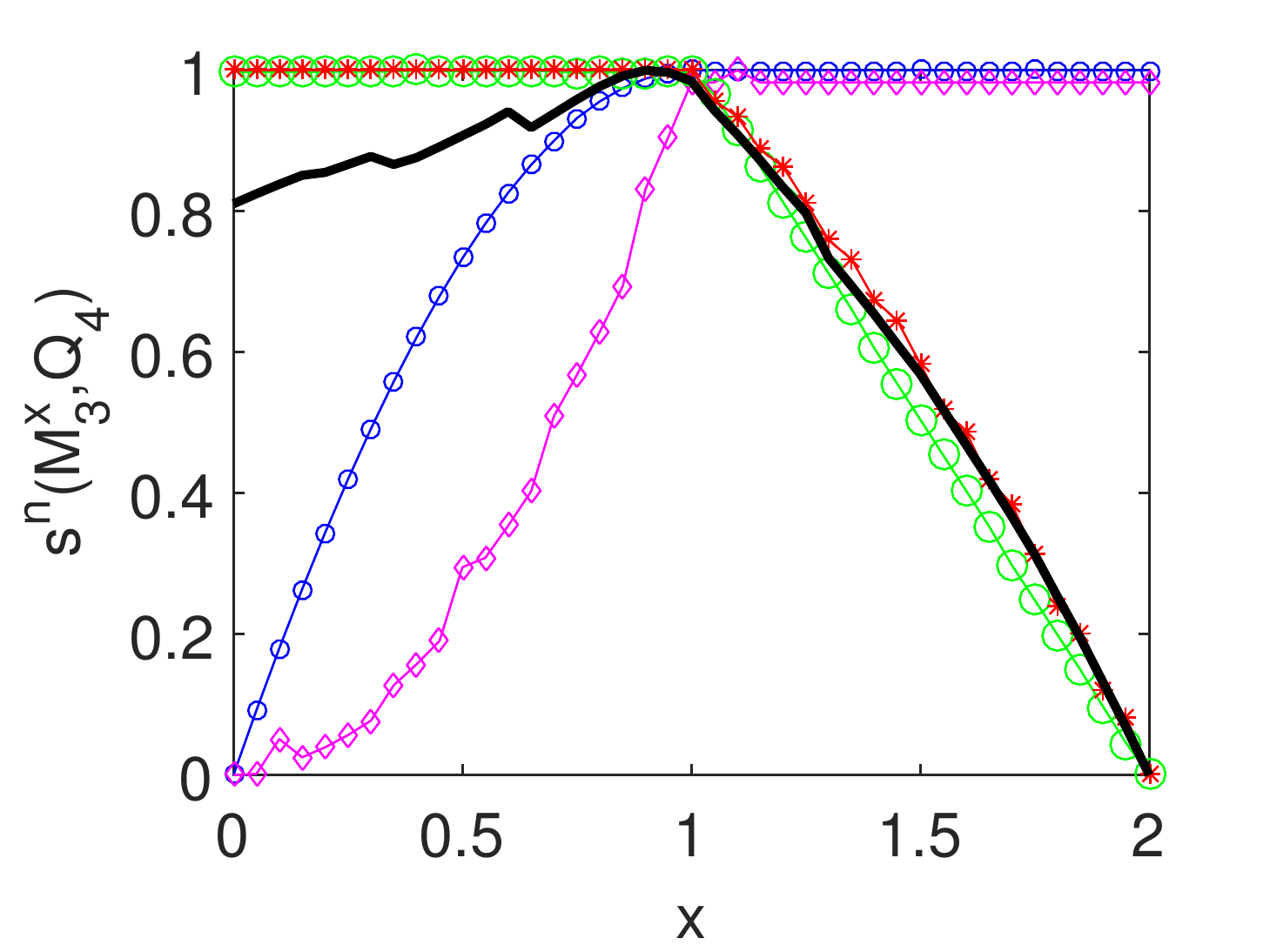}}
\hfill
\subfloat[]{\label{fig:q4m4}\includegraphics[height=1.3in]{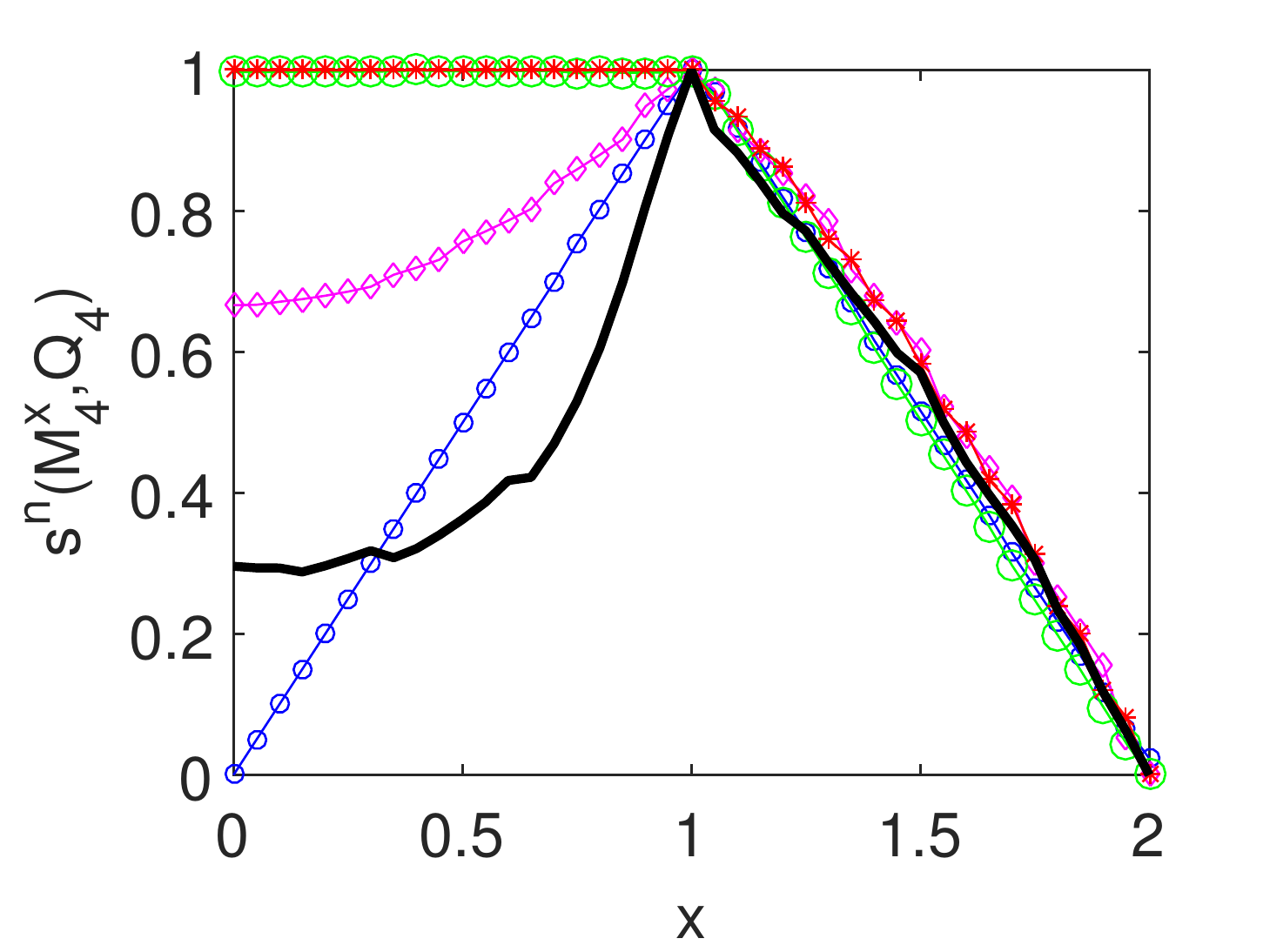}}
\caption{Metric comparison results. From left to right: the results of fitting Models $M^x_1$, $M^x_2$, $M^x_3$ and $M^x_4$ to the queries. From top to bottom: the results of fitting the models to Queries $Q_1$, $Q_2$, $Q_3$ and $Q_4$. The vertical axis ${s^n}( \cdot , \cdot )$ denotes the normalized similarity. We uniformly normalize the similarities into a range of $\left[ {0,1} \right]$. The legend for these figures are presented in \protect\subref{fig:q1m1}. The figures in diagonal, above diagonal and below diagonal represent the results of full fitting on complete data, partial fitting on over-complete data and partial fitting on incomplete data, respectively.}
\label{fig:comparisonresults}
\end{figure*}

The metrics used for comparison include negative SHD (-SHD), negative VD (-VD), negative OHD from query to model (-OHDQM), and inlier ratio (IR). VD has been used in \cite{talton_metropolis_2011} \cite{ritchie2015controlling}, while OHDQM has been used in \cite{ullrich2008semantic}. As the foundation of many BGMF methods such as \cite{fischler1981random} and \cite{isack2011energy-based}, IR is defined as:
\begin{equation}\label{eq:IR}
{s_{IR}(M,Q)} = \frac{{z(Q \cap M)}}{{z(Q)}}
\end{equation}
where $z( \cdot )$ denotes the size of discrete point set. The comparison results of fitting the models (Fig. \ref{fig:m1-4}) to the queries (Fig. \ref{fig:q1-4}) are shown in Fig. \ref{fig:comparisonresults}. To compute SHD, OHDQM and WMM, we uniformly sample points from the models with a resolution of 0.01, which is half of the query resolution. In these 16 experiments,  the weighting factor $h$ for WMM calculation is 2.5, and the resolution for VD calculation is 0.04. 

\begin{figure*}[ht!]
\centering
\subfloat[]{\label{fig:comparisontime}\includegraphics[width=0.6\columnwidth]{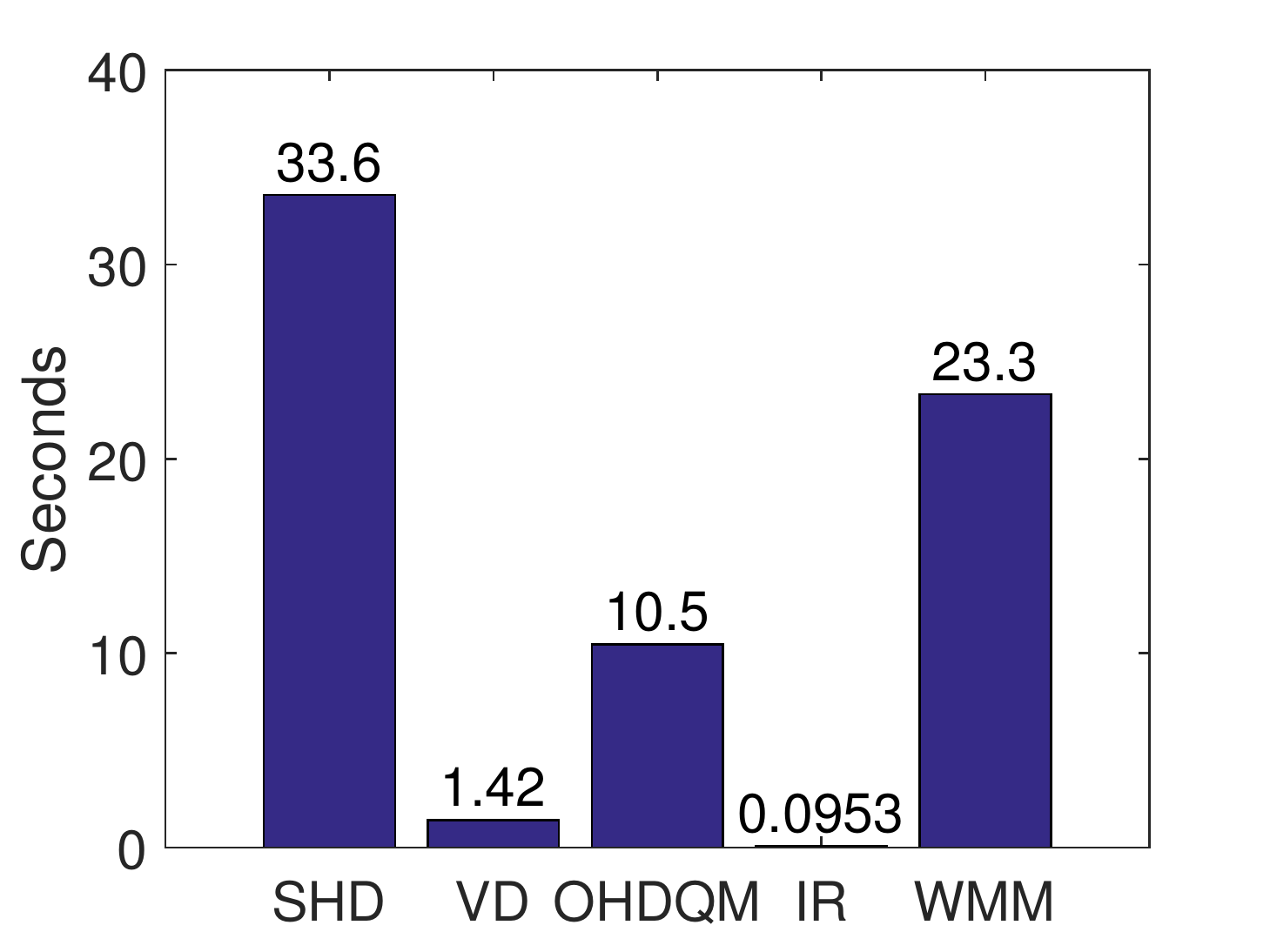}}
\hfill
\subfloat[]{\label{fig:vd}\includegraphics[width=0.6\columnwidth]{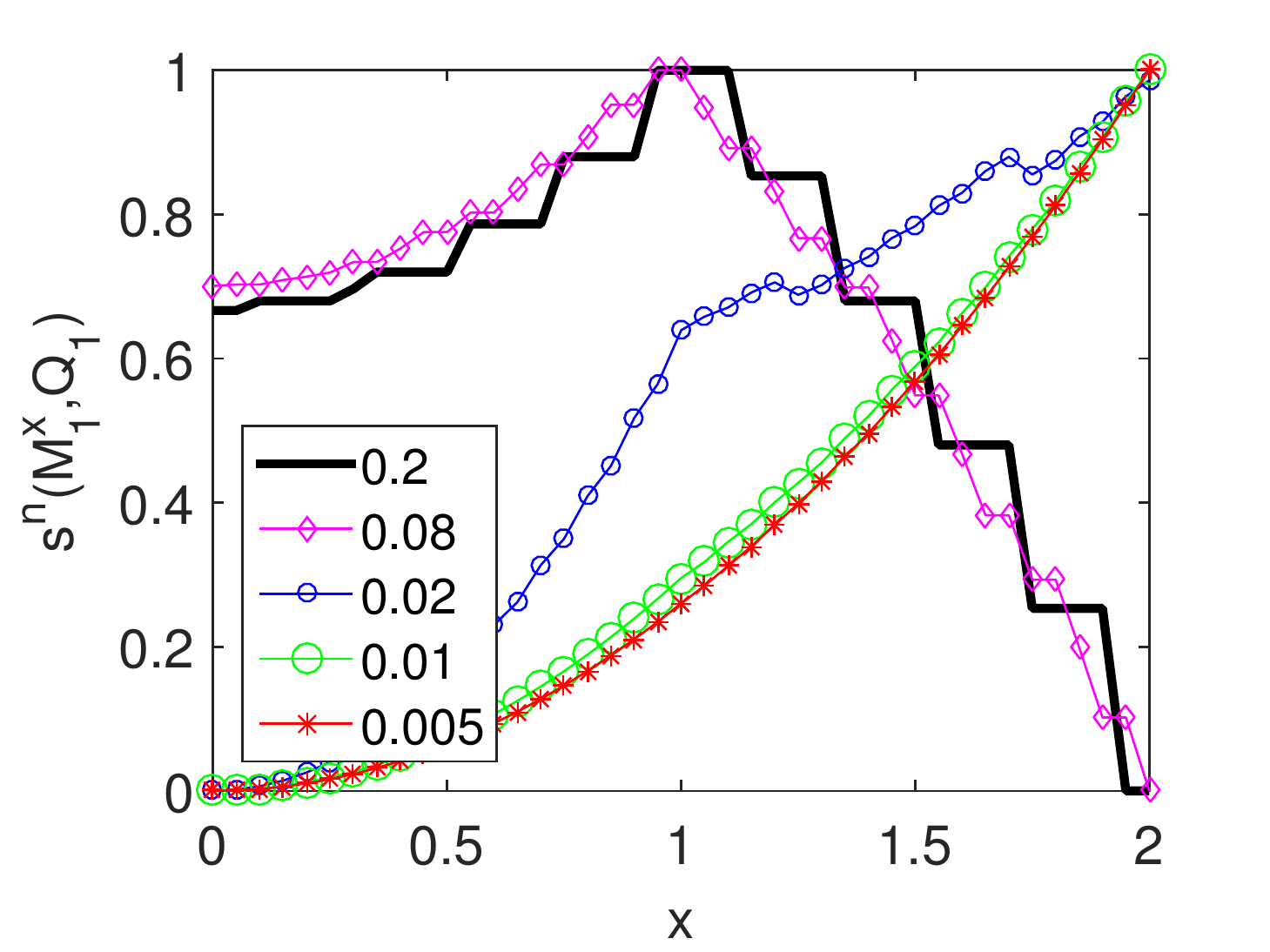}}
\hfill
\subfloat[]{\label{fig:factorh}\includegraphics[width=0.6\columnwidth]{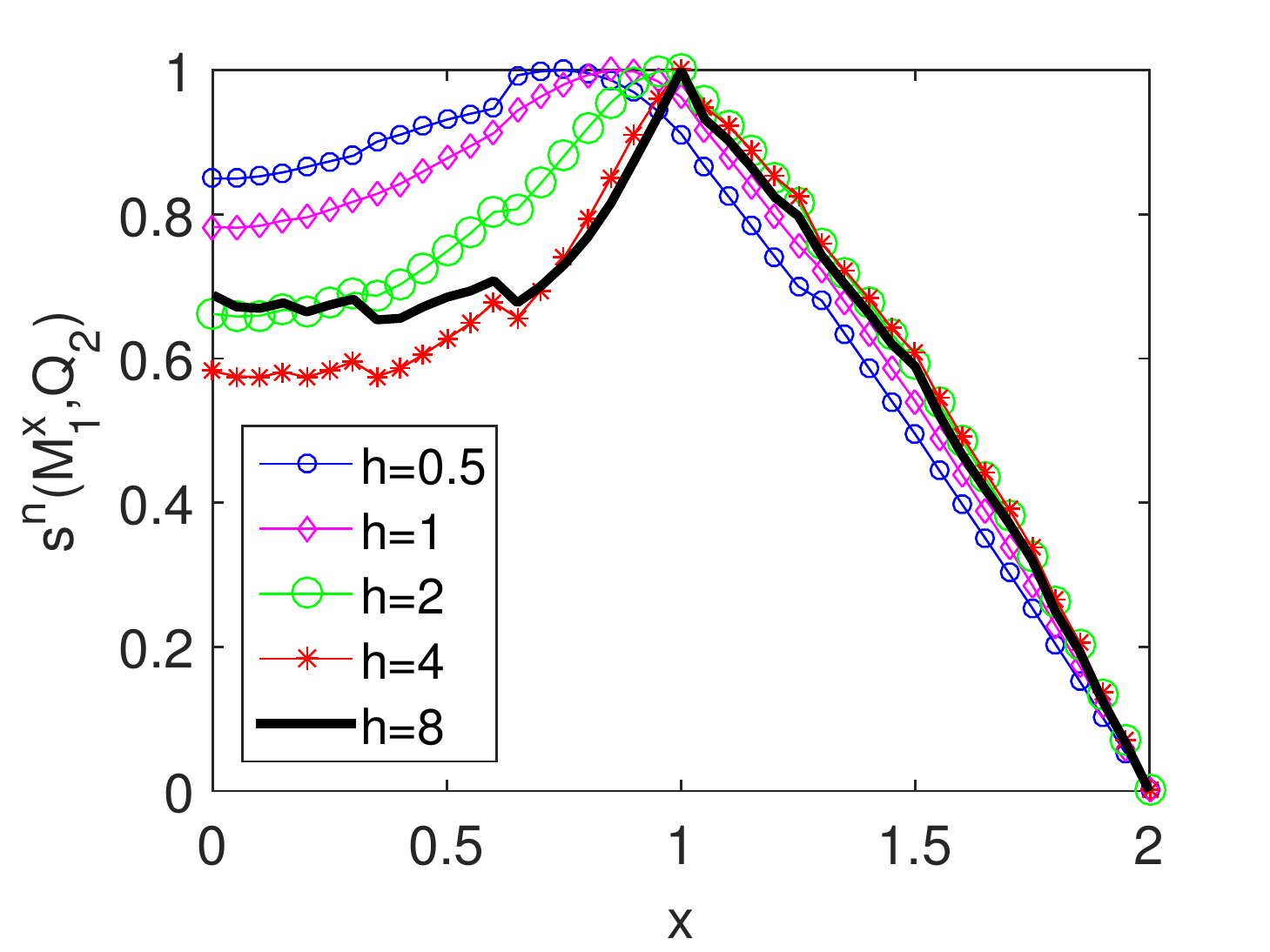}}
\caption{ \protect\subref{fig:comparisontime} Total time for similarity calculation on experiments shown in Fig. \ref{fig:comparisonresults}. \protect\subref{fig:vd} -VD similarities of fitting $M^x_1$ to $Q_1$ with resolutions 0.2, 0.08, 0.02, 0.01 and 0.005. \protect\subref{fig:factorh} WMM similarities of fitting $M^x_1$ to $Q_2$ with weighting factor $h$=0.5, 1, 2, 4 and 8.}
\label{fig:comparisonresults2}
\end{figure*}

As the target models of the queries are models with $x=1$, it is expected that the models with $x=1$ have the largest similarities. As shown in Fig. \ref{fig:comparisonresults}, our WMM is the only metric to achieve this goal for all experiments. SHD is successful for full fitting (Figs. \ref{fig:q1m1}, \ref{fig:q2m2}, \ref{fig:q3m3} and \ref{fig:q4m4}), but failed for partial fitting except Fig. \ref{fig:q2m3}. IR is failed to distinguish models with $x<1$ for all experiments except Fig. \ref{fig:q3m3}. The total computational time of these 16 experiments is shown in Fig. \ref{fig:comparisontime}, it can be seen that WMM is faster than SHD. 

It is worth noting that SHD, OHDQM and WMM always prefer to sample points from model with smaller resolution to obtain more accurate similarity. However, VD produces worse results with smaller resolution for a discrete query. Fine voxelization of a discrete query produces more empty voxels. Therefore, an empty model (e.g. $M^{x=2}_1$) is preferred, of that one example is shown in Fig. \ref{fig:vd}. This indicates that voxelization is not suitable for the fine fitting of point clouds.

It is interesting to find that Fig. \ref{fig:q1m3} resembles Fig. \ref{fig:q1m4}. Actually, before normalizing, the original similarities are different. As shown in Table \ref{tab:originalsim}, the WMM similarities are comparable across different model spaces for the same query.
This table along with Fig. \ref{fig:comparisonresults} demonstrates that, a model is most similar to itself than any other models using WMM. Finally, we take the experiment of fitting $M^x_1$ to $Q_2$ as an example to evaluate the effect of weighting factor $h$. As shown in Fig. \ref{fig:factorh}, WMM is very stable with respect to different values of $h$.

\begin{table}[ht!]
	\centering
		\begin{tabular}{|c|c|c|c|c|}\hline
			 \textbf{Model}&$M^{x=1}_1$&$M^{x=1}_2$&$M^{x=1}_3$&$M^{x=1}_4$\\\hline
			 \textbf{Query}&&&&\\
			 $Q_1$ &\textbf{1056.4}&792.288&528.195&264.094\\
			 $Q_2$ &239.651&\textbf{792.288}&528.195&264.094\\
			 $Q_3$ &96.4585&113.127&\textbf{528.195}&264.094\\
			 $Q_4$ &32.6646&34.4811&44.3055&\textbf{264.094}\\\hline
		\end{tabular}
	\caption{WMM similarities between the target models and queries. The diagonal elements represent the similarities between the ground-truth models and queries. It is shown that, for the same query, the similarity between the ground-truth model and the query is the largest among all similarities.}
	\label{tab:originalsim}
\end{table}

\subsection{Fitting Noisy Data}\label{sec:sphere}
\begin{figure}[ht!]
\centering
\subfloat[]{\label{fig:sphere-query1}\includegraphics[height=0.8in]{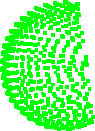}}
\hfill
\subfloat[]{\label{fig:sphere-query2}\includegraphics[height=0.8in]{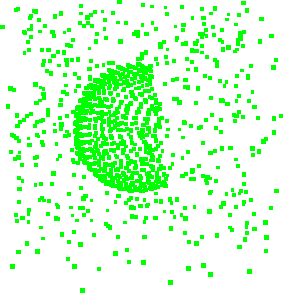}}
\hfill
\subfloat[]{\label{fig:sphere-query3}\includegraphics[height=0.8in]{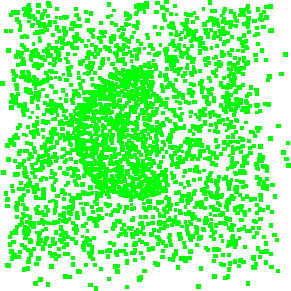}}
\hfill
\subfloat[]{\label{fig:sphere-query4}\includegraphics[height=0.8in]{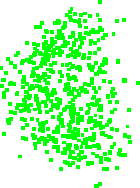}}
\caption{Queries. From left to right: Queries $Q_5$, $Q_6$, $Q_7$ and $Q_8$. $Q_5$ is a noise-free point cloud consisting of 549 points sampled from a unit sphere surface. $Q_6$ and $Q_7$ are generated by adding low-level and high-level uniform noise to $Q_5$, respectively. $Q_6$ consists of 1098 points, while $Q_7$ consists of 2985 points. The uniform noise is distributed within a cube with a length of 2. The cube and the unit sphere have the same center. $Q_8$ is generated by adding Gaussian noise with a standard deviation of 0.2 to $Q_5$.}
\label{fig:sphere-queries}
\end{figure}

\begin{figure}[ht!]
\centering
\subfloat[]{\label{fig:sphere-model1}\includegraphics[height=0.8in]{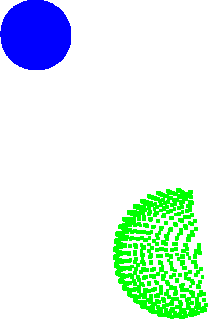}}
\hfill
\subfloat[]{\label{fig:sphere-model2}\includegraphics[height=0.8in]{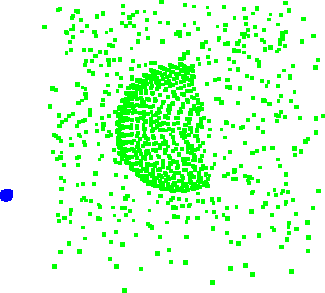}}
\hfill
\subfloat[]{\label{fig:sphere-model3}\includegraphics[height=0.8in]{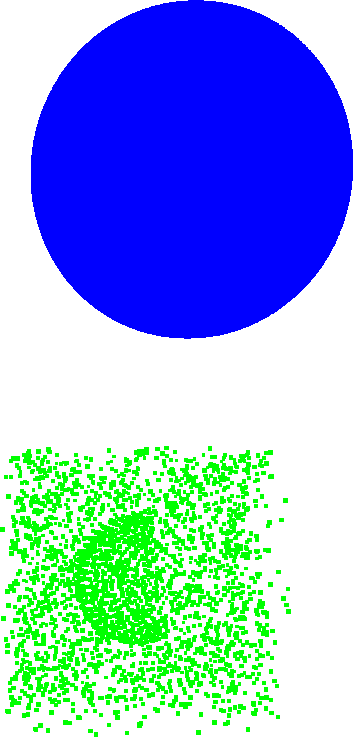}}
\hfill
\subfloat[]{\label{fig:sphere-model4}\includegraphics[height=0.8in]{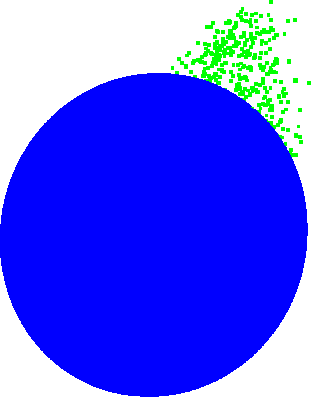}}\\
\subfloat[]{\label{fig:sphere-model5}\includegraphics[height=0.8in]{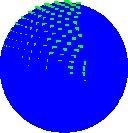}}
\hfill
\subfloat[]{\label{fig:sphere-model6}\includegraphics[height=0.8in]{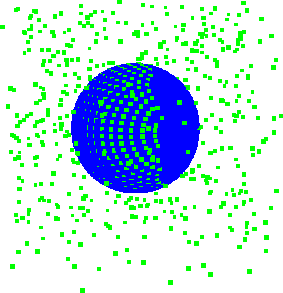}}
\hfill
\subfloat[]{\label{fig:sphere-model7}\includegraphics[height=0.8in]{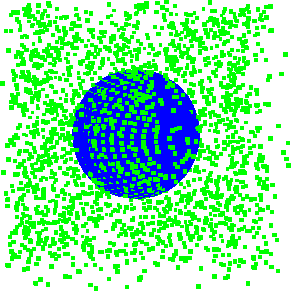}}
\hfill
\subfloat[]{\label{fig:sphere-model8}\includegraphics[height=0.8in]{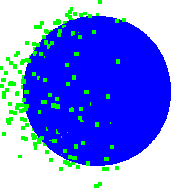}}
\caption{Fitted sphere models (blue) along with the queries (green, Fig. \ref{fig:sphere-queries}). Top row: randomly initialized models. Bottom row: final fitted models (Fig. \ref{fig:sphere-results}). From left to right: fitting the sphere to $Q_5$, $Q_6$, $Q_7$ and $Q_8$, respectively.}
\label{fig:sphere-models}
\end{figure}

We evaluate our method against uniform and Gaussian noise by fitting a sphere model $\mathcal{M}_5$ to 4 queries (Fig. \ref{fig:sphere-queries}). We use the method in \cite{marsaglia1972choosing} to sample points from the sphere surface and replace $\gamma $ (see Eq. \eqref{eq:divdinglevel}) by $2\pi R$, where $R$ is the radius of the sphere. The sphere model $\mathcal{M}_5$ has 4 parameters, including 3 location parameters and 1 radius parameter. Some sample models of $\mathcal{M}_5$ are shown in the top row of Fig. \ref{fig:sphere-models}. It is worth noting that all the model parameters involved in this paper are uniformly distributed. As a result, the optimization objective is reduced from posterior to likelihood and then rigid similarity. Besides, we stop model fitting casually in this paper.

\begin{figure*}[ht!]
\centering
\subfloat[]{\label{fig:sphere-result2}\includegraphics[width=0.6\columnwidth]{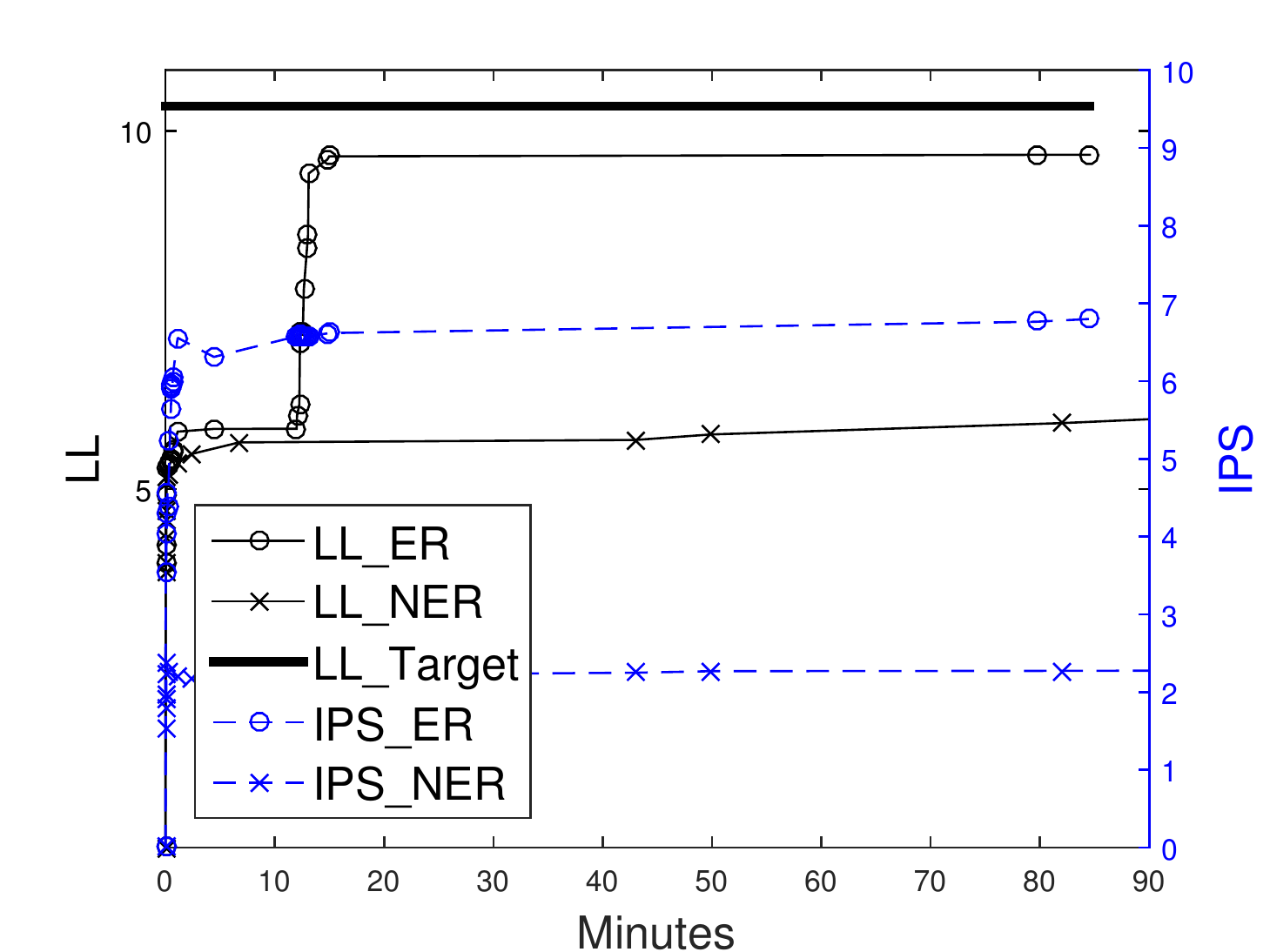}}
\hfill
\subfloat[]{\label{fig:sphere-result3}\includegraphics[width=0.6\columnwidth]{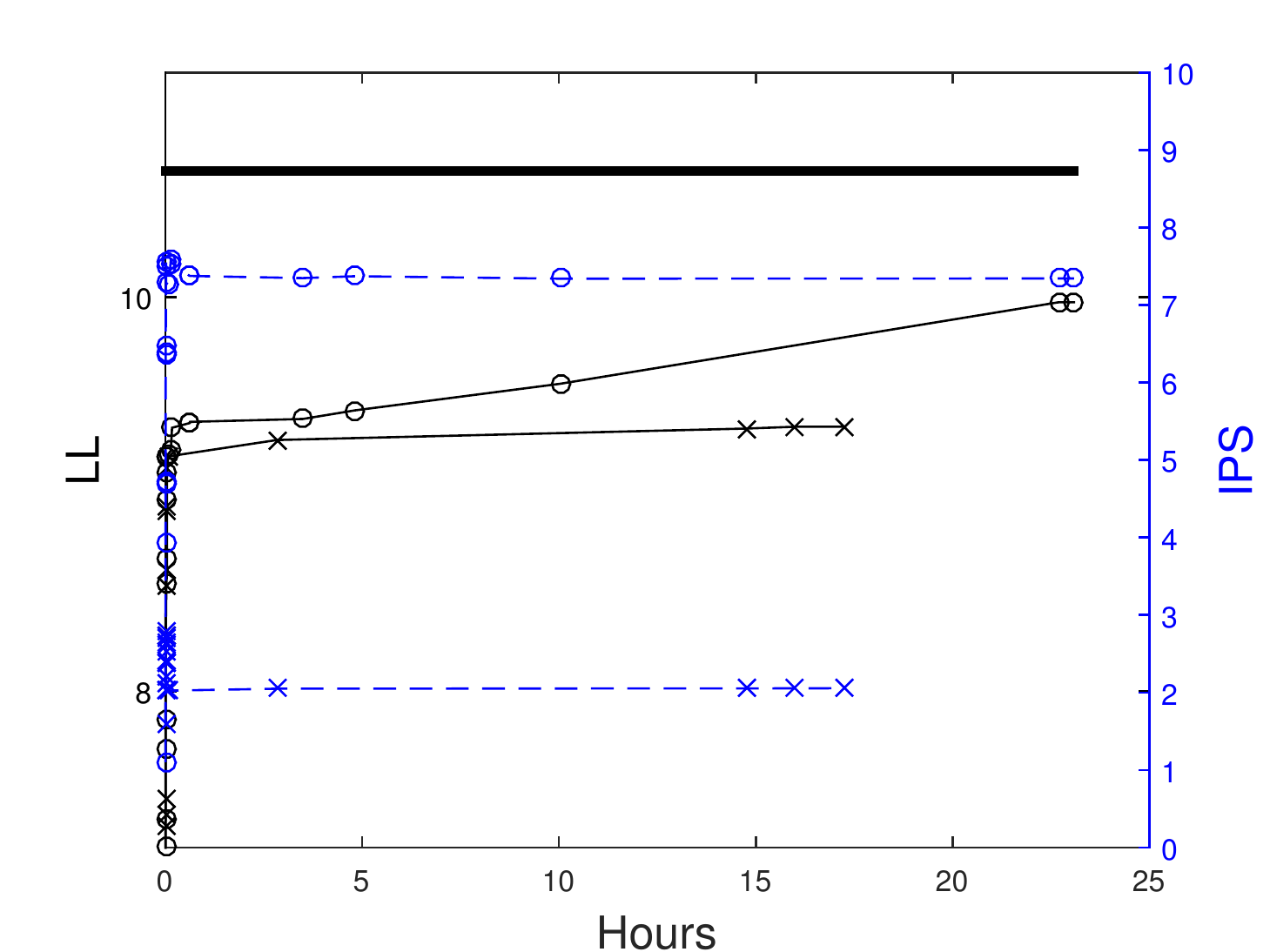}}
\hfill
\subfloat[]{\label{fig:sphere-result4}\includegraphics[width=0.6\columnwidth]{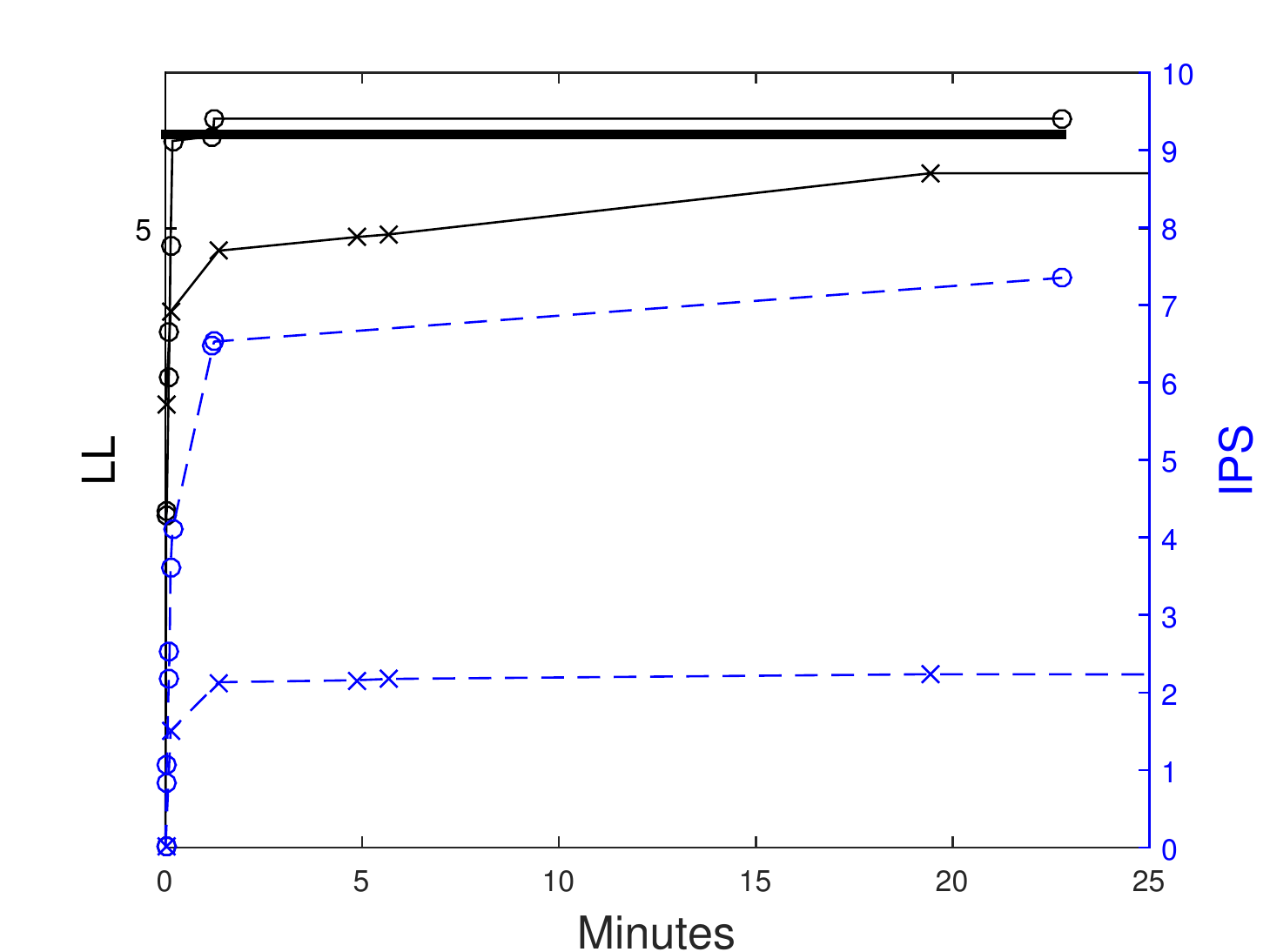}}
\caption{Sphere fitting results.  \protect\subref{fig:sphere-result2}, \protect\subref{fig:sphere-result3} and \protect\subref{fig:sphere-result4} are the results of fitting the sphere model $\mathcal{M}_5$ to $Q_6$, $Q_7$ and $Q_8$ (Fig. \ref{fig:sphere-queries}), respectively. LL, IPS and ER denote log likelihood, iterations per second, and early rejection, respectively. LL\_ER, LL\_NER, IPS\_ER and IPS\_NER denote the evolutions of LL with ER, LL without ER, IPS with ER, and IPS without ER, respectively. LL\_Target denotes the target log likelihood.
}
\label{fig:sphere-results}
\end{figure*}

The resolution of the noise-free query $Q_5$ (Fig. \ref{fig:sphere-query1}) is 0.2. We set $\delta=0.04$ and $h=10$ in these sphere fitting experiments, the results are presented in Figs. \ref{fig:sphere-models} and \ref{fig:sphere-results}. As shown in Figs. \ref{fig:sphere-result2} and \ref{fig:sphere-result3}, the target similarities (log likelihood) still remain the largest similarities after sufficient evolution time. This indicates that our method is robust to uniform noise. Similarly, Fig. \ref{fig:sphere-result4} shows that our method is also robust to Gaussian noise. Our method can only be slightly affected by Gaussian noise.

However, uniform noise influences the efficiency of our method. Particularly, tens of minutes have been consumed to generate the results shown in Figs. \ref{fig:sphere-result2} and \ref{fig:sphere-result4}, however, several hours have been consumed to obtain a desirable model shown in Fig. \ref{fig:sphere-result3}. This is because high-level uniform noise introduces many local maxima to the objective function, making it difficult to find the global maximum. The IPS indicator in Fig. \ref{fig:sphere-results} shows the influence of our early rejection strategy. It can be observed that,  the optimization process is accelerated by about 3 times using early rejection. These experiments additionally demonstrate that our method is able to fit non-planar models.

\subsection{Fitting Models with length-varying parameters} \label{sec:recursive}

\begin{figure*}[ht!]
\centering
\subfloat[]{\label{fig:recursive-query1}\includegraphics[width=0.45\columnwidth]{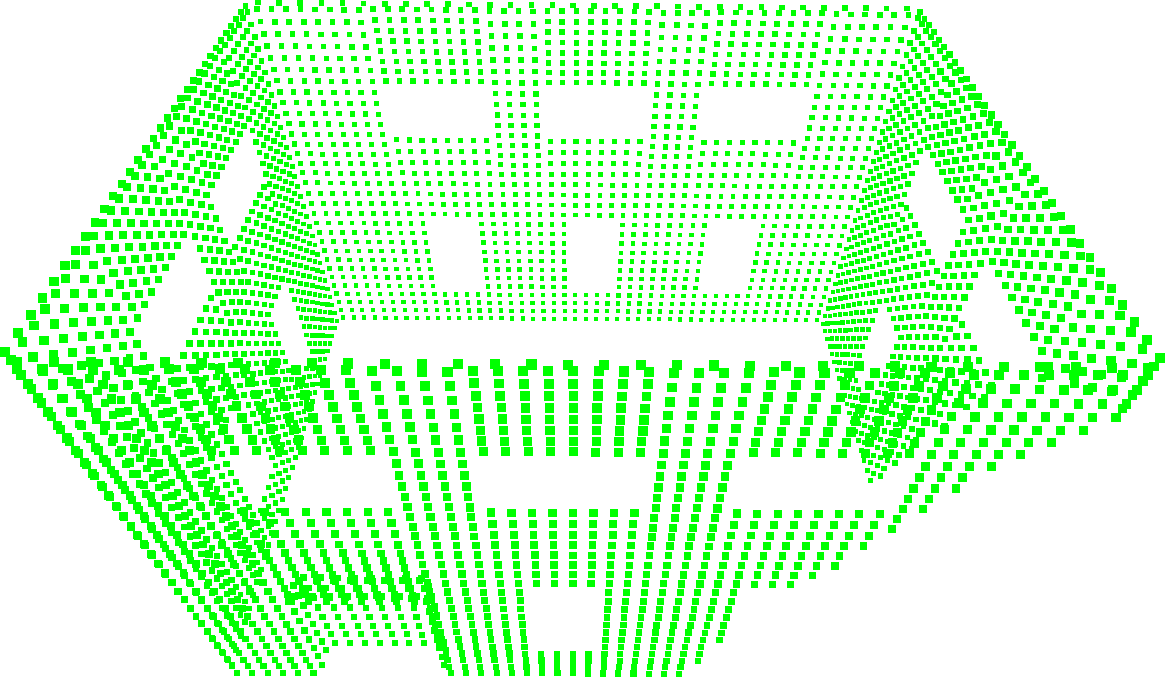}}
\hfill
\subfloat[]{\label{fig:recursive-query2}\includegraphics[width=0.45\columnwidth]{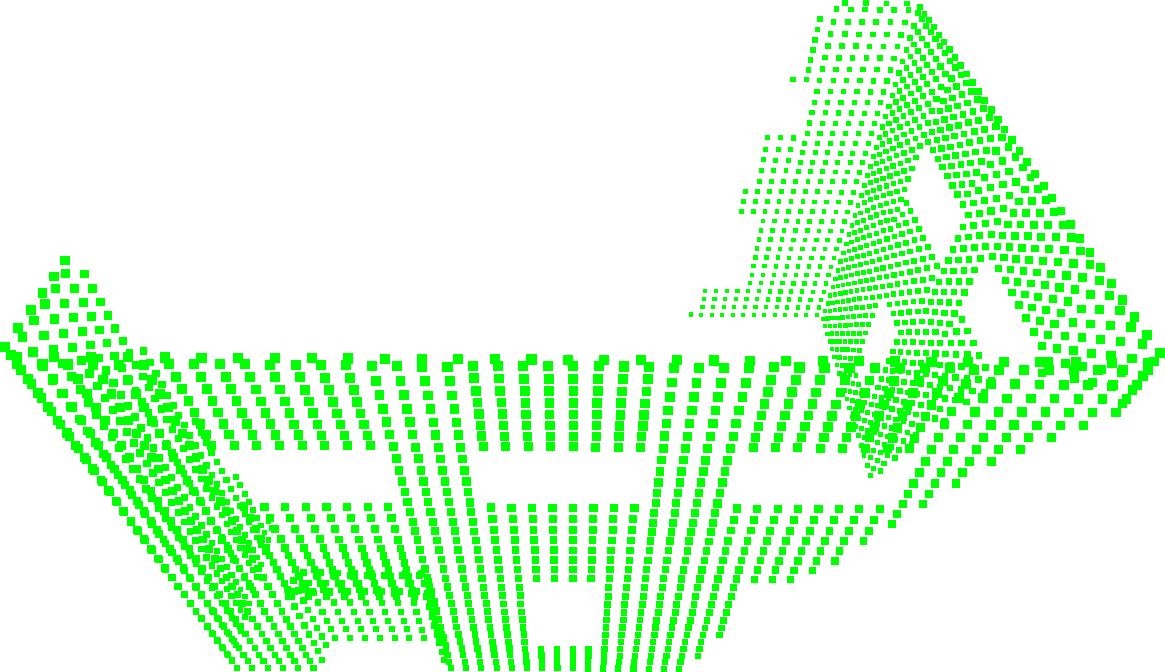}}
\hfill
\subfloat[]{\label{fig:recursive-query3}\includegraphics[width=0.45\columnwidth]{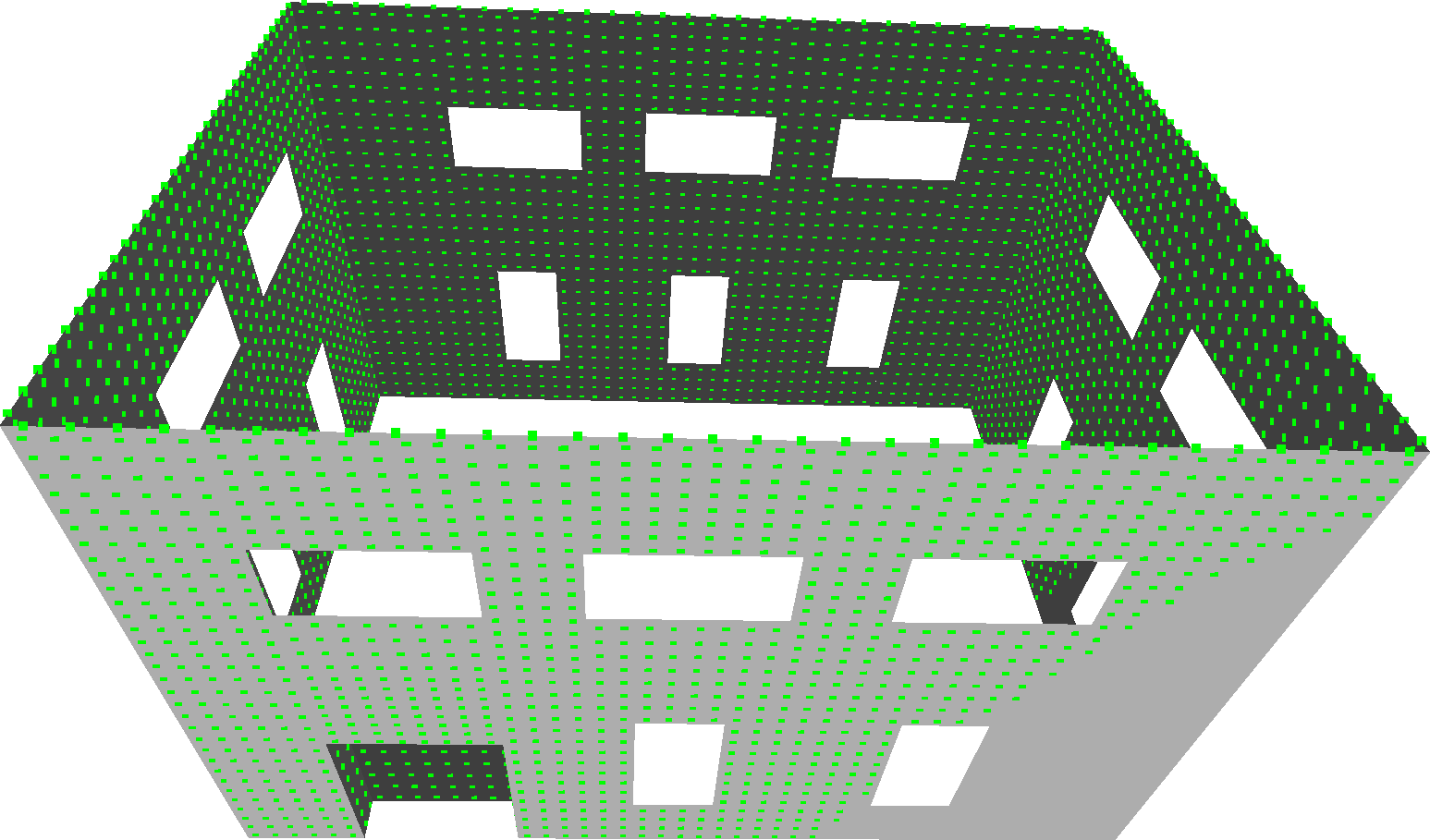}}
\hfill
\subfloat[]{\label{fig:recursive-query4}\includegraphics[width=0.45\columnwidth]{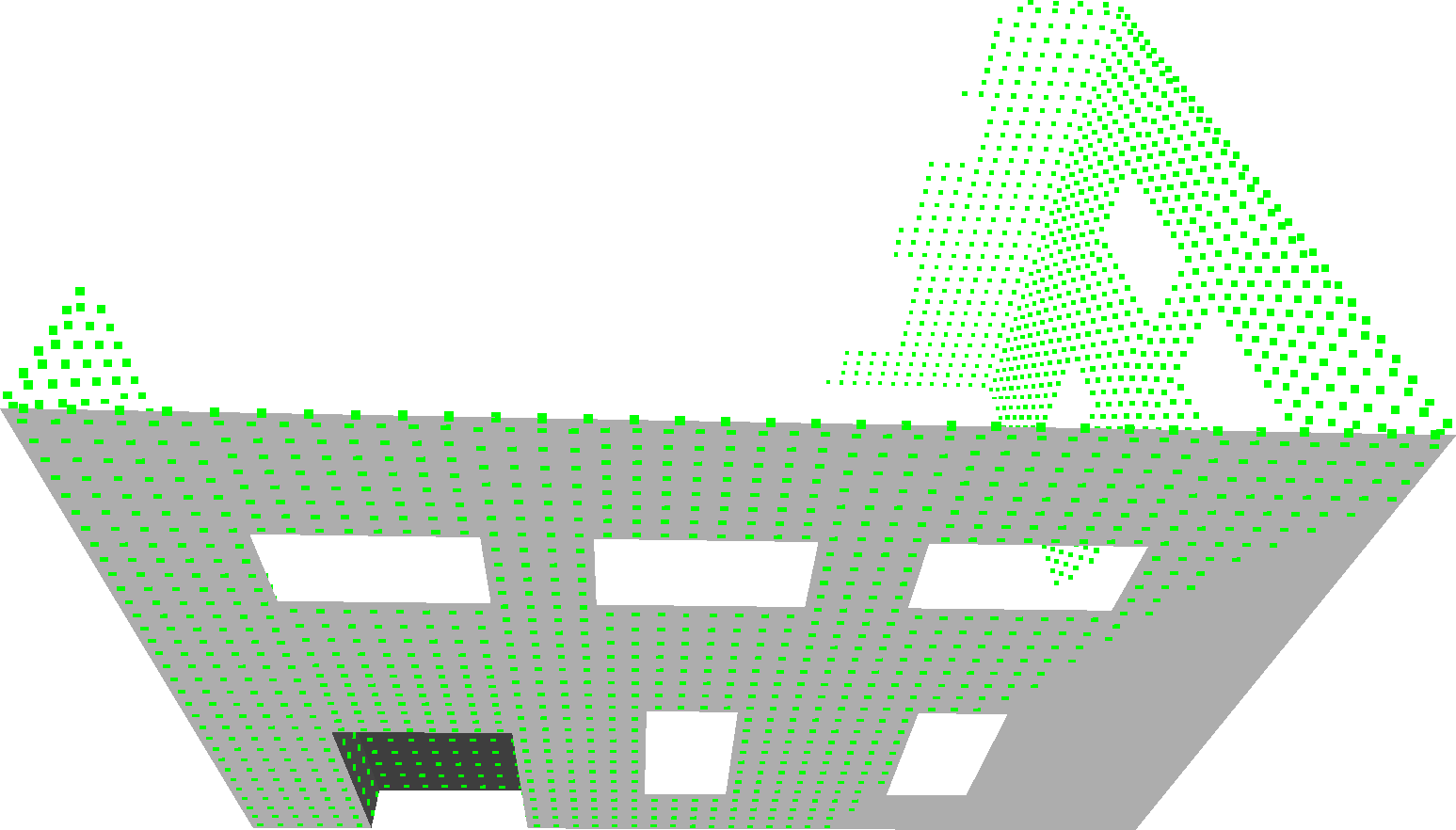}}
\caption{Queries and target models. \protect\subref{fig:recursive-query1} Query $Q_9$, \protect\subref{fig:recursive-query2} Query $Q_{10}$. \protect\subref{fig:recursive-query3} and \protect\subref{fig:recursive-query4} are models in Model spaces $\mathcal{M}_6$ and $\mathcal{M}_7$, respectively. Both $Q_9$ and $Q_{10}$ are part of the point cloud which is uniformly sampled from \protect\subref{fig:recursive-query3} with resolution 0.2. \protect\subref{fig:recursive-query4} is a part of \protect\subref{fig:recursive-query3}. Consequently, \protect\subref{fig:recursive-query3} and \protect\subref{fig:recursive-query4} are target models of $Q_9$ (and $Q_{10}$) in $\mathcal{M}_6$ and $\mathcal{M}_7$, respectively. $Q_9$ consists of 4204 points, while $Q_{10}$ consists of 2452 points.}
\label{fig:recursive-queries}
\end{figure*}

\begin{figure*}[ht!]
\centering
\subfloat[]{\label{fig:recursive-sample1}\includegraphics[height=0.8in]{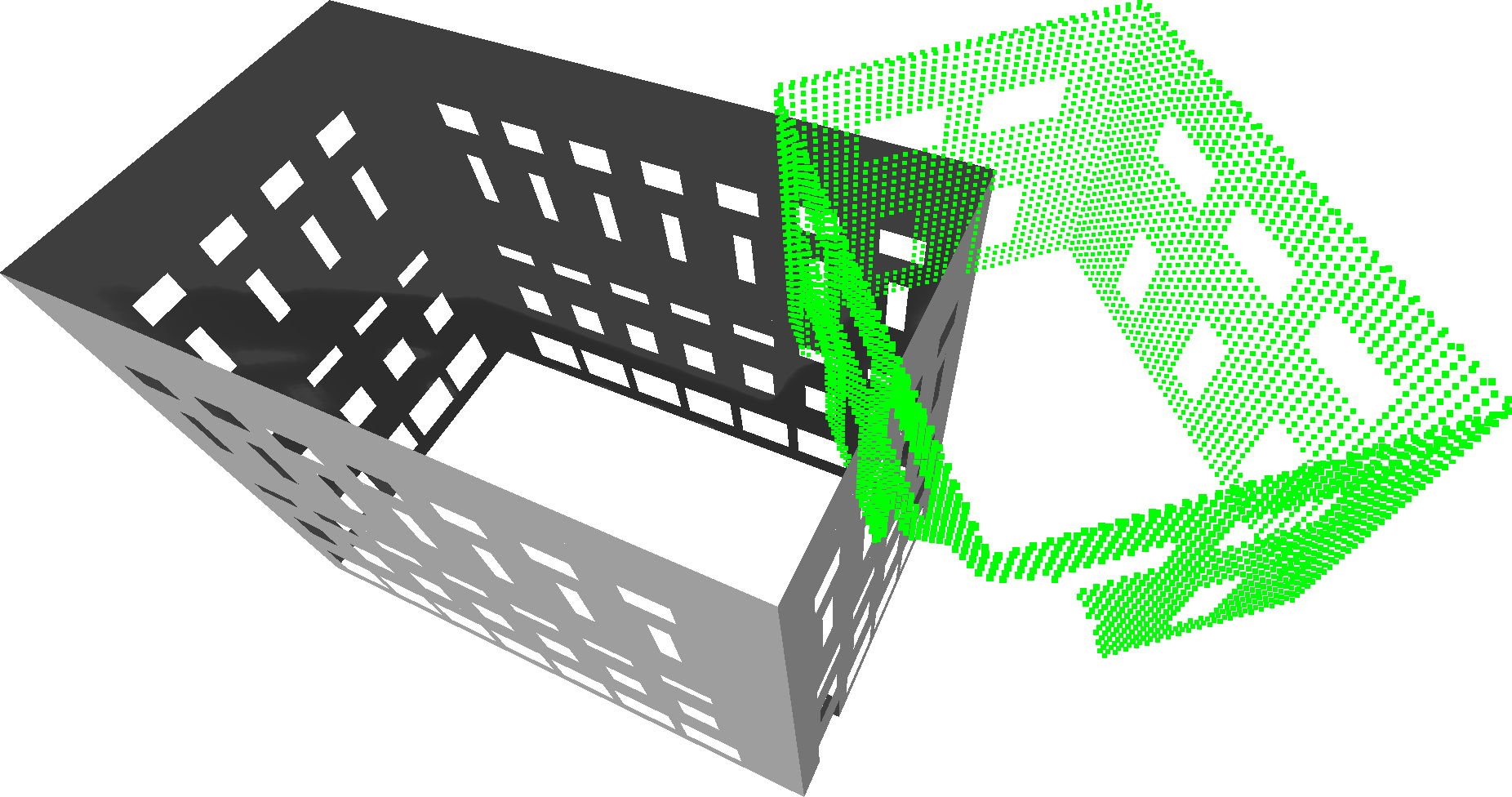}}
\hfill
\subfloat[]{\label{fig:recursive-sample2}\includegraphics[height=0.8in]{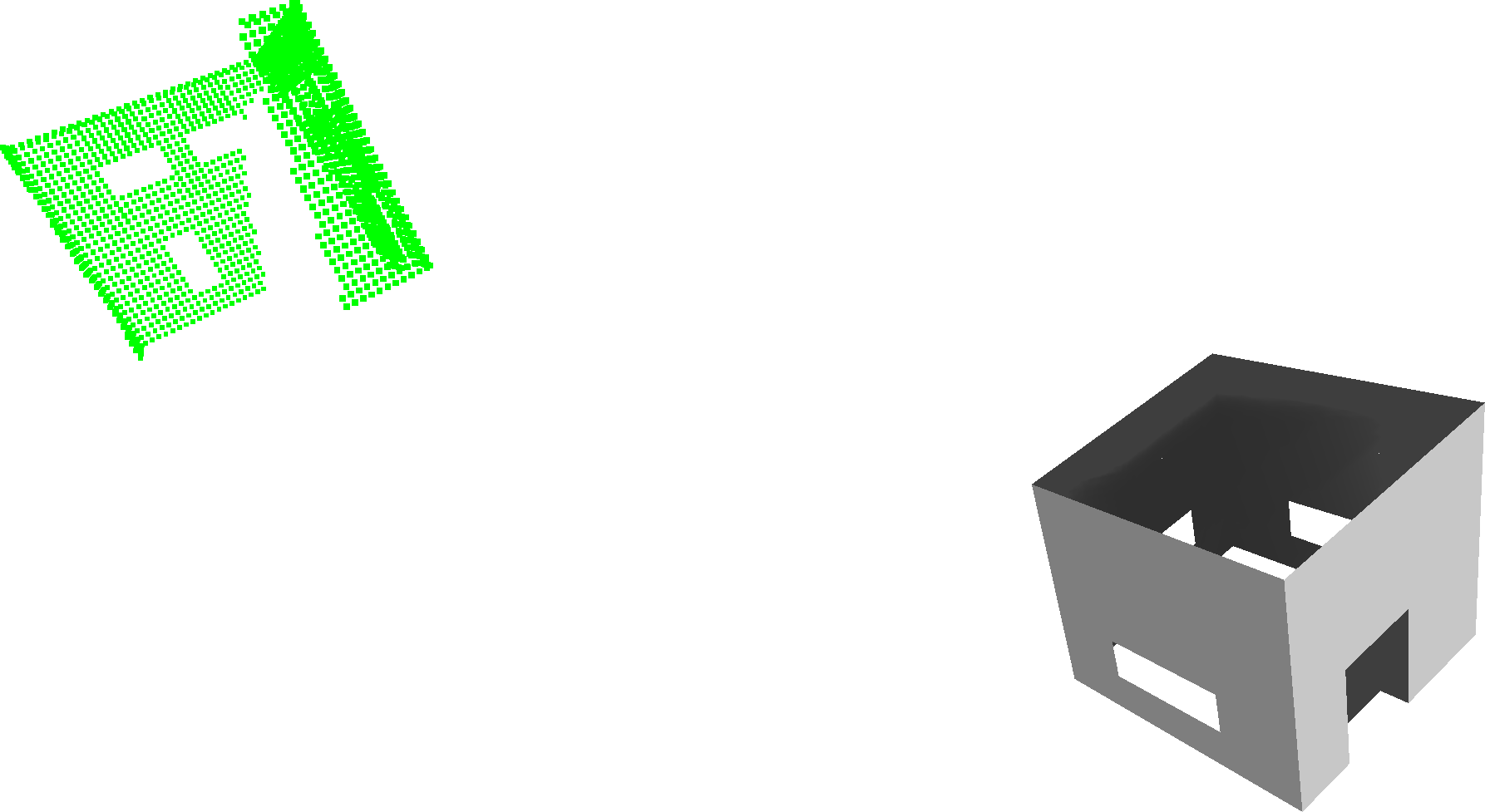}}
\hfill
\subfloat[]{\label{fig:recursive-sample3}\includegraphics[height=0.8in]{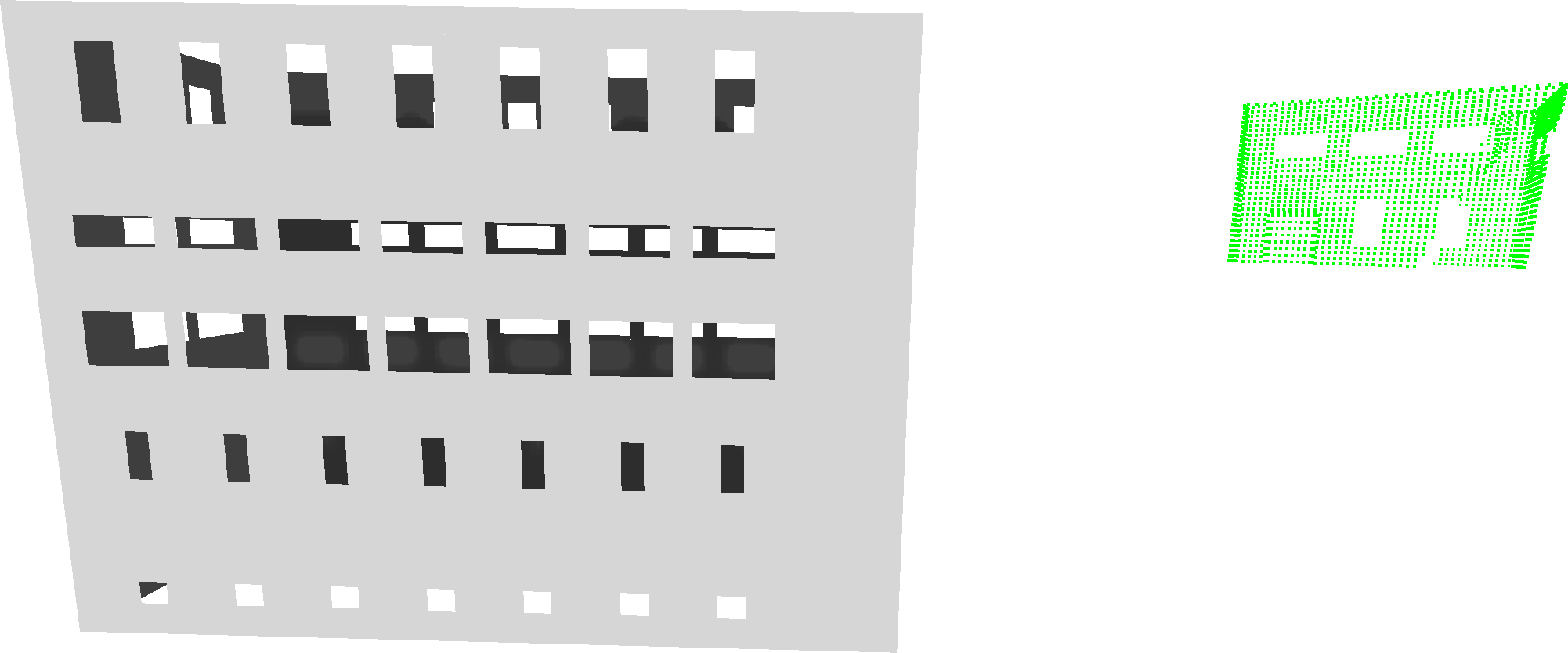}}
\hfill
\subfloat[]{\label{fig:recursive-sample4}\includegraphics[height=0.8in]{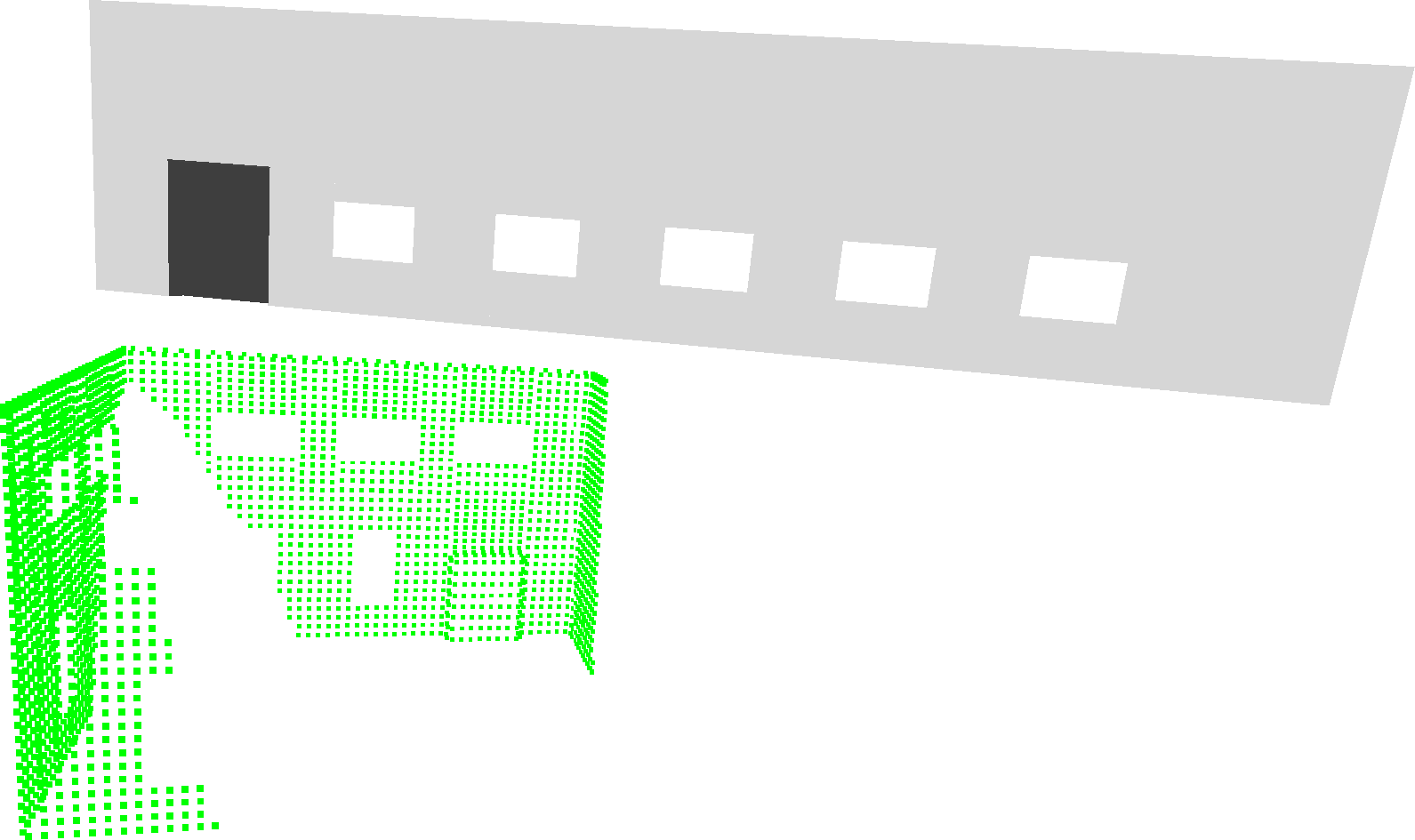}}
\caption{Some sample models of $\mathcal{M}_6$ or $\mathcal{M}_7$ along with $Q_9$ or $Q_{10}$.}
\label{fig:recursivesamplemodels}
\end{figure*}

Fitting a model with varying number of parameters is more difficult than fitting a model with a fixed number of parameters. In this paper, we investigate two models $\mathcal{M}_6$ and $\mathcal{M}_7$ (Figs. \ref{fig:recursive-queries} and \ref{fig:recursivesamplemodels}) with varying numbers of parameters, which are based on the CGA shape grammar \cite{muller2006procedural}. $\mathcal{M}_6$ and $\mathcal{M}_7$ are models of buildings. The models in $\mathcal{M}_6$ consist of 4 facades, while the models in $\mathcal{M}_7$ consist of 1 facade. Let $n$ be the number of floors, $\mathcal{M}_6$ has $7+2n$ parameters, i.e., 1 parameter for rotation, 3 parameters for location, 3 parameters for mass size (height, length and width), $2n$ parameters for window size. $n$ depends on the height of building. Different from $\mathcal{M}_6$, $\mathcal{M}_7$ does not have the width parameter. Windows in the same floor have the same size, but may have different sizes on different floors. Some sample models of $\mathcal{M}_6$ and $\mathcal{M}_7$ are shown in Fig. \ref{fig:recursivesamplemodels}.

\begin{figure*}[ht!]
\centering
\subfloat[]{\label{fig:recursive-model1}\includegraphics[height=0.8in]{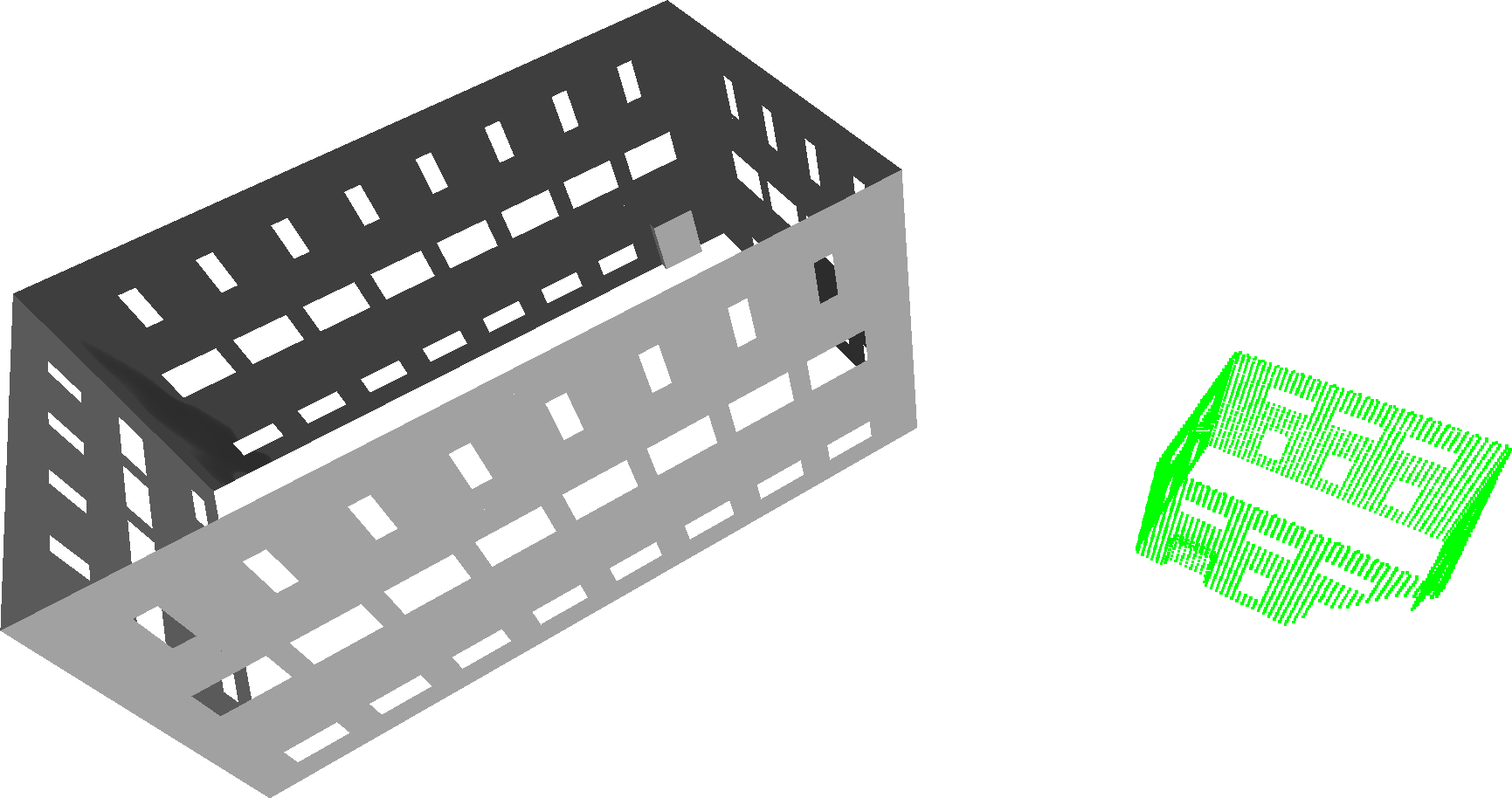}}
\hfill
\subfloat[]{\label{fig:recursive-model2}\includegraphics[height=0.8in]{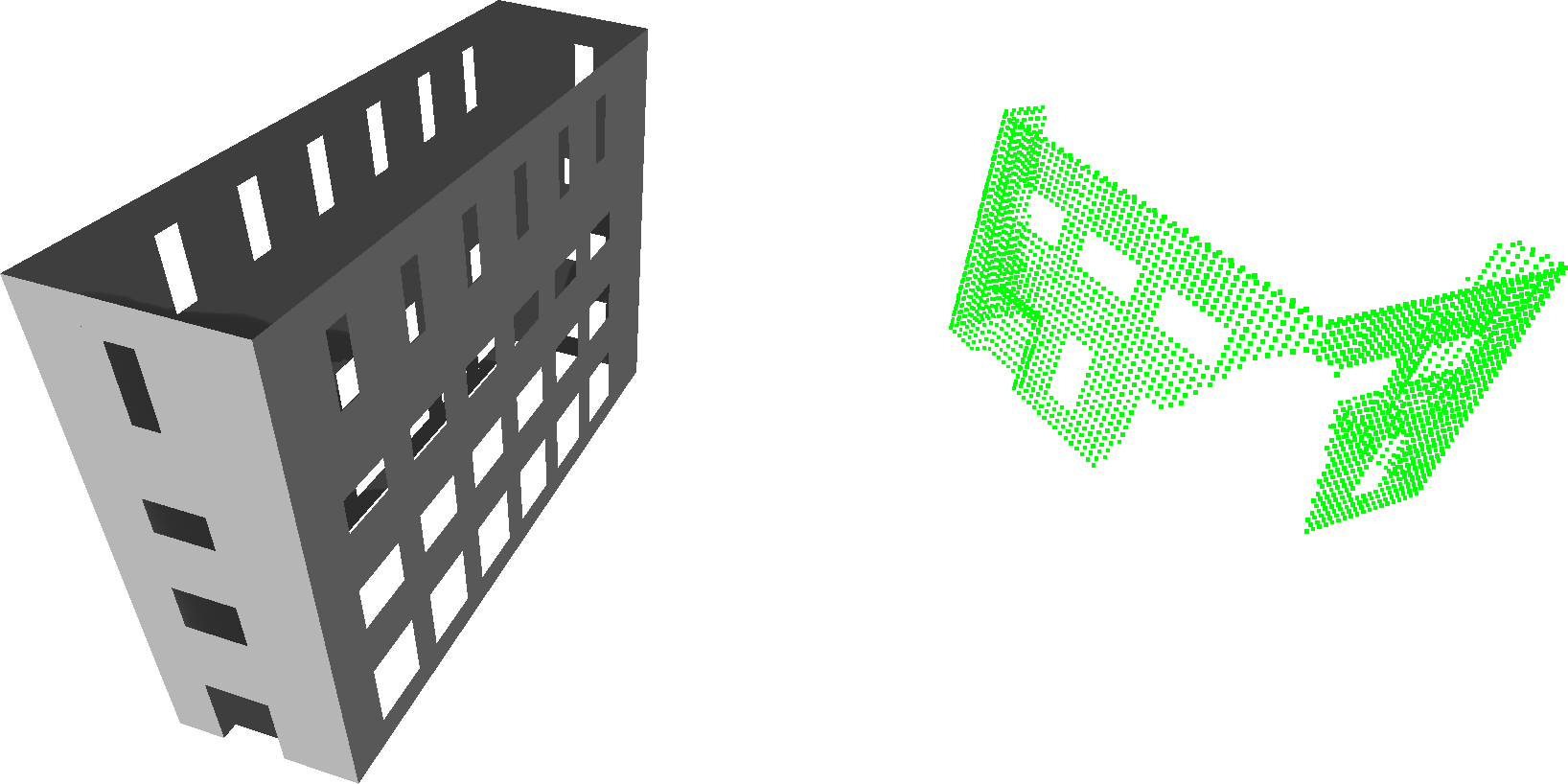}}
\hfill
\subfloat[]{\label{fig:recursive-model3}\includegraphics[height=0.8in]{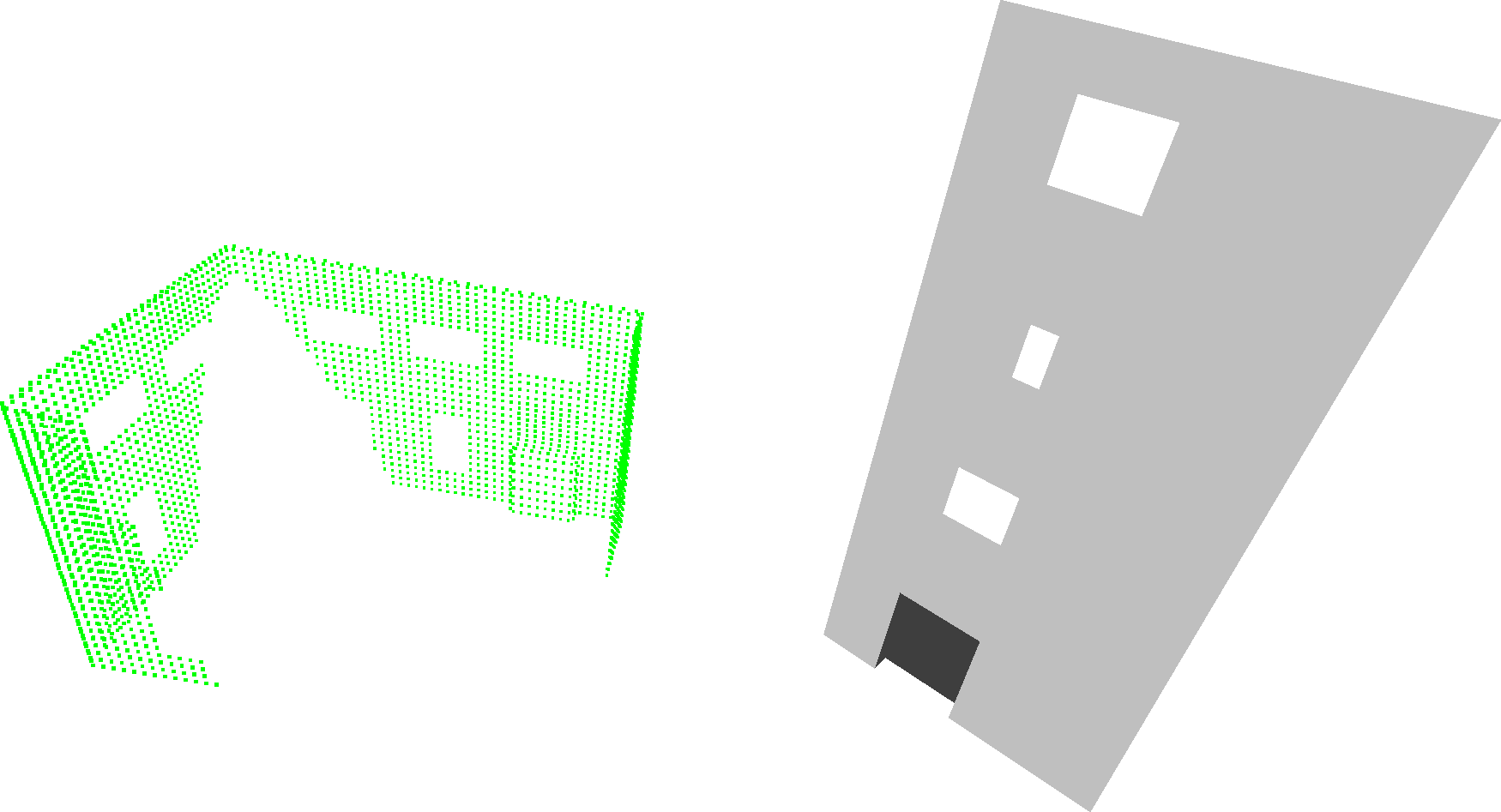}}
\\
\subfloat[]{\label{fig:recursive-model4}\includegraphics[height=0.8in]{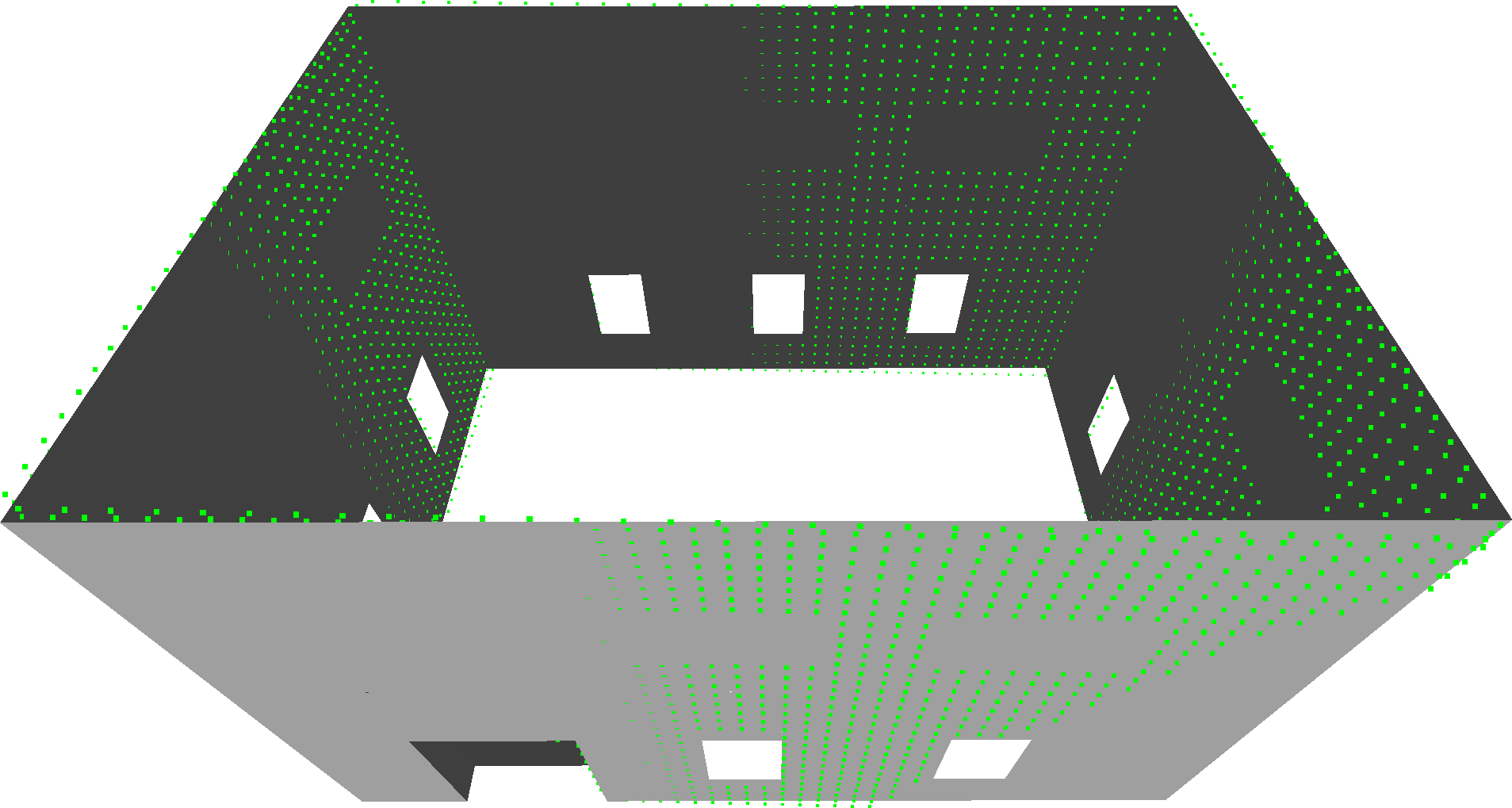}}
\hfill
\subfloat[]{\label{fig:recursive-model5}\includegraphics[height=0.8in]{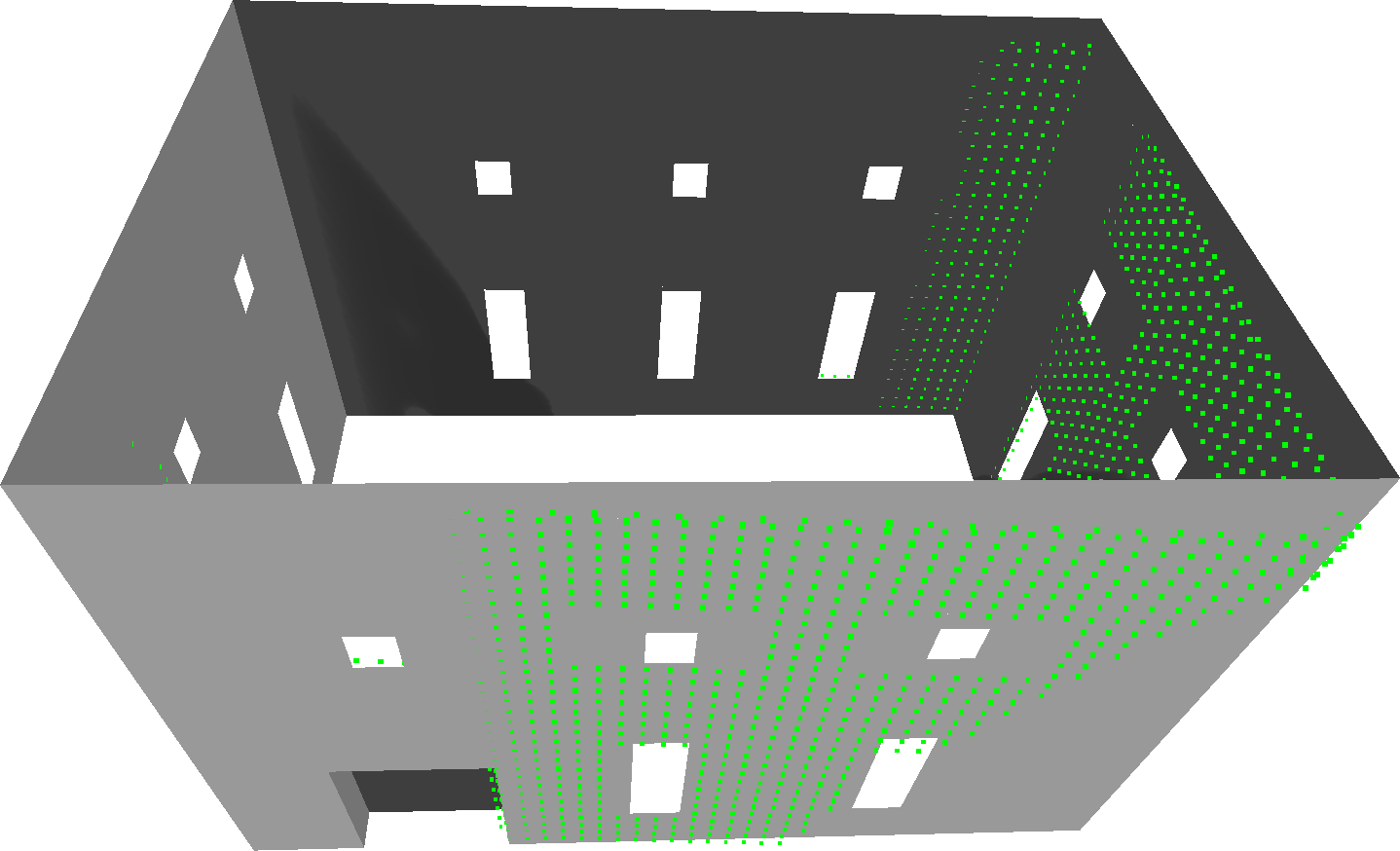}}
\hfill
\subfloat[]{\label{fig:recursive-model6}\includegraphics[height=0.8in]{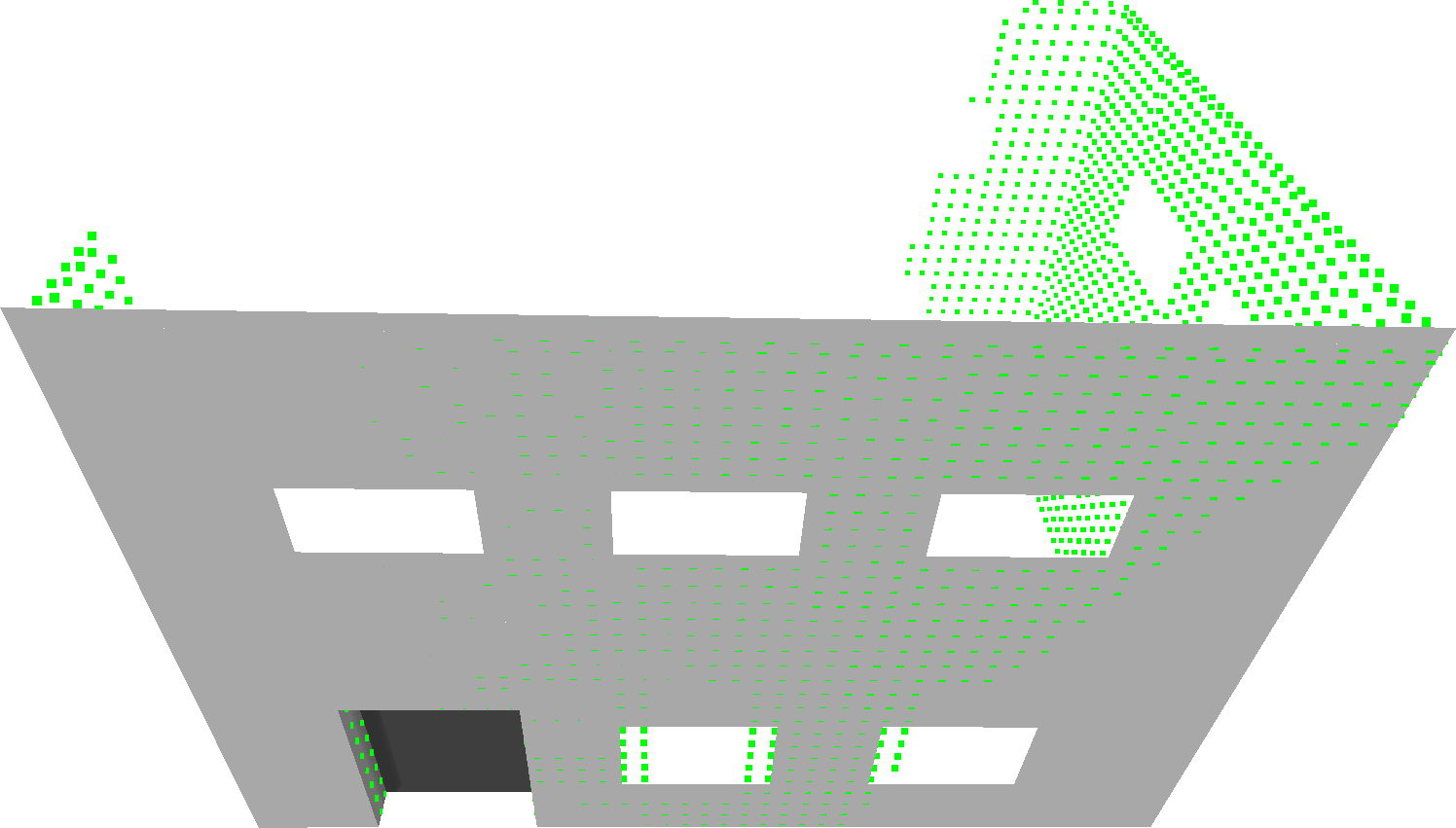}}
\\
\subfloat[]{\label{fig:recursive-model7}\includegraphics[height=0.8in]{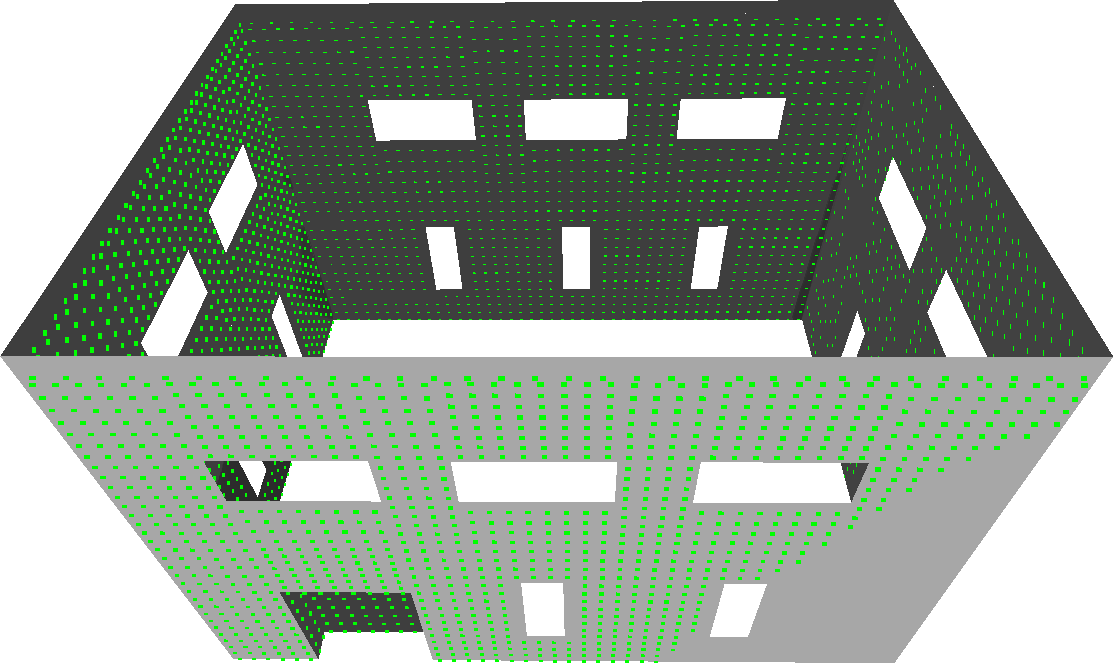}}
\hfill
\subfloat[]{\label{fig:recursive-model8}\includegraphics[height=0.8in]{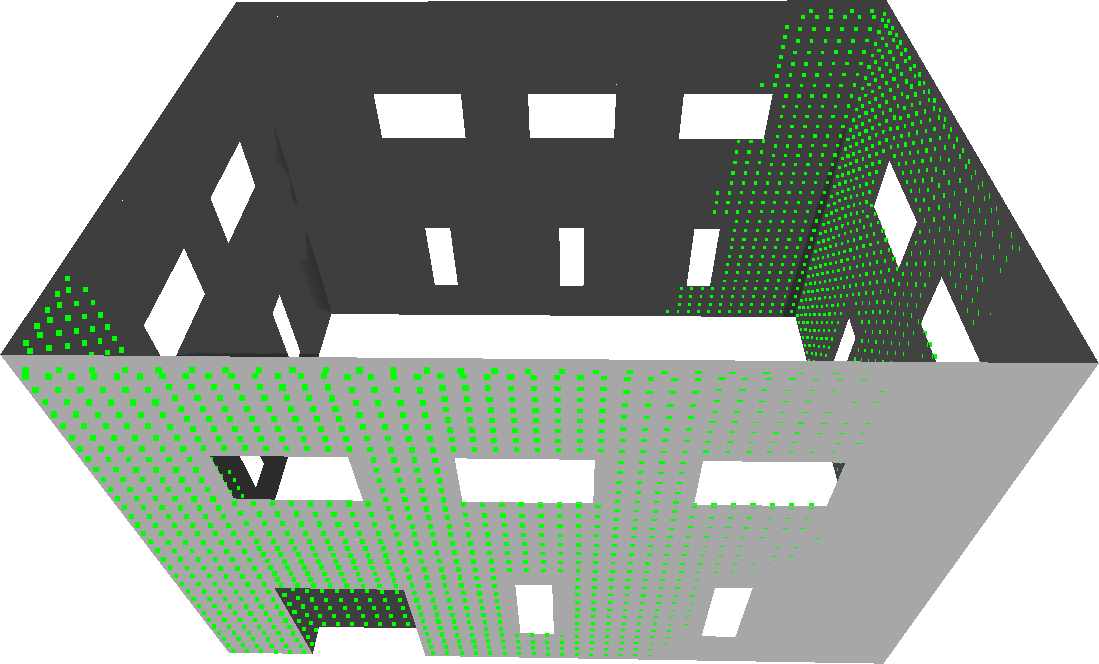}}
\hfill
\subfloat[]{\label{fig:recursive-model9}\includegraphics[height=0.8in]{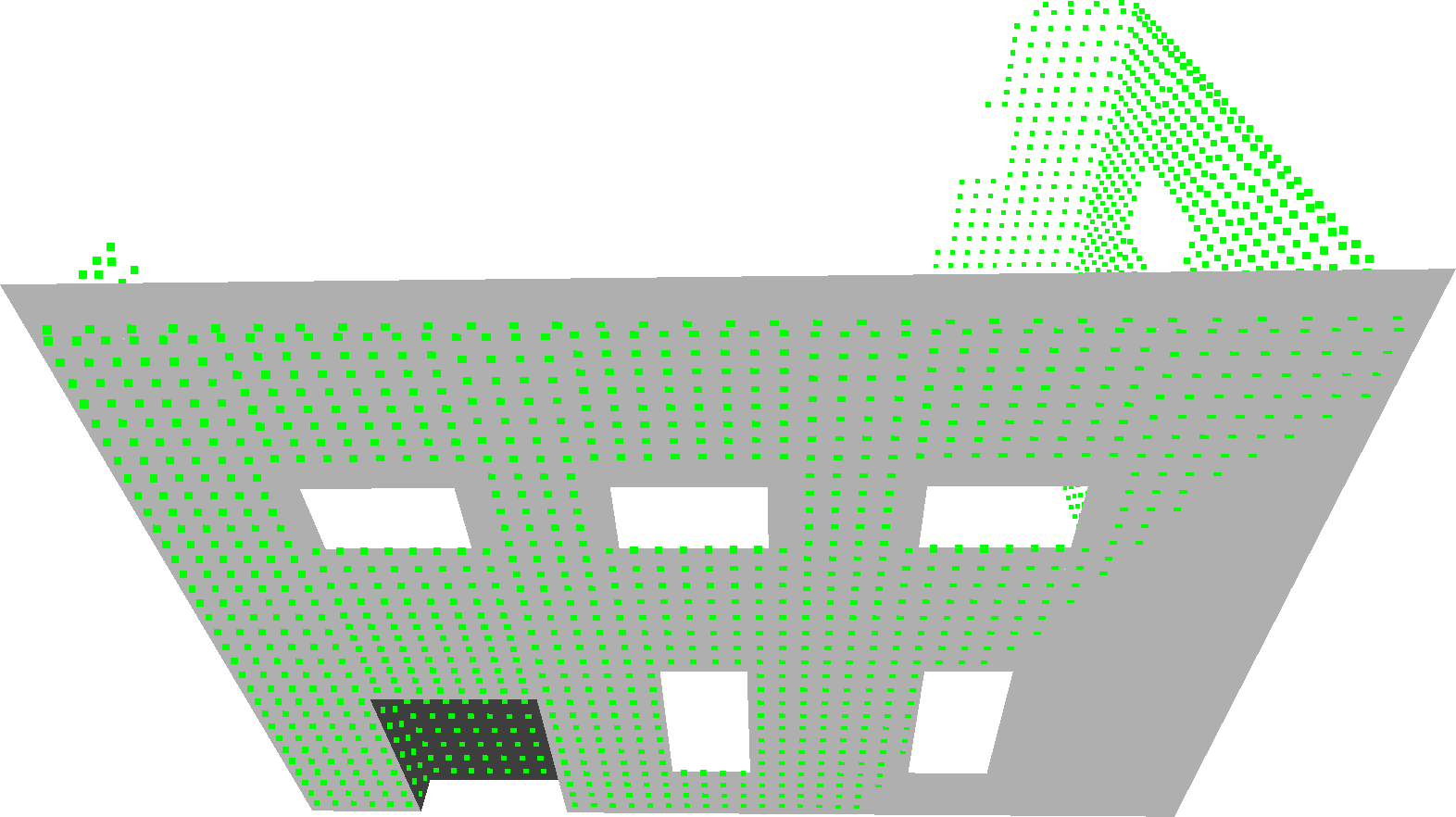}}
\caption{Fitted models. From left to right: results of fitting $\mathcal{M}_6$ to $Q_9$, $\mathcal{M}_6$ to $Q_{10}$, and $\mathcal{M}_7$ to $Q_{10}$. From top to bottom: the randomly initialized models, fitted models after 5000 iterations, and final fitted models.}
\label{fig:recursive-models}
\end{figure*}

\begin{figure*}[ht!]
\centering
\subfloat[]{\label{fig:recursive-log1}\includegraphics[width=0.6\columnwidth]{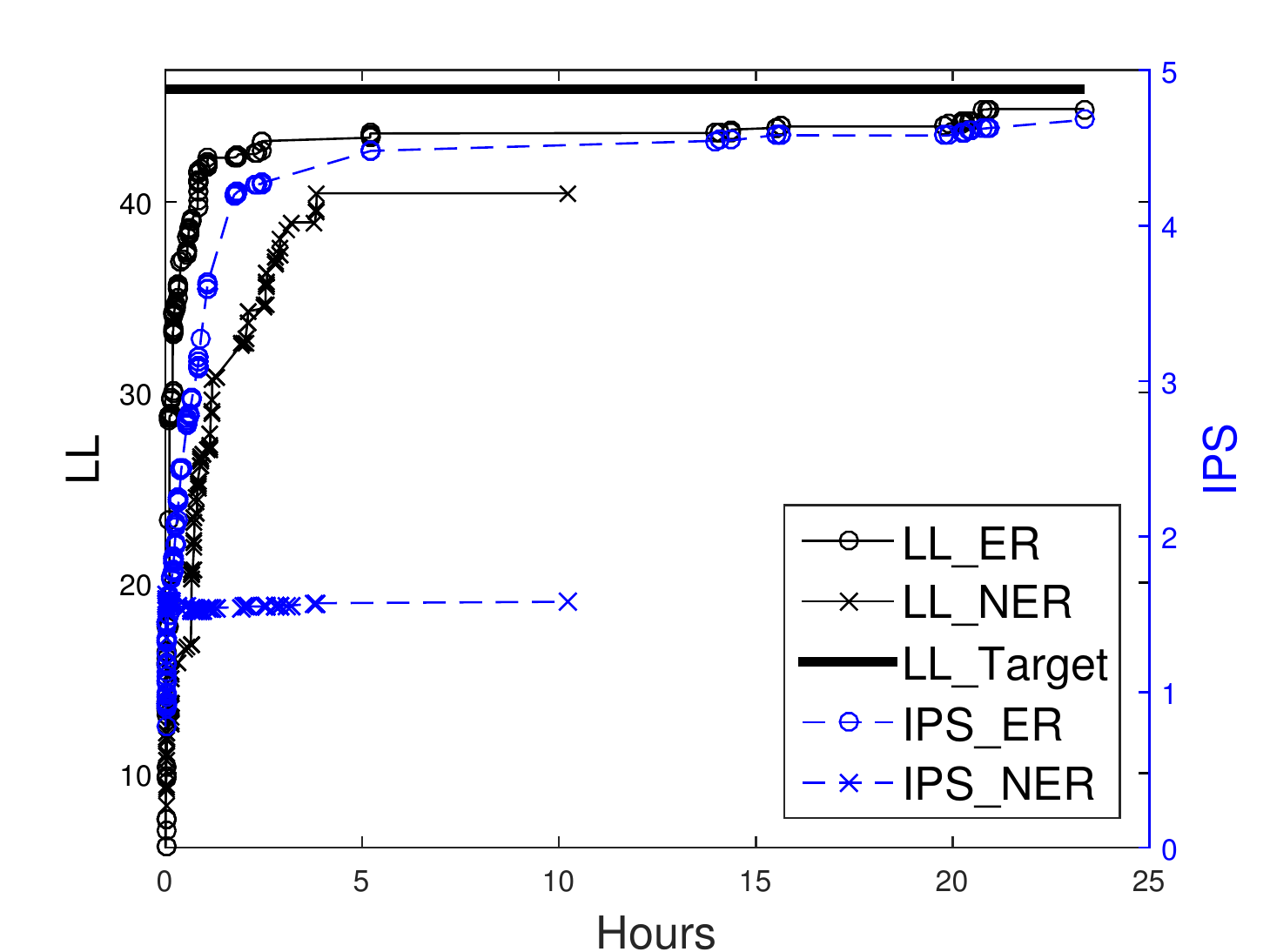}}
\hfill
\subfloat[]{\label{fig:recursive-log2}\includegraphics[width=0.6\columnwidth]{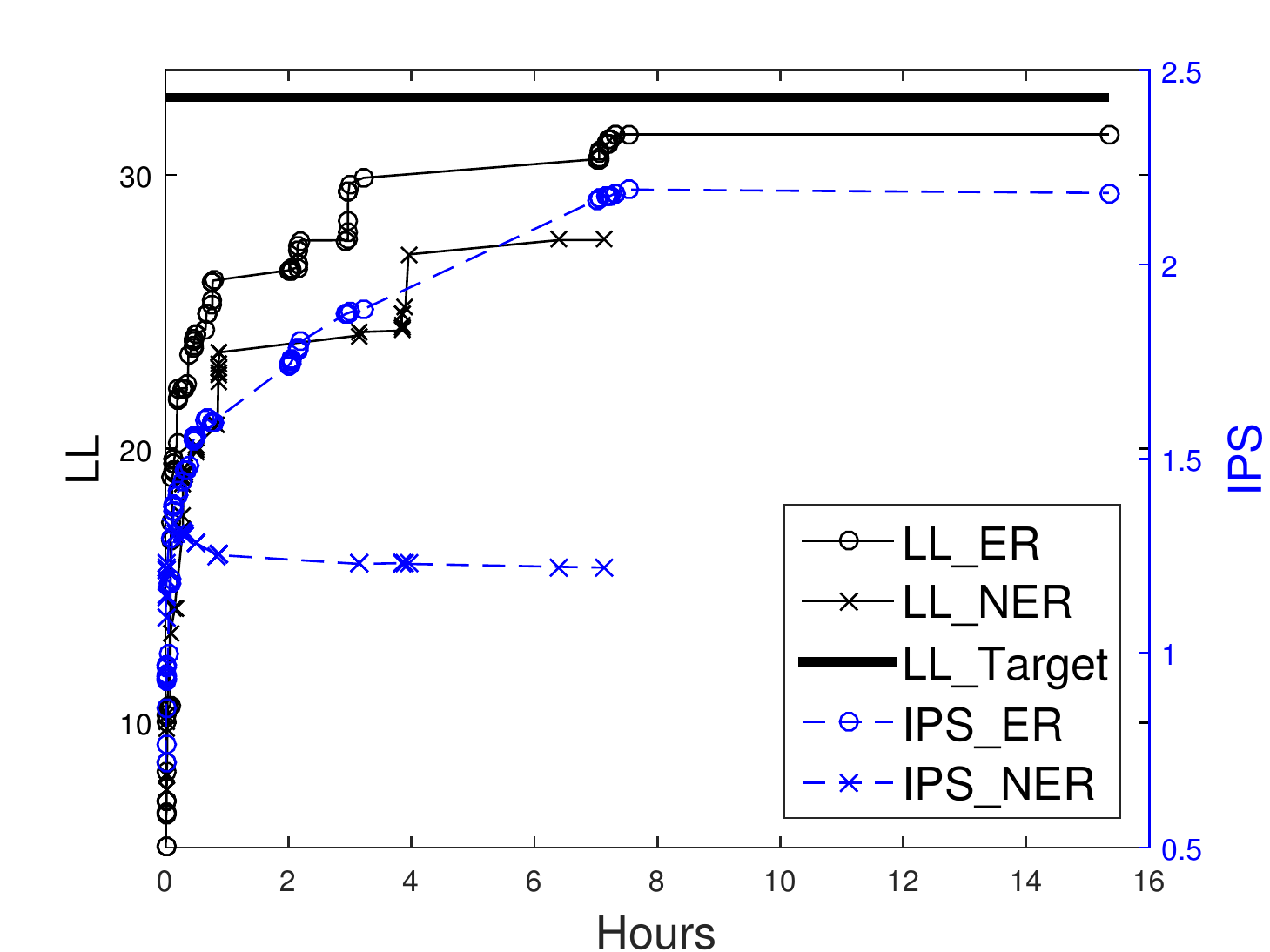}}
\hfill
\subfloat[]{\label{fig:recursive-log3}\includegraphics[width=0.6\columnwidth]{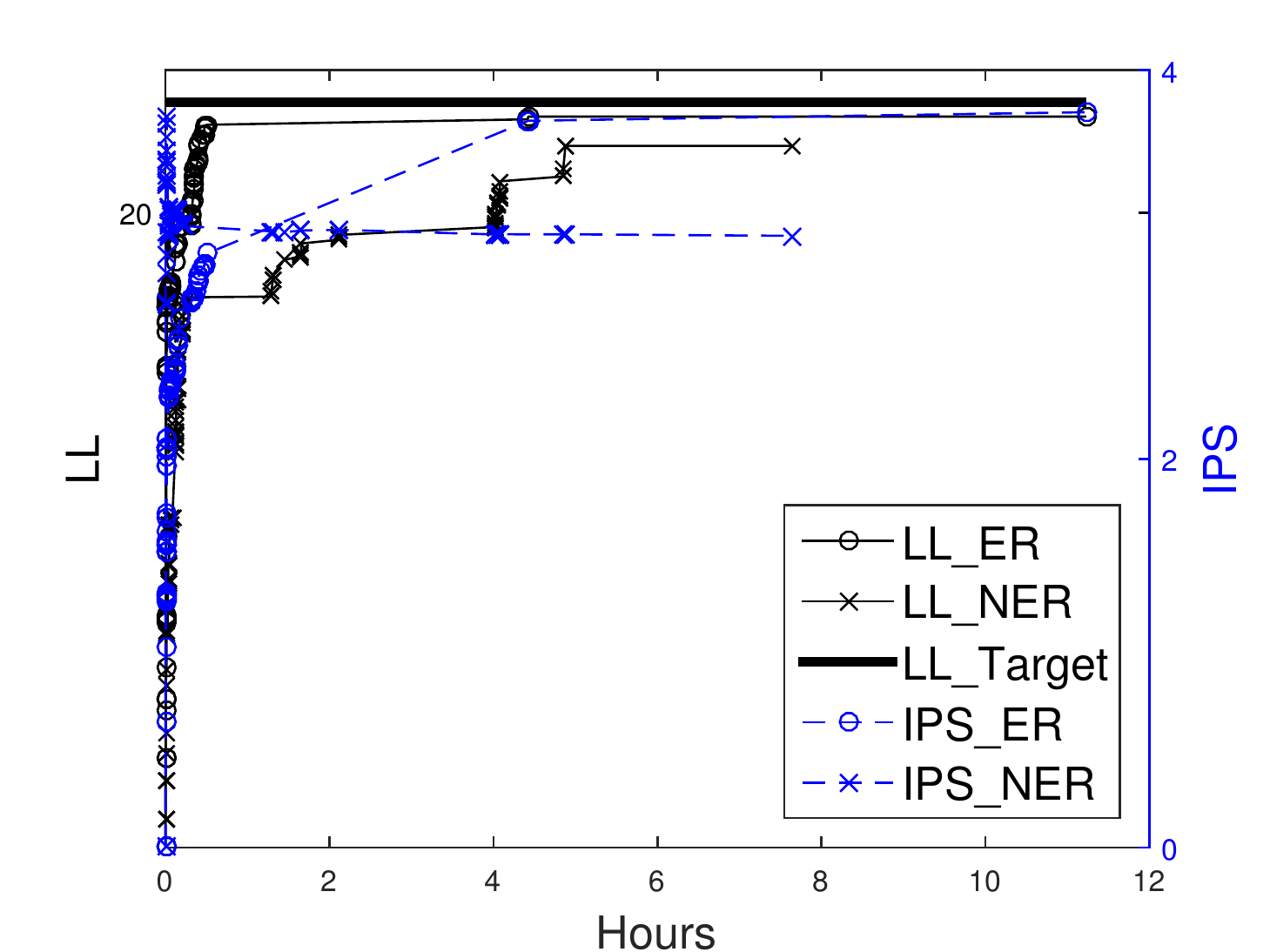}}
\caption{Fitting results. \protect\subref{fig:recursive-log1}, \protect\subref{fig:recursive-log2} and \protect\subref{fig:recursive-log3} are results of fitting $\mathcal{M}_6$ to $Q_9$, fitting $\mathcal{M}_6$ to $Q_{10}$, and fitting $\mathcal{M}_7$ to $Q_{10}$, respectively.}
\label{fig:recursivelogs}
\end{figure*}

The results of fitting $\mathcal{M}_6$ to $Q_9$ (Fig. \ref{fig:recursive-query1}) and $Q_{10}$ (Fig. \ref{fig:recursive-query2}), fitting $\mathcal{M}_7$ to $Q_{10}$ are shown in Figs. \ref{fig:recursive-models} and \ref{fig:recursivelogs}. In these 3 experiments, we set $h=2.5$, $\delta=0.1$. As shown in Fig. \ref{fig:recursive-models}, after 5000 iterations, the mass parameters are correctly estimated, while the window parameters are incorrectly estimated. Finally, all the parameters are correctly estimated. These 3 experiments show that our method is able to fit incomplete data with holes. The experiment of fitting $\mathcal{M}_7$ to $Q_{10}$ additionally demonstrates that our method is able to fit hybrid-complete data. 

%

\subsection{Fitting Laser Scanning Data}\label{sec:laser}
\begin{figure*}[ht!]
\centering
\subfloat[]{\label{fig:xiangan-original}\includegraphics[height=0.9in]{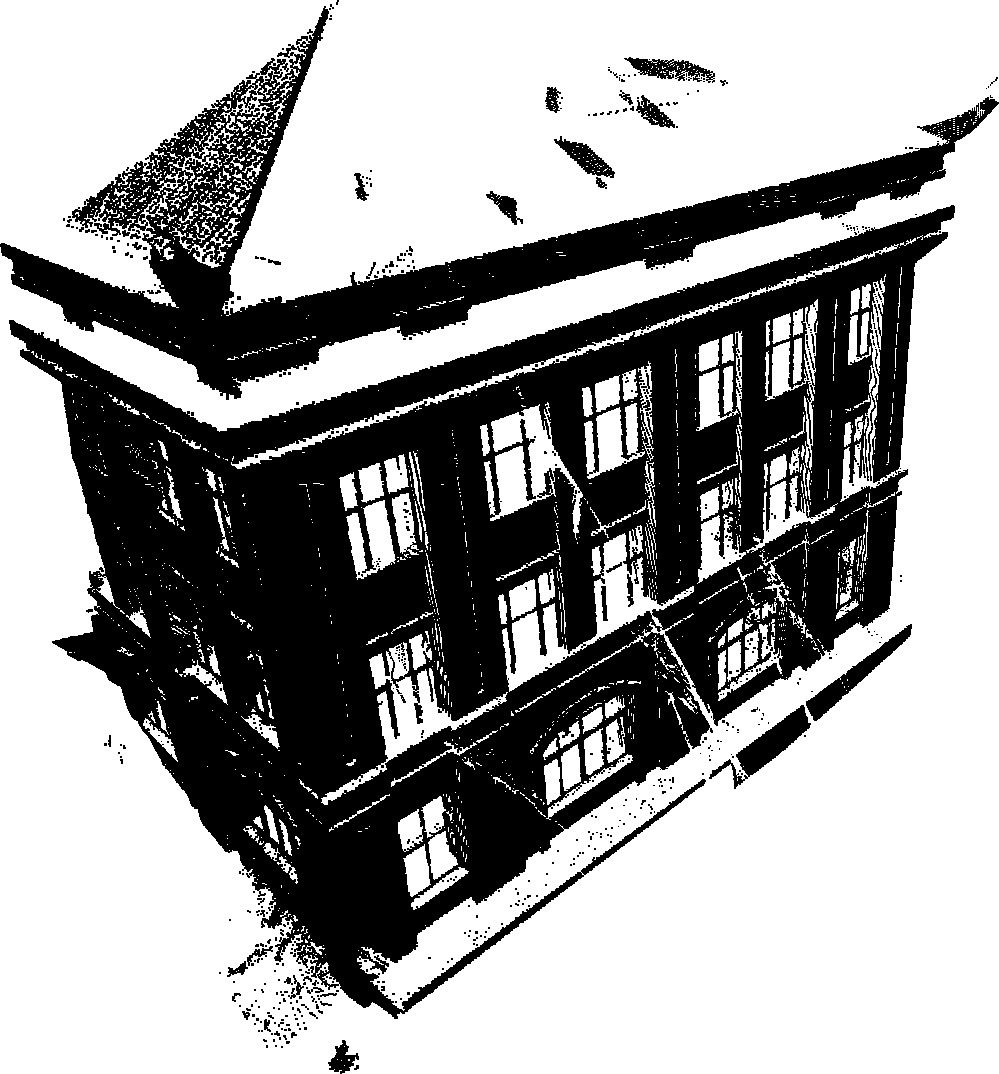}}
\hfill
\subfloat[]{\label{fig:xiangan-query}\includegraphics[height=0.9in]{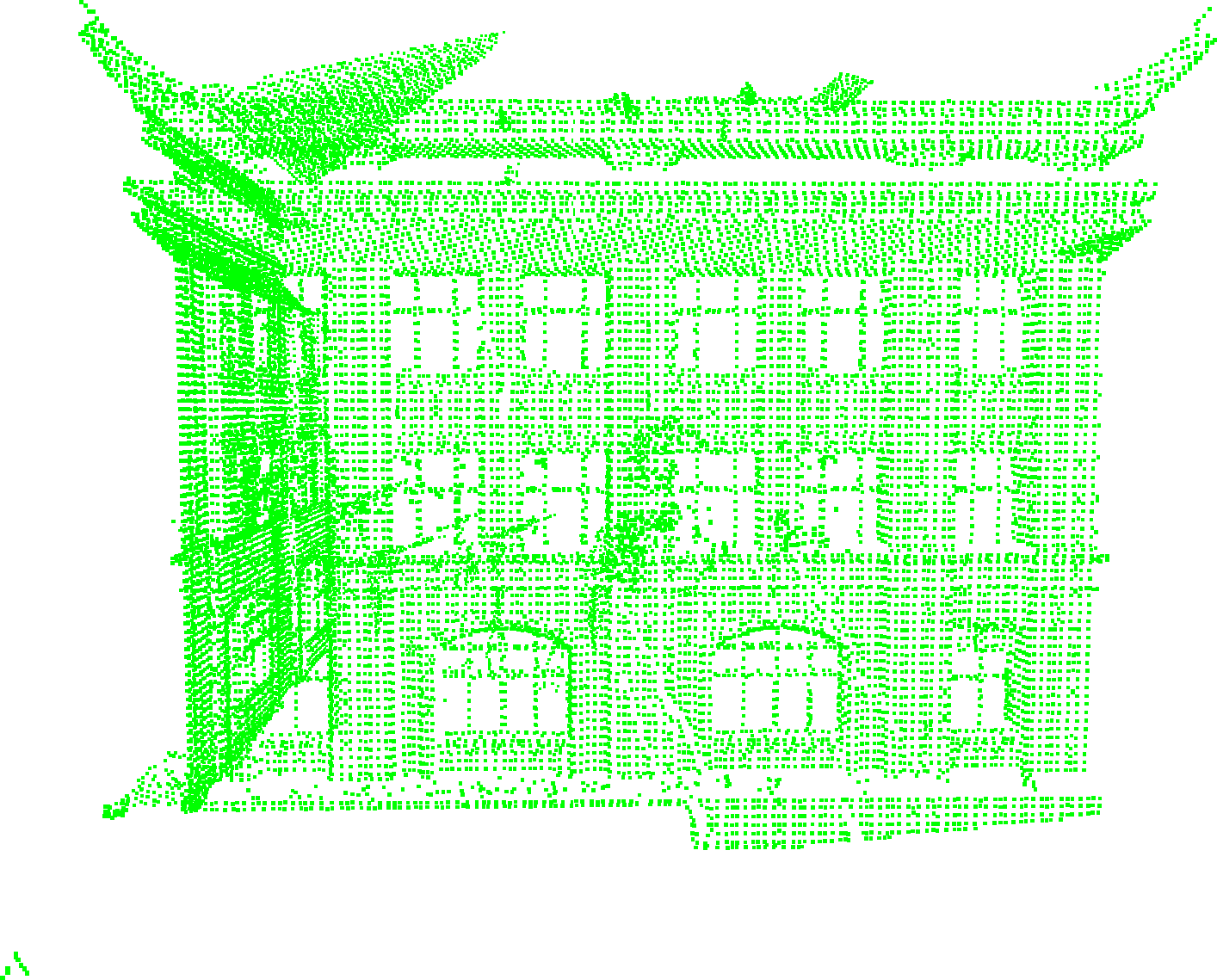}}
\hfill
\subfloat[]{\label{fig:xiangan-model}\includegraphics[height=0.9in]{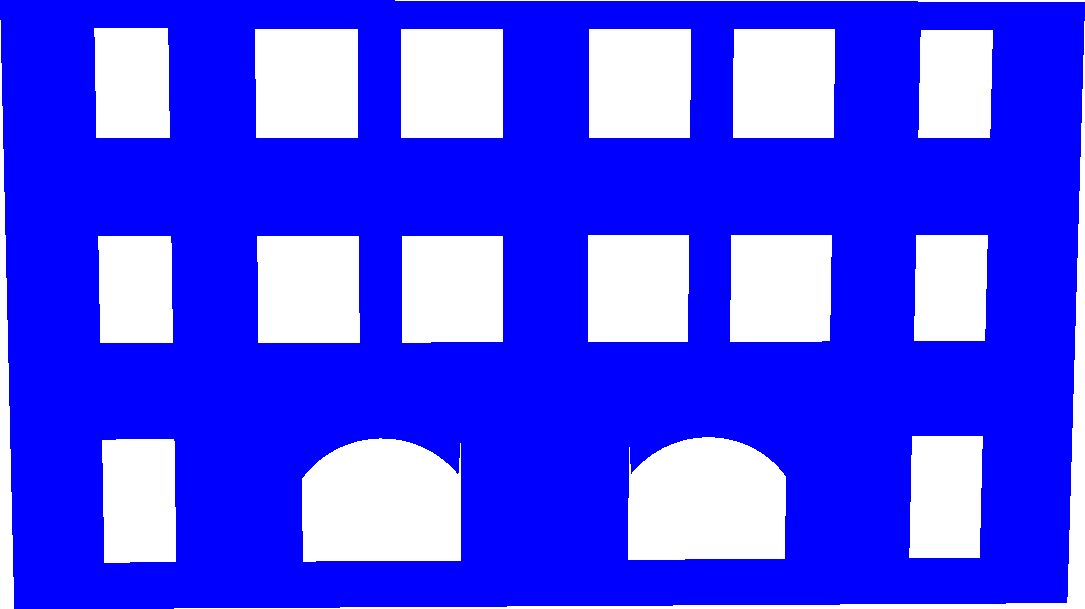}}
\hfill
\subfloat[]{\label{fig:xiangan-overlap}\includegraphics[height=0.9in]{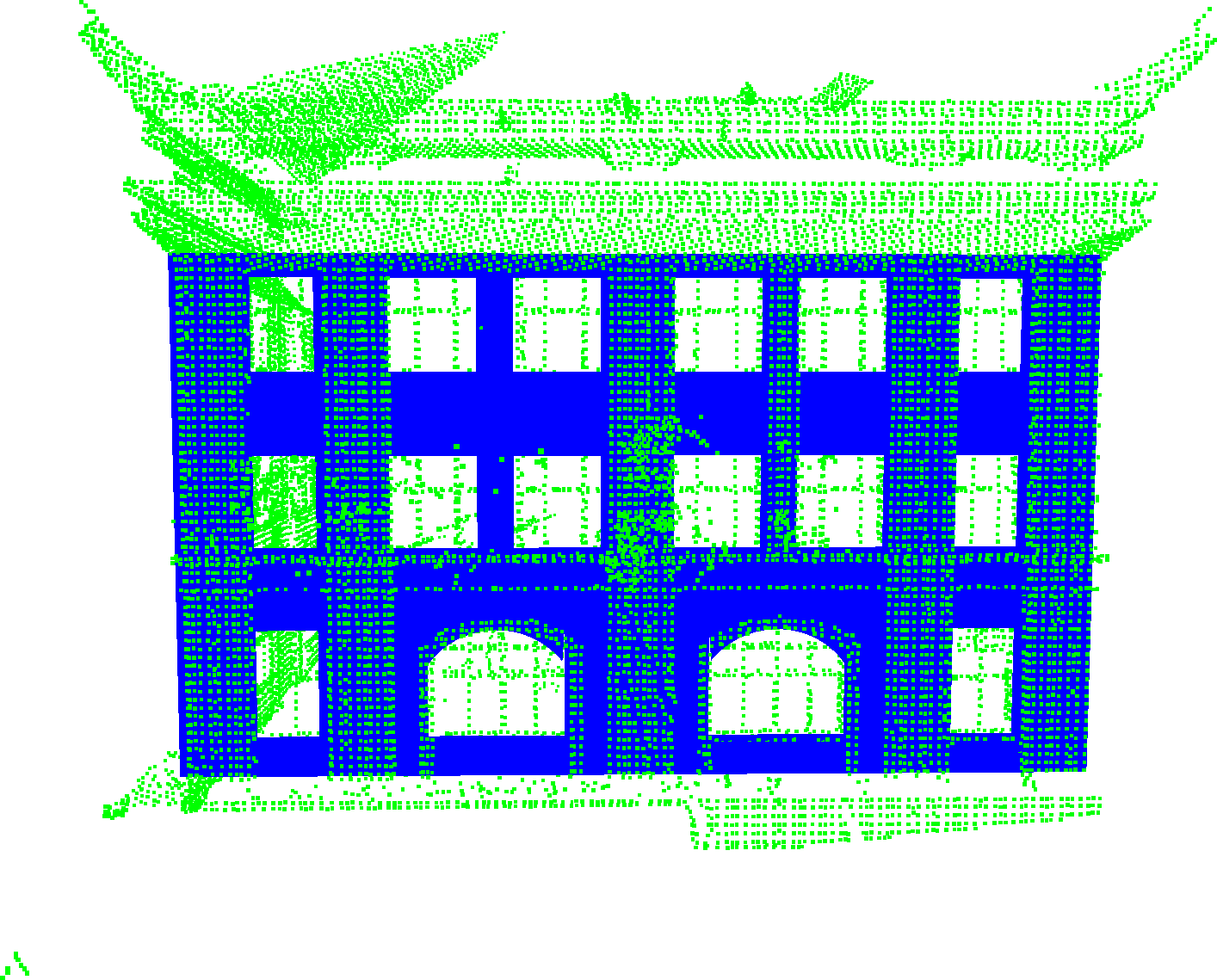}}
\caption{Fitted model. \protect\subref{fig:xiangan-original} An original point cloud consisting of 385793 points. \protect\subref{fig:xiangan-query} A query $Q_{11}$ consisting of 23266 points. $Q_{11}$ is generated by downsampling the original point cloud \protect\subref{fig:xiangan-original} with a resolution of 0.2. \protect\subref{fig:xiangan-model} Final fitted model (after 84540 iterations and 36146.0820 seconds) for fitting $\mathcal{M}_8$ to $Q_{11}$. \protect\subref{fig:xiangan-overlap} Overlap between the query $Q_{11}$ and the final model shown in \protect\subref{fig:xiangan-model}.}
\label{fig:xiangan}
\end{figure*}

We also conducted experiments for fitting real-world laser scanning point clouds, which were collected by mobile laser scanners \cite{guan2014using} \cite{yu2016bag-of-visual-phrases}. The results of fitting a facade model $\mathcal{M}_8$ to a facade point cloud $Q_{11}$ (Fig. \ref{fig:xiangan-query}) are shown in Figs. \ref{fig:xiangan} and \ref{fig:xianganevolution}. $\mathcal{M}_8$ has 15 parameters. Instead of WMM, we use SMM (Eq. \eqref{eq:rs}) to perform this experiment and set $\delta=0.08$. In contrast to WMM, no argument $h$ is needed to calculate SMM. We tested WMM and found that WMM is not very effective for $Q_{11}$ due to that the holes in $Q_{11}$ are corrupted. That is, there are undesired points within the holes. These corrupted holes are incorrectly recognized as missing data by WMM. In other words, although WMM is able to distinguish the difference between uncorrupted holes and missing data (as shown in Section \ref{sec:recursive}), WMM is unable to distinguish the difference between corrupted holes and missing data. Fortunately, as shown in the results, SMM is able to deal with corrupted holes.

\section{Conclusion}\label{sec:conclude}
We have proposed a novel rigid geometric similarity metric to measure the similarity between geometric models. Based on the proposed metric, we presented the first method to rigidly fit arbitrary geometric model to non-complete data. We formulate the fitting problem as a Bayesian inference problem and employ MCMC technique to perform the inference. We also proposed a novel technique to accelerate the inference process. Our method has been demonstrated on various geometric models and non-complete data. Experimental results show that our metric is effective for fitting non-complete data, while other metrics are ineffective. It is also shown that our method is robust to noise. In summary, our method is able to fit non-complete data without holes (Section \ref{sec:sphere}), non-complete data with uncorrupted holes (Section \ref{sec:recursive}) or over-complete data with corrupted holes (Section \ref{sec:laser}).

We believe our work bridges the gap between inverse procedural modeling and geometric model fitting. However, several issues still remain open. For example, due to the curse of dimensionality, the fitting problem becomes intractable if the  geometric model has a large number of parameters. New techniques such as deep learning \cite{nishida_interactive_2016} \cite{ritchie2016neurally} are expected to address this problem. Besides, it is also challenging for our method to fit incomplete data with corrupted holes, because a corrupted hole may incorrectly be recognized as missing data. This problem may be addressed using some preprocessing process, e.g., filtering out the undesired points within each hole.

\section{Acknowledgements}
This work was supported by the National Natural Science Foundation of China under Grants 41471379, 61602499 and 61471371, and by Fujian Collaborative Innovation Center for Big Data Applications in Governments. The authors would like to thank the reviewers for their comments.


\bibliographystyle{acmsiggraph}
\bibliography{ppgmf.bib}
\end{document}